# High-areal-capacity Na-ion battery electrode with uncompromised energy and power densities by simultaneous electrospinning-spraying fabrication

Dr Mengzheng Ouyang [1, 11, 12, *], Dr Zhenyu Guo [2, 11], Dr Luis E Salinas-Farran [1], Dr Yan Zhao [3], Dr Siyu Zhao [4, 5], Kaitian Zheng [2, 6], Dr Hao Zhang [5, 7], Dr Mengnan Wang [2], Dr Guangdong Li [1, 8], Feiran Li [1], Prof. Xinhua Liu [9], Prof. Shichun Yang [9], Prof. Fei Xie [10], Prof. Paul R. Shearing [5], Prof. Maria-Magdalena Titirici [2], Prof. Nigel P. Brandon [1]

1. Department of Earth Science and Engineering, Imperial College London, London, UK
2. Department of Chemical Engineering, Imperial College London, London, UK
3. Department of School of Energy and Engineering, Jiangsu University, Zhenjiang, China
4. Department of Chemical Engineering, University College London, London, UK
5. Department of Engineering Science, University of Oxford, Oxford, UK
6. School of Chemical Engineering and Technology, Tianjin University, Tianjin, China
7. Department of Chemical Engineering, Massachusetts Institute of Technology, Cambridge, MA, USA
8. School of Chemistry and Chemical Engineering, Beijing Institute of Technology, Beijing, China
9. School of Transportation Science and Engineering, Beihang University, Beijing, China
10. Institute of Physics, Chinese Academy of Science, Beijing, China

[11] These authors contributed equally

[12] Lead contact

* Correspondence: mo113@ic.ac.uk M.O.

## Summary

Sodium-ion batteries (SIBs) are cost-effective alternatives to lithium-ion batteries (LIBs), but their low energy density remains a challenge. Current electrode designs fail to simultaneously achieve high areal loading, high active content, and superior performance. In response, this work introduces an ideal electrode structure, featuring a continuous conductive network with active particles securely trapped in the absence of binder, fabricated using a universal technique that combines electrospinning and electrospraying (co-ESP). We found that the particle size must be larger than the network's pores for optimised performance, an aspect overlooked in previous research. The free-standing co-ESP $Na_2V_3(PO_4)_3$ (NVP) cathodes demonstrated state-of-the-art 296 mg cm$^{-2}$ areal loading with 97.5 wt.% active content, as well as remarkable rate-performance and cycling stability. Co-ESP full cells showed uncompromised energy and power densities (231.6 Wh kg$^{-1}$/7152.6 W kg$^{-1}$), leading among reported SIBs with industry-relevant areal loadings. The structural merit is analysed using multi-scale X-ray computed tomography, providing valuable design insights. Finally, the superior performance is validated in the pouch cells, highlighting the electrode's scalability and potential for commercial application.

## Keywords

Sodium-ion batteries, cathode, electrode design, electrospinning, electrospraying, pouch cell, high areal loading, high rate performance, X-ray computed tomography.

## Introduction

Sodium-ion batteries (SIBs) have emerged as a cost-efficient and sustainable alternative of lithium-ion batteries (LIBs)[1]. However, their application is significantly hindered by the lower energy density of existing cathode materials[2]. $Na_2V_3(PO_4)_3$ (NVP) is recognised as one of the most promising cathode candidates due to its high working voltage, high Na$^+$ conductivity and superior cycling stability[3]. Yet, it suffers from low electron conductivity and a limited theoretical capacity of 117 mAh g$^{-1}$. Additionally, most reported SIBs have areal loadings far below industrial demands, with high-areal-loading SIB electrodes typically around 10 mg cm$^{-2}$ [4-6] and a maximum of 60 mg cm$^{-2}$ [7], compared to up to 170 mg cm$^{-2}$ for LIBs[8,9]. This discrepancy is due to the underdeveloped state of cathode materials and structures, further widening the energy density gap between SIBs and LIBs.

There are three effective strategies to enhance an electrode's energy density[10,11]: (i) applying high active materials areal loading, (ii) eliminating the current collector, and (iii) increasing the active materials content (weight ratio of active materials in the whole electrode). All these strategies require carefully designed electrode microstructures. Conventional structures, with randomly-aligned polymeric binders of reasonable weight content, are not strong enough to support such electrodes[9,12-14]. Their highly tortuous electron and ion transportation pathways significantly lower the electrochemical performances[15].

No reports to date have detailed a high-performance electrode design successfully implementing all three strategies[10,16] (free-standing electrode with >50 mg cm$^{-2}$ areal loading and > 95 wt.% active content). Those implement one or two of these strategies often sacrifice power densities and stabilities to achieve high energy densities.[9,17].

The challenges stem from the inherent trade-offs in existing electrode structures:

- An electrode with high active content struggles to achieve high areal loading/free-standing structure [9,10,16], or to achieve good performance, due to the insufficient structure support and electron conductivity. Vice versa, free-standing and high-areal-loading electrodes usually have active contents below 80 wt.%[18-20].
- Achieving high electron and ion conductivity simultaneously is challenging[21,22]. A high content of carbon black/binders will inevitably reduce the porosity and increase pore tortuosity.

Clearly, it is difficult to further enhance cell performances based on the existing electrode structures. Any novel structures need to meet the following prerequisites: a highly conductive network with both horizontal and vertical robustness, a low-tortuosity pore network that can access all the particle surface, and active particles that are evenly distributed and firmly attached to the conductive network.

While electrospinning is an ideal and scalable method to fabricate such networks[23,24], an optimal approach for introducing active particles has not been identified. Particles introduced within the electrospun fibres[25] would leave unnecessarily high porosity (>90%) and thus low energy density[23,26]. Introducing the particles into the electrospun network[27] would exclude the use of commercially-available large particles, and would require additional binder/conductive additives, leading to inferior performance[28,29]. Both routes have resulted in electrodes with lower active contents than the conventional electrodes, with no obvious improvement in energy/power density.

In this study, we synthesised an ideal Na-ion battery electrode structure by introducing the active particles through electrospraying simultaneously with electrospinning, a method termed co-electrospinning-electrospraying, or co-ESP. Both methods are highly scalable techniques[30].

While previous efforts of combining electrospinnig and spraying did not fabricate electrodes with state-of-art performance[31,32], this work shows that the overlooked particle size effect and the absence of binder/conductive additives are keys to achieving good performance. When the electrosprayed particles are significantly larger than the pores of the electrospun fibre network, they are strongly bound through spatial constrictions without binders, promoting the interphase contact while exposing the particle surfaces to electrolyte. This allows a carbon nanotube-embedded carbon nanofibre (CNTF) network to function as the conductive additive, binder and current-collector with only 2.5 wt.% content.

The synthesised high-active-content, free-standing and binder-free electrodes for Na-ion batteries met all the prerequisites and showed one of the best performances at high-areal loading among all reported Li-ion and Na-ion battery electrodes. With 97.5 wt.% carbon-coated $Na_2V_3(PO4)_3$ (NVPC) content, the electrodes demonstrated record-high stable areal loading (up to 296 mg cm$^{-2}$ for NVPC, 120 mg cm$^{-2}$ for hard carbon) and rate performance (200 C at 4 mg cm$^{-2}$ and 5 C at 296 mg cm$^{-2}$). The electrodes exhibited low polarisation, high capacity retention and cycling stability, and state-of-art energy density/power densities across all areal loadings, in both half-cells and full cells. With the assist of multiscale synchrotron-based X-ray computed tomography, we found that the superior performance was linked with the ideal pore structures, high electron accessibility, and hierarchically porous particles. Finally, pouch cells with capacities up to 200 mAh were assembled using co-ESP electrodes, demonstrating their scaling-up potential.

## Results and discussion

### *The co-ESP fabrication of electrodes*

**Figure 1a** illustrates the schematics of the co-electrospinning-spraying (co-ESP) set-up. The electrospinning slurries were mixtures of polyacrylonitrile (PAN), and carbon nanotube (CNT) in a dimethylformamide (DMF) solvent, where PAN served as the electrospinning carrier and carbon precursor. The electrospraying slurries were mixtures of polyethylene oxide (PEO) and commercial NVPC particles in DMF, with PEO functioning as both the electrospraying carriers and dispersant.

The areal loadings (thicknesses) of the electrodes were controlled by the total volume of slurries, and the active contents were controlled by the volume ratios of electrospinning/spraying slurries. The detailed fabrication process is summarised in the Method section and illustrated in **Figure S1,** with an accompanying video provided in **Video S1.**

In contrast to conventional electrosprayed battery electrodes, which use metal salt solutions[33], we employed a highly concentrated suspension. This accelerated the electrospraying of NVPC by over tenfold [34], aligning its rate with the electrospinning to achieve the desired active content. This change also ensured a comparable fabrication time with the conventional slurry-casting and drying process.

Following co-ESP, the mats were calcined to remove the PEO and pyrolyse the PAN. The resulting electrodes, shown in **Figure 1b**, exhibited remarkable flexibility and a fabric-like texture. Notably, our lab-scale process is capable of fabricating electrodes with up to 600 cm$^2$ per batch (**Figure 1b**), enough for 300 CR2032 coin cells, highlighting co-ESP's strong potential for scaling up.

Regarding morphology, **Figure 1c** shows that the co-ESP electrode consisted of a percolating, inter-supported network of CNT-embedded CNF (CNTF), which homogeneously encapsulated the NVPC cathode particles. The morphologies of NVPC particles and co-ESP NVPC electrodes, both as-fabricated and post-pyrolysis, are detailed in **Figure S2**. The fibres, with diameters of approximately 150 nm and lengths of up to 1 cm (**Figure S3**) [35], are two orders of magnitudes longer than the CNTs typically used in battery electrodes[9,36]. Such a high length-to-diameter ratio is beneficial to the formation of robust supporting backbones at low mass contents[37,38], while leaving sufficient porosity. CNT was embedded into the PAN-derived CNF, with a weight content of 40 wt.% [39,40] (**Figure S4**, **S5**), enhancing the network's conductivity by an order of magnitude (**Figure S6**). The Raman spectrum of co-ESP NVPC and hard carbon electrodes indicated a higher degree of graphitisation than typical PAN-derived carbon pyrolysed at the same temperature (**Figure S7)**.

*The size effect of co-ESP electrodes*

Notably, the co-ESP method allowed for the introduction of particles larger than the pores of the fibre network (**Figure 1e**). This enabled strong particle binding through the spatial constraints of the network, thus eliminating the need for binders and conductive additives. The average NVPC particles size was 20 µm, significantly larger than the average pore size of the CNTF network (2 µm) (**Figure S8, S9**). Consequently, this network fulfils the roles of conductive additive, binder, and current collector, while also providing adequate porosity for electrolyte immersion. In contrast, when particles smaller than the pore sizes were introduced, as shown in **Figure 1d**, they were poorly bound to the network, while resulting in insufficient electrical contact (**Figure 1f**).

To confirm the importance of particle sizes, small NVPC particles were produced by mildly ball-milling the pristine NVPC particles (**Figure 1d**). The average diameter of ball-milled particles (300 nm) was well below the average pore size of electrospun fibre network (2 µm) (**Figure S9**). The ball-milling did not induce any crystal structure changes or impurities (**Figure S10**), which can occur under harsher milling condition[41]. The crystalline sizes remained unchanged, confirming that the ball-milling only broke down the secondary particles[42]. Na-ion half-cells were assembled with pristine and ball-milled co-ESP NVPC cathodes of 18 mg cm$^{-2}$, 97.5 wt.% active loading (1.5 wt.% CNF/1 wt.% CNT) (**Figure 2a**).

Both electrodes showed near theoretical initial discharge capacities (~110 vs 117 mAh g$^{-1}$) under 0.1C. However, the ball-milled electrode showed a lower initial coulombic efficiency (ICE) of 93.8% compared to 97.7% of the pristine electrode (**Figure S11**) likely due to more sodium being consumed in the formation of the cathode-electrolyte interface (CEI) on the ball-milled particles[43]. Additionally, ball-milled electrode demonstrated poorer rate performance (**Figure 2b**). The pristine co-ESP NVPC electrode retained 72.1% and 36.8% of its theoretical capacity at 5C and 20C, respectively, while the ball-milled electrode retained 47.8% and 0%, respectively. The conductivities of the electrodes before and after ball-milling were similar (**Figure S12**), indicating the particle sizes did not affect the percolation of CNTF network. The conventional electrode with ball-milled NVPC showed similar voltage profile with pristine particles (**Figure S13**), suggesting the electrochemical properties did not deteriorate with ball-milling.

After 100 cycles at 0.2C, the ball-milled electrode showed 77.4 % capacity retention, notably lower than the 99.6 % retention of the pristine electrode **(Figure 2c)**. Post-cycling, the ball-milled electrode was partly disintegrated, with small NVPC particles detaching the CNTF network, whereas the pristine electrode's morphology remained intact (**Figure S14**). The robustness of the co-ESP electrodes were also demonstrated through sonication (**Video S2, S3**). Powders detached from the ball-milled electrode from the start, while the pristine electrode remained intact throughout the sonication process. Nevertheless, both electrodes exhibited significantly better rate performance than the conventional slurry-casted electrode (**Figure 2g**).

The poorer rate performance and cycling stability of the ball-milled electrodes are attributed to the weaker binding of smaller particles.

Unlike the pristine particles (**Figure 2d**), the ball-milled particles were much smaller than the pore of the CNTF network and were not bound by spatial constrictions (**Figure S8, S9**). These particles were only loosely attached to the CNTF network, resulting in high contact resistance. Additionally, not all particles were directly in contact with the conductive network, requiring electrons to traverse multiple particle-particle interfaces to reach these particles, further increasing resistance[44,45] (**Figure 2e**). Since the ion diffusion and insertion/extraction should be quicker as particles became smaller[44], the poor rate performance of the ball-milled electrode was likely because of the poor electronic conduction.

Therefore, CNTF network is especially ideal for capsulating commercial electrode materials of Li and Na-ion batteries, most of which have secondary particle sizes of 5-50 μm[46]. To demonstrate this, we have fabricated co-ESP LiFePO4-C LIB cathode, SiOx-C and graphite LIB anodes, as shown in **Figure S15.** The voltage profiles of these electrodes are shown in **Figure S16**. All electrodes showed high areal capacities, high specific capacities, and small overpotential.

Previous attempts to fabricate battery electrodes using combinations of electrospin and spray have not demonstrated competitive areal loading, active content, rate performance, or cycling stability compared to other state-of-the-art techniques [31,32]. This is likely because nano-sized particles were used. Additional binder/conductive additives were added to stabilise the particles, which further deteriorated the performance and significantly reduced the active content. There was also no in-depth analysis of the merit of the co-ESP method and the resulting structures in previous works, which we are aiming to do here.

Given the significantly superior performance, we will only use pristine micron-sized NVPC in the co-ESP NVPC electrodes for the remainder of this work.

*Performance of co-ESP electrodes with different CNTF content*

Given the critical role of the CNTF network, it is important to investigate its minimum content required to provide sufficient electron conductivity and structural support, thereby achieving the highest energy density.

We synthesised electrodes with active contents of 90 wt.%, 97.5 wt.%, and 99 wt.% (**Table S1**), all with areal loadings of ca. 18 mg cm$^{-2}$. Electrodes with active content above 99% were too fragile for use in cells.

As shown in **Figure S17,** increasing the active content resulted in sparser fibres. In the 99 wt.% electrode not all particles were in direct contact with the CNTF network.

At 0.1 C, 97.5 wt.% and 90 wt.% electrodes showed similar discharge capacity and polarisation, while the 99 wt.% electrode had lower discharge capacity and discharge plateau (**Figure 2f**). The capacity difference became more pronounced with increasing C-rate. At 10C, the 99 wt.% electrode showed no discharge capacity (**Figure 2g**). In contrast, both the 97.5 wt.% and 90 wt.% electrodes maintained decent capacities even at 20C. All electrodes demonstrated superior rate performance compared to conventional electrodes.

Apart from lower rate performance, 99 wt.% electrode also showed lower cycling stability than its lower active content counterparts (**Figure 2h**). These results suggest that 1 wt.% CNTF neither provided sufficient electron accessibility for high-rate cycling nor supported the electrode structure adequately. The through plane conductivity of 99 wt.% electrode (1.5 S m$^{-1}$) was lower than 90 wt.% (18.5 S m$^{-1}$) and 97.5 wt.% (4.2 S m$^{-1}$), but was still sufficient according to previous work[21]. However, macroscopic conductivity does not reflect the electron accessibility of individual particles. As shown in **Figure S17**, NVPC particles not in direct contact with the CNTF network could not contribute their capacities at high rates. In contrast, active particles in 97.5 wt.% and 90 wt.% electrodes were all connected to the CNTF network, (**Figure 2i**) and had sufficient electron accessibility, shifting the rate-limiting process to ion transportation[21]. This explains why further reducing the active content beyond 97.5 wt.% showed only marginal improvement.

Thus, 97.5 wt.% is the optimal active content for the co-ESP NVPC electrode, balancing performance, electrode robustness, and active content. We will use this composition throughout the rest of this work. This active content brings 24.4% higher energy/power density than a commercial 25 mg cm$^{-2}$ electrode (90 wt.% NVPC, 5 wt.% PVDF, 5 wt.% carbon black, 15 μm aluminium foil, **Figure 2j**) even without considering the performance benefit of co-ESP electrodes.

Previous efforts to achieve high areal loading and superior rate performance often gave up controls over the active content[10,18], resulting in overall loss in the energy/power density. The co-ESP electrode, however, has demonstrated an uncompromised solution, maintaining high active content alongside excellent performance, while having an active content significantly higher than other high-loading electrodes[9,47].

*Morphology benefit of co-ESP NVPC electrodes*

Co-ESP NVPC electrodes have demonstrated remarkable rate performance and stability with high active contents. To further understand the morphology benefits, synchrotron-based micro- and nano-computed tomography (micro- and nano-CT) were employed to characterise multiscale features of the electrodes (**Figure 3a-f**). Micro-CT was used to resolve the morphology of particles and external pores (**Figure 3a, b**), with the electrodes characterised in a compressed state to mimic their condition in cells. The nano-CT was used to resolve the particles internal pores (**Figure 3c**), though it could not resolve all the CNTFs due to their small diameter relative to the resolution (150 nm vs 47 nm) and the low visibility of carbon materials in the presence of NVPC[48].

From **Figure 3a**, it is evident that the co-ESP NVPC electrode comprises well-distributed large NVPC particles and interconnected pores. A single NVPC particle resolved by the nano-CT (**Figure 3c**) revealed an internal porosity of 39.4% with an average pore size of 582 nm, based on an average of 20 particles. All internal pores were percolated and accessible from the particle surface.

Upon compression at 4 MPa, the thickness of the electrodes reduced significantly, as did the volume of external pores (**Figure 3b**). The thickness of the co-ESP electrodes was extracted from their compression curve (**Figure S18**). Uncompressed, these electrodes were more than four times thicker than conventional slurry-casted cathodes (NVPC: PVDF: CB = 90:5:5) of the same loading (**Figure 3d**), resulting in only 25% of the electrode active density (weight of active materials per volume). After assembly in cells, their thickness reduced by 70%, reaching 91% active density of conventional electrodes.

This high compressibility is an essential feature of the co-ESP NVPC electrode. As shown in the cross-section images (**Figure S3**), the CNTFs tend to align in-plane due to the layered deposition of electrospun fibres. The intrinsic flexibility of CNTFs granted the co-ESP electrode the ability to maintain structural integrity even after 70% strain (**Figure 3b, d**), enhancing the contact between the electrode and cases/metal tabs without a current collector. For half cells with the areal loadings higher than 150 mg cm$^{-2}$, sufficient electric connection between the electrode and the coin cell case can be achieved without the springs and spacers.

The volume compositions of different electrode components after compression are summarised in **Figure 3e**. Note that the external porosity only considered the pores outside the NVPC particles, while total porosity included pores both outside and inside the NVPC particles, acquired from micro-CT and nano-CT respectively.

After compression, the electrode had 55.8% total porosity, with external porosity at 27.1%. The porosity was uniform across the cell thickness (**Figure S19**). The total porosity is higher than most reported electrodes, although previous works rarely considered pores inside particles[26,49]. Therefore, external porosity is a better parameter for comparison. Active particles comprised 94.1 vol.% of the solid phase, indicating efficient utilisation of the electrode volume.

Other microstructural parameters after compression are summarized in **Figure 3f**, with detailed summaries available in **Table S2**. When uncompressed, the external pore tortuosity of the co-ESP NVPC electrode was ca. 1.1, identical across the x, y, and z directions, indicating a homogeneous pore structure. After compression along the z direction, the z-tortuosity increased to 2.0. Conventional electrodes typically have pore tortuosities of 5-8 due to the presence of carbon binder domain (CBD), which refers to the composite cluster of binder and conductive additive[49,50]. Although CNTFs could not be resolved, they are not expected to significantly increase tortuosity because of their low volume content (<3 vol.%) and small diameters relative to external pores (150nm vs 10 μm, **Figure S3**). The low z-tortuosity ensures smooth through-plane transportation of Na-ions.

In the conventional electrodes the presence of highly tortuous CBD has been proven to be the main reason of sluggish ion transportation in the pore [28,29,51] (**Figure 3g**). Even without binder, the nano-sized conductive additives would also significantly increase the pore tortuosity [52,53]. CBD is also the reason that the ion transportation and electron conduction cannot be simultaneously increased in conventional electrodes[50,54].

The absence of CBD or any nanosized conductive additives in our co-ESP NVPC electrodes significantly accelerated ion transportation, which is the rate-limiting process when electron conduction is sufficient[21]. Additionally, the surface of NVPC particles were fully accessible to sodium ions, maximizing the Na-ion intercalation/insertion interface (**Figure 3h**)[55].

The ion transportation was further enhanced by hierarchical internal pores in the NVPC particles (**Figure 3c**), which shortened the diffusion pathway of Na-ions in the NVPC solid by approximately 20-fold. This leveraged the fact that sodium diffusion in the electrolyte is at least five orders of magnitude faster than in the solid NVPC phase[56,57].

For electron conduction, the percolating CNTF network ensures electronic access to all NVPC particles, while the coated carbon on the NVPC guarantees uniform electron accessibility across the particles. Conventional electrodes require at least 5 wt.% CBD to ensure carbon black percolation[58]. In contrast, due to the intrinsic interconnecting nature of the CNTF network, there is no theoretical percolation threshold. Good macroscopic electronic conductivity was achieved with even less than 1 wt.% CNTF loading, despite the insufficient electron accessibility of individual particles.

The CNTF network also binds the particles through spatial constriction, providing a much stronger binding force than the van der Waals bond of conventional binders[9]. On the other hand, larger particles applied stress to the fibres caging them, securing their electrical contact.

Thus, in the co-ESP electrode, high conductivity, fast ion transportation, and robust structure are achieved simultaneously (**Figure 3d, e** and **Figure 2i, j**) with only 2.5 wt.% inactive content. Especially, both electron conduction and ion conduction pathways are minimised, creating an ideal electrode structure as predicted by previous work[44]. Since ion diffusion and electron conductivity have been found to co-limit the performance[21], this results in the outstanding rate performance and stability of the co-ESP NVPC electrodes.

## High areal loading, high-performance half-cells enabled by CNTF network

The co-ESP NVPC electrodes exhibited an intrinsic ability to support high-areal-loading active materials thanks to the inter-supportive nature of the CNTF network. Unlike conventional electrodes, there is no fundamental limitation to the thickness and the areal loadings of co-ESP electrodes. The thickness and areal loading were controlled by adjusting the total amount of raw materials in the co-ESP fabrication process, as detailed in **Table S3**.

Half-cells were assembled with co-ESP NVPC cathodes of up to 49.6 mg cm$^{-2}$ areal loading (**Figure 2a**). Cross-sectional images of co-ESP NVPC electrodes with different areal loadings are shown in **Figure S20**. When the areal loading exceeded 50 mg cm$^{-2}$, the half-cells began to exhibit serious over-charging after 15-20 cycles (**Figure S21**), preventing further increases in areal loading in half-cells. This was previously attributed to the degradation of metallic anode[9,59], which is confirmed by the intact structure of the co-ESP NVPC cathodes (**Figure S14**), and the severe degradation of Na metal anodes (**Figure S22**) after cycling. The addition of fluoroethylene carbonate (FEC) to the electrolyte slowed the degradation but could not eliminate it (**Figure S23**)[60,61]. Thus, the co-ESP electrodes' ability to hold higher areal loading will be demonstrated in full cells.

All co-ESP NVPC electrodes showed state-of-the-art rate performance and cycling stability for their respective areal loadings. At the lowest areal loading of 4.3 mg cm$^{-2}$, co-ESP NVPC electrodes showed decent capacity even at 200C, while the 49.6 mg cm$^{-2}$ electrodes were usable at 10C (**Figure 4a-d**). The 4.3 mg cm$^{-2}$ cell was cycled at 50C for 5000 cycles with 84.8% capacity retention, and the 49.6 mg cm$^{-2}$ cell was cycled at 0.2C for 200 cycles with 97.5% capacity retention (**Figure 4e**).

High areal loading co-ESP NVPC electrodes did not exhibit significantly higher degradation rate than their low loading counterparts, a common issue in other high-areal-loading cathodes[9,62]. This can be attributed to the following factors:

1. The CNTF network physically bound the NVPC particles and absorbed their volume changes during charge/discharge cycles, avoiding the loss of electron accessibility;
2. The absence of CBD prevented pore clogging by the formation of CEI, preserving fast ion transportation;
3. The combination of the above factors prevented the NVPC particles disintegration caused by the inhomogeneous sodiation.

Thus, the co-ESP electrodes overcame the three main causes of the accelerated degradation in high-areal-loading electrodes[28].

The detailed electrochemical test result across all areal loadings (**Figure S24**) were summarised into **Figure 4f** and **g**, demonstrate how specific capacities and areal capacities changed with C-rate and areal loading. When cycled at 0.1C, the specific capacities remained almost invariant until the areal loading reached 23.9 mg cm$^{-2}$. Increasing the loading further to 49.6 mg cm$^{-2}$ reduced the capacity to 89.2% of the theoretical capacity. In comparison, other Na-ion cathode works did not report more than 80% capacity retention for areal loadings above 10 mg cm$^{-2}$ [7,63]. The average discharge voltage is summarised in **Figure S25**.

On the other hand, the 4.3 mg cm$^{-2}$ co-ESP NVPC's retained 71.8% of the theoretical capacity at 100C, and 36.4% at 200C. Increasing the areal loading caused the specific capacities to reduce more pronouncedly with increasing C-rate, as expected for all high-areal-loading electrodes[9,18]. However, the 11.3 mg cm$^{-2}$ electrode still retained 25.6% of the theoretical capacity at 50 C, while the 49.6 mg cm$^{-2}$ electrode retained 33% at 10C, demonstrating one of the best rate performances at this level of areal loading among all Na and Li-ion batteries[16,64-66] (detailed comparisons in **Table S4, 5**).

**Figure 4g** shows that 49.6 mg cm$^{-2}$ electrode retained 3.9 and 1.9 mAh cm$^{-2}$ areal capacity at 2C and 10C, respectively, demonstrating that high areal capacity and high charge/discharge current were achieved simultaneously. These state-of-art rate performances and stabilities confirmed the structural merit of co-ESP electrodes.

Electrochemical impedance spectroscopy (EIS) of the electrodes showed low Ohmic resistances across all samples (**Figure S26**). While Ohmic resistance increased with areal loading, the polarization resistance decreased, consistent with previous report[67]. This reduction is mainly due to the reduced charge-transfer resistance for Na intercalation/extraction, facilitated by the higher number of active sites.

The data in **Figure S24** were summarised in two Ragone plots (**Figure 4h, i**). Previously reported Na-ion battery half-cell data from the literatures were included in the same figures[4-7,63,68-71]. Note that most literature did not disclose sufficient information to calculate the energy/power densities of the whole cells, such as electrode porosities and the weight of electrolytes and separators. For a better comparison, the half-cell energy/power densities in this work considered only the weight of cathodes, including the current collector. Literatures values were recalculated under the same standard, as presented in **Supplementary Appendix 3**. The combination of energy/power densities of co-ESP electrodes leads by a noticeable margin among reported Na-ion battery electrodes, attributing to a combination of high capacity retention, high active content, and superior rate performance.

Previously, the highest reported areal loading of Na-ion batteries' cathodes was 60 mg cm$^{-2}$, where no cycling or rate data were presented[7]. The second highest was 48.9 mg cm$^{-2}$, which showed much inferior rate performance and cycling stability than the 49.6 mg cm$^{-2}$ co-ESP NVPC electrode [63].

The half-cells data were also compared with high-areal-loading lithium-ion half-cells (**Table S5**). Although the NVPC co-ESP electrodes' energy densities do not match those of lithium cobalt oxides (LCO) and lithium nickel manganese cobalt oxides (NMC) cathodes, they are comparable to lithium iron phosphate (LFP) cathodes, the second most widely used cathode material in electric vehicles.

*Ultra-high loading full cells and pouch cells*

Na-ion full cells (coin cells) were assembled using co-ESP NVPC cathodes and co-ESP hard carbon (HC) anodes (**Figure 5a**). The high-performance glucose-derived hard carbon (HC) was synthesised through a facile method as detailed in the previous literature[72]. The voltage-capacity profile and the cycling stability of the co-ESP HC half-cell is demonstrated in **Figure S27**. The composition of fabrication raw materials is presented in **Table S6.** The PAN-derived CNF, embedded CNT and HC were all active sodium storage materials[73-75], in which the HC contributed >97% of the total capacity.

The morphology of the co-ESP HC anode is shown in **Figure S28**, where nano-sized HC particles are agglomerated into secondary particles of average size of 9.9 μm, bound by the CNTF networks through spatial constriction, similar to NVPC particles.

By maintaining the cathode/anode mass ratios at 2.5:1, we assembled full cells with cathode areal loading ranging from 25.4 mg cm$^{-2}$, an industry-relevant areal loading, to a record high of 296 mg cm$^{-2}$. The maximum areal loading of full cells significantly exceeded the half-cells due to the absence of the problematic Na metallic anode. A 296 mg cm$^{-2}$ loading was realised by layering two 148 mg cm$^{-2}$ co-ESP NVPC cathodes. In comparison, the conventional electrodes would crack and delaminate from the current collector at areal loadings over 50 mg cm$^{-2}$ (**Figure S29**). Detailed full cell compositions are shown in **Table S7, 8**, with a typical conventional cell composition provided in **Table S9**. This represents the highest areal loading to date that can cycle stably among reported Na-ion batteries (**Table S10**) and Li-ion batteries full cells (**Table S11)**.

The ICE of the 25.4 mg cm$^{-2}$ full cell was 88.9% (**Figure S30**). Due to the high ICE of the co-ESP NVPC half-cell, the lower ICE was mainly attributed to the irreversible sodium intercalation to the HC anode and the formation of solid-electrolyte interphase (SEI)[76], which can be enhanced through electrolyte optimisations[77]. The ICE decreased with increasing areal loading, reaching 77% at 296 mg cm$^{-2}$, likely due to the increased irreversible sodium plating, a common issue for thick anodes[78].

The co-ESP full cells exhibited superior rate performance (**Figure 5b, c**). Detailed electrochemical testing results are available in **Figure S31.** The average discharge voltage is shown in **Figure S32** The 25.4 mg cm$^{-2}$ full cells delivered 58% of theoretical capacity at 10C, and 17% capacity at 50C. The full cell with 298 mg cm$^{-2}$ cathode loading exhibited 76.5% of theoretical capacity at 0.1 C. Even at 2 C it still delivered 10 mAh cm$^{-2}$, much higher than a typical conventional electrode (<3 mAh cm$^{-2}$).

The full cells also exhibited great stability. The 298 mg cm$^{-2}$ loading cell exhibited 73.4% capacity retention after 200 cycles at 0.2 C (**Figure 5d, Figure S31**). A 60.7 mg cm$^{-2}$ full cell had 92.1 % capacity retention after 200 cycles at 0.2 C and 79.6 % after 1000 cycles at 1 C.

The EIS of full cells (**Figure S33**) followed similar trend as half-cells (**Figure S26**): Ohmic resistance increased with areal loading, while polarization resistance drastically decreased.

The energy/power densities of co-ESP full cells were summarised in Ragone plots **(Figure 5e, f),** compared with previously reported Na-ion full cells[5,7,63,68,69,71]. The co-ESP full cell exhibited maximum gravimetric energy and power densities of 231.6 Wh kg$^{-1}$ and 7152.6 W kg$^{-1}$, respectively. While the maximum areal energy and power densities were 77.7 mWh cm$^{-2}$ and 248.4 mW cm$^{-2}$, respectively. These performance metrics lead among all reported SIB electrode designs works by significant margin. Thus, uncompromised power and energy were achieved for the co-ESP full cells. Detailed comparisons are provided in **Table S9.**

Similar with half-cells, the gravimetric energy/power densities of all full cells in **Figure 5e** were calculated as in **Supplementary Appendix 3**. The gravimetric energy/power densities of the co-ESP full cells, considering the electrolyte and separator, are also presented in **Figure S34** to better compare with the industry standard. The results were also compared with Li-ion battery full cells in **Table. S11**. The energy densities of co-ESP SIB full cells are on par with reported LIBs, while the power densities are notably higher.

To demonstrate the scale-up potential of co-ESP electrodes, we have assembled pouch cells with ~70 and 200 mAh capacities, which also showed great rate performances and stabilities (**Figure 5g**). Both cells were composed of 100 mg cm$^{-2}$ co-ESP NVPC cathodes and 40 mg cm$^{-2}$ co-ESP HC anodes.

The 70 mAh pouch cell delivered 24.3 mAh at 1C and 11.8 at 2C charge/discharge rate, showing 33.9% and 16.5% capacity retention, respectively (**Figure 5h**). This remarks the best capacity retention among previous reports of pouch cells with similar areal loading[13]. Cycling stability testing were performed on the 200 mAh cell, exhibiting 80.8% capacity retention after cycling at 0.2C for 300 cycles (**Figure 5i**). The detailed voltage profiles and rate performance of the pouch cells are shown in **Figure S35**. Considering the mass and volume of the whole cell, co-ESP pouch cell delivered an unprecedented gravimetric energy density of 147 Wh kg$^{-1}$, and a volumetric energy density of 307 Wh L$^{-1}$, one of the highest among all reported SIB pouch cells. The calculation method is detailed in **Supplementary Appendix 3**.

*Future remarks*

We have shown that co-ESP SIB full cells can deliver comparable energy densities and much superior power densities compared to existing LFP-based LIBs while using commercial particles. We have also shown the scaling-up potential of co-ESP method, by producing 600 cm$^2$ of co-ESP NVPC mat in one batch on a lab-scale electrospinning-electrospraying machine (**Figure 1b**). An industry-scale electrospinning/spraying machine has a production capability of over 20,000,000 m$^2$ per year[79], equivalent to 12 GWh capacity, assuming a mid-of-range areal loading 60 mg cm$^{-2}$. These suggest the co-ESP SIBs could be rational alternatives for cheaper and quicker electric vehicles in the future.

However, the current need for a calcination step in fabricating co-ESP electrodes, which is not part of the standard process for conventional electrodes, presents a major barrier to wider application. Integrating the calcination step into battery manufacturing will be costly and energy-intensive. While there have been effort to directly electrospin conductive fibres[80], their conductivities are far from enough for battery electrodes. Therefore, it is necessary to explore novel techniques to electrospin conductive fibres, which will be the focus of our next stage of research.

*Conclusion*

In this study, we developed an electrode fabrication technique by concurrently electrospinning CNTF conductive backbones and electrospraying carbon-coated $Na_3V_2(PO_4)_3$ (co-ESP NVPC) onto identical substrates.

Our in-depth 2D and 3D morphology characterisations revealed that the NVPC particles are bound to the CNTF network by spatial constriction, ensuring all pores remain fully accessible for unhindered Na-ion transport. The electrospun CNTF network, while constituting merely 2.5 wt.% of the content, adeptly serves as a binder, conductive additive, and current collector, ensuring electronic connectivity to all NVPC particles.

Due to the fast species transportation and robust structure, the co-ESP NVPC electrodes exhibited superior rate-performance and stability. Notably, particles larger than the pore of CNTF network proved to have better performance than their smaller counterparts as the electrosprayed species. The sturdy CNTF networks facilitated the production of extremely high-areal-loading electrodes with up to 296 mg cm$^{-2}$ areal loading.

Both coin cells and pouch cells, with co-ESP electrodes showed state-of-art and uncompromised energy and power densities, even comparable to lithium-ion batteries, demonstrating the merit of co-ESP method.

Finally the co-ESP is a promising fabrication method to greatly enhance the energy/power density of battery electrode. It is applicable to a variety of commercial cathode and anode materials of Na-ion batteries and Li-ion batteries and is scalable, demonstrating its commercialisation potential.

## Experimental procedures

*Preparation of electrode active materials*

The carbon-coated $Na_3V_2(PO_4)_3$ (NVPC) was purchased from Guangdong Canrd New Energy Technology Co.,Ltd. The NVPC particles are coated with ~1 wt.% carbon on their surface. Hard carbons powder was synthesised by the facile and scalable method reported by literature[72]. The method is that firstly hydrothermal carbonization of 30g D-glucose (d-(+)-glucose, ≥ 99.5%, Sigma-Aldrich) with 270 ml deionized water in an autoclave reactor (50% fill volume) and heated to 230 °C for 12 h; and then the resulting powder was heated at 80 °C under vacuum till fully dried, and finally pyrolysis at 1500 °C (ramping rate 5 °C/min from room temperature) for 2h under a continues 500 ml/min $N_2$ gas flow.

## Electrodes fabrication by co-electrospinning-spraying

The co-ESP electrodes were fabricated through a simultaneous electrospinning-electrospraying method followed by calcination, or called co-ESP method. The fabrication set-up is modified from Bioinicia LE-50 electrospinning machine. The schematic of the set-up is shown in **Figure 1a**, which consists of two sets of syringes, syringe pumps, and high voltage power supplies for electrospinning and electrospraying respectively. The electrospinning syringe is horizontally placed, and the electrospraying syringe is vertically placed, with a grounded aluminium roller collector placed in the centre. Two high voltage power supplies apply adjustable high voltage to two syringes respectively. During fabrication, the rotation speed of the cylindrical collector was set to 50 rpm to avoid any fibre orientation. Both syringes were moving side to side parallel with the collector to ensure uniform thickness.

The fabrication of NVPC/carbon nanotube-carbon nanofibre cathode (NVPC/CNTF) involved sinmultaneously electrospinning polyacrylonitrile (PAN)/carbon nanotube (CNT) DMF slurry, and electrospraying polyethylene oxide/NVPC DMF slurry. Composition of electrospinning slurry: 5 w/v% polyacrylonitrile (PAN, Goodfellow), 1 w/v% carbon nanotube (MTI) are dissolved/dispersed in dimethylformamide (DMF, Sigma-Aldrich) solvent. Composition of electrospraying slurry: 2 w/v% PEO (Sigma-Aldrich) and 100 w/v% NVPC are dissolved/dispersed in DMF. The total volume and volume ratio of these two slurries were adjusted to fabricate electrodes with different areal loading and active contents. During the fabrication, the distance of electrospinning syringe to roller collector was fixed at 15 cm, distance of electrospraying syringe to the roller 10 cm. The feeding rate of electrospinning slurry is set to 2 mL h$^{-1}$, feeding rate of electrospraying slurry changes according to the volume ratio of two slurries in order to synchronise the two processes. The voltage applied to electrospinning syringe was adjusted between 10 to 15 kV to ensure a continuous and drop-free spinning process. Similarly, electrospraying voltage was adjusted between 15 to 20 kV.

The produced NVPC-PEO/CNT-PAN composite mats were peeled off from the roller collector and calcined in 1 % $H_2/N_2$ atmosphere at 850 °C for 5 h, to eliminate PEO and carbonise PAN fibre to carbon fibre (CNF). Finally, co-ESP NVPC cathodes were acquired.

The sub-micron sized NVPC particles were made through ball-milling. 10 g NVPC and 10 mL tert-butanol were put in a ball-milling jar. Zirconia balls were used as the milling ball with ball-to-powder ratio of 10:1, which were mixed with 1:1:1 weight ratio of 1-mm, 5-mm and 10-mm zirconia balls. The ball-mill was performed at 100 rpm for 6 hours before freeze drying and collecting the powder. Co-ESP balled-milled NVPC electrodes were made using the ball-milled powder through the same co-ESP process, aiming to acquire the same active content.

To fabricate hard carbon (HC) /CNTF anode, same co-electrospinning-spraying method as above was used. The only difference was that the electrospraying slurry is changed to 2 wt.% PEO-50 wt.% HC-DMF. The as-prepared HC/PEO/CNF/PAN composite mats were calcined at 1100 °C for 5 h to acquire HC/CNT-CNF anodes.

The co-ESP LiFePO$_4$/C (MTI), SiO$_x$/C (MTI), and graphite (MTI) electrodes were fabricated using the same co-ESP method as above. The active contents of these electrodes were controlled to be 95 wt%.

*Electrodes fabrication by conventional slurry casting*

Regarding the hard carbon slurry, 90 wt. % hard carbon and 10 wt. % pre-prepared sodium carboxymethyl cellulose (CMC, Mw ~ 250 000, Sigma) binder solution (5 wt.%) in water were well mixed. Electrodes were coated from slurries onto battery-grade Al foil (17 μm in thickness, MTI) followed by drying at room temperature and ambient environment for 6 hours followed by drying in a vacuum oven for 18 hours. Regarding the NVP slurry, 90 wt. % NVP powder, 4 wt. % Super P carbon additive (Sigma) and 6 wt. % pre-prepared poly(vinylidene fluoride) binder (Mw ~ 534,000, Sigma) solution (5 wt.% in N-Methyl-2-pyrrolidone, Sigma) were well mixed. Electrodes were coated from slurries onto battery-grade Al foil (17 μm in thickness, MTI) followed by drying at 80 °C 6 hours followed by drying in a vacuum oven at 80 °C for 18 hours. The mass loading of the resulting electrodes is between ~4 mg cm$^{-2}$ for anode, and ~16 mg cm$^{-2}$ for cathode.

*Materials characterisation*

The morphologies of the electrodes were examined by field emission SEM (Zeiss LEO Gemini 1525 FEGSEM), with an acceleration voltage of 5 kV. The particle and pore size distributions were acquired through image analysis in ImageJ.

The TEM images were acquired by JEOL STEM 2100Plus, with an acceleration voltage of 200 kV.

The phase of the electrodes and raw materials were characterized by X-ray powder diffraction (XRD, X'Pert³ Powder, Malvern Panalytical). The Raman spectrum was performed on a Renishaw inVia confocal Raman microscope, using 532 nm laser.

Thermogravimetric analysis (TGA) was performed on a Netzsch STA449C. TGA experiments were performed in air and nitrogen atmosphere, under 5 °C min$^{-1}$ ramping rating.

Microscope assisted nano-scale x-ray tomography (nano-CT) was performed in Diamond Synchrotron I-13-2 beamline. The energy was of X-ray was 8eV and 1950 images were acquired by continuously rotating the sample 180 degrees using an integration time of 1.8s per radiograph. The spatial resolution was 47 nm and field of view was 150*150*150 µm. Image reconstruction was carried out using a bespoke routine implemented by the I-13-2 beamline scientists.

Micron-scale X-ray tomography (micron-CT) was performed at the European Synchrotron Radiation Facility (ESRF) beamline ID19. The energy of X-ray was 16 keV and 1800 images were acquired by continuously rotating the sample 180 degrees using an integration time of 1s per radiograph. The linear resolution was of 350 nm, with a large field of view of 1*1*1 mm. Image reconstruction was carried out using a bespoke routine implemented by the ID19 beamline scientists.

The reconstructed images from CTs were analysed in the Avizo software, from which the microstructural parameters including volume contents, porosity, pore tortuosity and porosity distribution were extracted[81,82].

The in-plane conductivities were measured on a Ossila four-point conductivity tester. The through-plane conductivities were measure by a potentiostat (Metrohm Autolab PGSTAT302N).

The strain-stress curve of co-ESP NVPC electrodes were acquired from the compressive mechanical testings, which was performed on a ZwickRoell ZwickiLine universal testing machine. Sample sizes was 2*2 cm$^2$.

*Electrochemical characterisation*

The electrochemical properties of the electrodes were examined in CR2032 coin cells, assembled in an argon-filled glove box with water and oxygen content both lower than 1 ppm. For high-areal loading co-ESP NVP electrodes, no spacer or spring is needed when assembling the cells. All cells were tested on a Biologic Ultra-precision battery cycler at 25 °C. When assembling half-cells, electrodes (co-ESP electrodes and conventional electrodes) were used directly as cathodes. Sodium metal was rolled and punched into 12 mm-diameter round chips and used as anodes. Celgard 2400 polypropylene membranes were used as separators. 1 M NaPF$_6$ (Canrd) in ethylene carbonate (EC)/ diethyl carbonate (DEC) (EC/DEC = 1:1, v%, Sigma-Aldrich) with 10% fluoroethylene carbonate (FEC) (Sigma-Aldrich) was used as the electrolyte. The amount of electrolyte is controlled to 3:1 mass ratio relative to the electrode.

The NVPC half-cells were cycled in a voltage range of 2.0-3.8 V. The HC half-cells were cycled in a voltage range of 0.005-2.5 V.

The full cells were assembled using NVPC electrodes (co-ESP electrodes and conventional electrodes) as the cathodes, and HC electrodes (co-ESP electrodes and conventional electrodes) as the anodes. Celgard 2400 polypropylene membranes were used as separators. 1 M $NaPF_6$ in ethylene carbonate (EC)/dimethyl carbonate (DEC) (EC/DEC = 1:1, v%) was used as the electrolyte. The full cells were cycled in a voltage range of 0.5-3.8 V. In the cycling of NVPC half-cells and full cells, the charging processes were done in a constant current-constant voltage (CC-CV) mode. In the CC stage, cells were charged to 3.8 V under constant current. Then in the CV stage, the cells were charged with 3.8 V voltage until the current reach 0.05C.

The impedance measurements of all cells were performed on the Biologic Ultra-precision battery cycler with a frequency range of 0.1 to 10 kHz, an amplitude of 10mV was used. All measurements were conducted in a fully discharged state.

The Na-ion monolayer pouch cell was constructed using 40 mg $cm^{-2}$ co-ESP hard carbon as the anodes and 100 mg $cm^{-2}$ co-ESP NVPC as the cathodes. Both electrodes were pressed under 30 MPa to flatten the electrodes, ensuring their good contact with the separator. The electrolyte consisted of 1M $NaPF_6$ in a solvent mixture of EC and DEC, mixed in a 1:1 volume ratio. A Celgard 2400 separator was positioned between the anode and cathode and soaked in the electrolyte solution to ensure sufficient ionic transport during cell operation. The thickness of aluminium plastic film is 90 μm, weight 14 mg $cm^{-2}$.

The assembly of the pouch cell was performed layer-by-layer in an ambient environment because both hard carbon and NVP are resistant to humidity. The separator was sandwiched between the anode and cathode, and three ends of the layered assembly were sealed first, before being placed in an antechamber of a glovebox for drying at 80°C for 18 hours. After drying, the pouch cell was transferred to the glovebox where oxygen and moisture levels were maintained below 5 ppm, and allowed to cool to room temperature. Then, an electrolyte of 1M $NaPF_6$ in EC/DEC was injected into the pouch cell, followed by vacuum sealing the final end. The edges of the pouch were heat-sealed at 180°C for 4 seconds to securely encapsulate the electrodes and electrolyte. The sealed cell rested at room temperature for 12 hours before any electrochemical testing to ensure good wettability of both electrodes and the separator.

For pre-sodiation, the pouch cell was constructed with a piece of Na metal positioned against the co-ESP hard carbon electrode, separated by a piece of Celgard 2400. A three-electrodes set-up as described in a previous work was applied to monitor the voltages of cathode and anode individually [72]. The assembly was completed inside a glovebox ($O_2$ and $H_2O$ levels less than 0.5 ppm). The assembled pouch cell rested for 18 hours before undergoing formation cycles (constant current mode at 30 mA $g^{-1}$, voltage window from 10 mV to 2.0 V). After pre-sodiation, the pouch cell was opened inside a glovebox to prevent exposure to oxygen and humidity. The pre-sodiated co-ESP hard carbon electrode was then transferred from one pouch cell to another. It is important to note that during this transfer, a negligible amount of carbon content was lost due to mechanical forces, which could not be measured.

The three-electrode pouch cells were tested on Biologic channels, while or the three-electrode pouch cells were tested on Land CT3002A channels.

## Acknowledgement

M.O. would like to thank the funding support from EPSRC (EP/W032589/1). Z.G. and M.T. would like to thank the grants from Faraday institute (NEXGENNA, reference number: FIRG064).

S.Z. acknowledges funding from the Faraday Institution (EP/S003053/1) under the Degradation project (FIRG060).

The authors would like to acknowledge Diamond Light Source for time on Beamline I13-2 under Proposal MG-34782, especially we would like to thank Dr Leonard Turpin for his assistance in finishing the session.

The authors would like to acknowledge the European Synchrotron Radiation Facility (ESRF) for provision of synchrotron radiation facilities under proposal MA-5933 and we would like to thank Dr. Benoit Cordonnier for assistance and support in using beamline ID19.

## Author Contributions

M. Ouyang and Z. Guo contributed equally to this work. M. Ouyang conceived the project, designed the experiments, developed the novel fabrication methods and supervised the project. M. Ouyang, G. Li, F. Li fabricated the novel electrodes. M. Ouyang, Z. Guo and K. Zheng assembled the coin cells and conducted electrochemical tests. Z. Guo and K. Zheng assembled the pouch cells and conducted electrochemical tests. M. Ouyang, L.E.S Farran and S. Zhao performed the CT characterisation. L.E.S Farran performed image processing and analysis. M. Ouyang, H. Zhang, Z. Guo, M. Wang and F. Li performed the physical characterisations. Y. Zhao contributed to the experimental design and the development of electrode fabrication. X. Liu, S. Yang, M. Titirici and N. Brandon contributed to the experimental design and the manuscript modification. M. Ouyang and Z. Guo wrote the original manuscript. All authors analysed the data and proof-read the manuscript.

## Declaration of interests

The authors declare no competing interests.

# Figure legends

**Figure 1. Fabrication and 2D morphology of the co-esp NVPC/CNTF electrodes. a.** Schematic diagram of co-electrospinning-electrospraying fabrication set-up. **b.** Photographic pictures of 600 cm$^2$ as-spun NVPC/PAN electrode (above) and calcined 20 cm$^2$ co-esp NVPC/CNF electrode (below, containing weight ratio of CNT: CNF: NVPC of 1:1.5:97.5). SEM images of co-esp electrodes with **c.** pristine micron-sized NVPC particles and **d.** ball-milled nano-sized NVPC particles. The schematic diagrams of the NVPC/CNTF co-ESP electrode **e.** composed of pristine micron-sized NVPC particles; **f.** composed of ball-milled NVPC particles.

**Figure 2. The performance of co-ESP NVPC cathodes with different particle sizes and active content: a.** Schematic diagram of a sodium-ion battery half cell; Half cell performance of co-ESP NVPC cathode consist of pristine and ball-milled NVPC: **b.** Rate performance and **c.** 0.2C cycling stability; The schematic diagrams of the morphology and electron transportation path of NVPC/CNTF co-ESP electrode with **d.** pristine NVPC particles and **e.** ball-milled NVPC particles; Half cell performance of co-ESP NVPC with different active contents: **f.** The third discharge curve and **g.** rate performance and **h.** 0.2 C cycling stability; **i.** Electric conductivity of co-ESP NVPC electrode with different CNTF content component's; **j.** composition of co-ESP and conventional slurry-casted NVPC cathode with 25 mg cm$^{-2}$ areal loading

**Figure 3. Physical properties and 3D morphology of co-ESP NVPC electrodes:** The schematics and 3D reconstruction of NVPC/CNTF electrodes, reconstructed from micro-CT scans: **a.** Uncompressed; **b.** compressed; **c.** fine structure of a single NVPC particle (cross-section indicated in yellow); **d.** summary of structural parameters acquired from XCT; **e.** the volume ratio of different components in a compressed NVPC/CNTF electrode; **f.** the thickness of compressed and uncompressed NVPC/CNTF electrodes with different areal loadings and conventional electrodes (current collector included); The schematics of sodium ion transportation in the pores of **g.** conventional electrodes and **h.** co-ESP electrodes

**Figure 4. The performance of co-ESP NVPC cathodes with different areal loading:** 4.3 mg cm$^{-2}$ cathode's **a.** rate performance and **b.** voltage profile; 49.6 mg cm$^{-2}$ cathode's **c.** rate performance and **d.** voltage profile; **e.** Cycling stability of different areal loading half cells; **f.** the change of specific discharge capacity with areal loading and cycling rate; **g.** The change of areal capacity with areal

current; Ragone plots of **h.** gravimetric energy density versus power density and **i.** areal energy density versus areal power density, including the data acquired from previous sodium-ion battery half cells for comparison.

**Figure 5 The performance of sodium ion batteries full cells and pouch cells made of co-ESP NVPC cathodes and co-ESP HC anodes: a.** Schematic diagram of a sodium-ion battery full cell; **b.** the change of specific discharge capacity with areal loading and cycling rate; **c.** The change of areal capacity with areal current; **d.** Cycling stability of different areal loading full cells; Ragone plots of **e.** gravimetric energy density versus power density and **f.** areal energy density versus areal power density, including the data acquired from previous sodium-ion battery full cells for comparison; pouch cell performance with 100 mg cm$^{-2}$ cathode loading: **g.** Schematic diagram of a sodium-ion battery pouch cell; **h.** Rate performance and **i.** Cycling performance of 0.2 Ah, 100 mg cm$^{-2}$ cathode loading pouch cell.

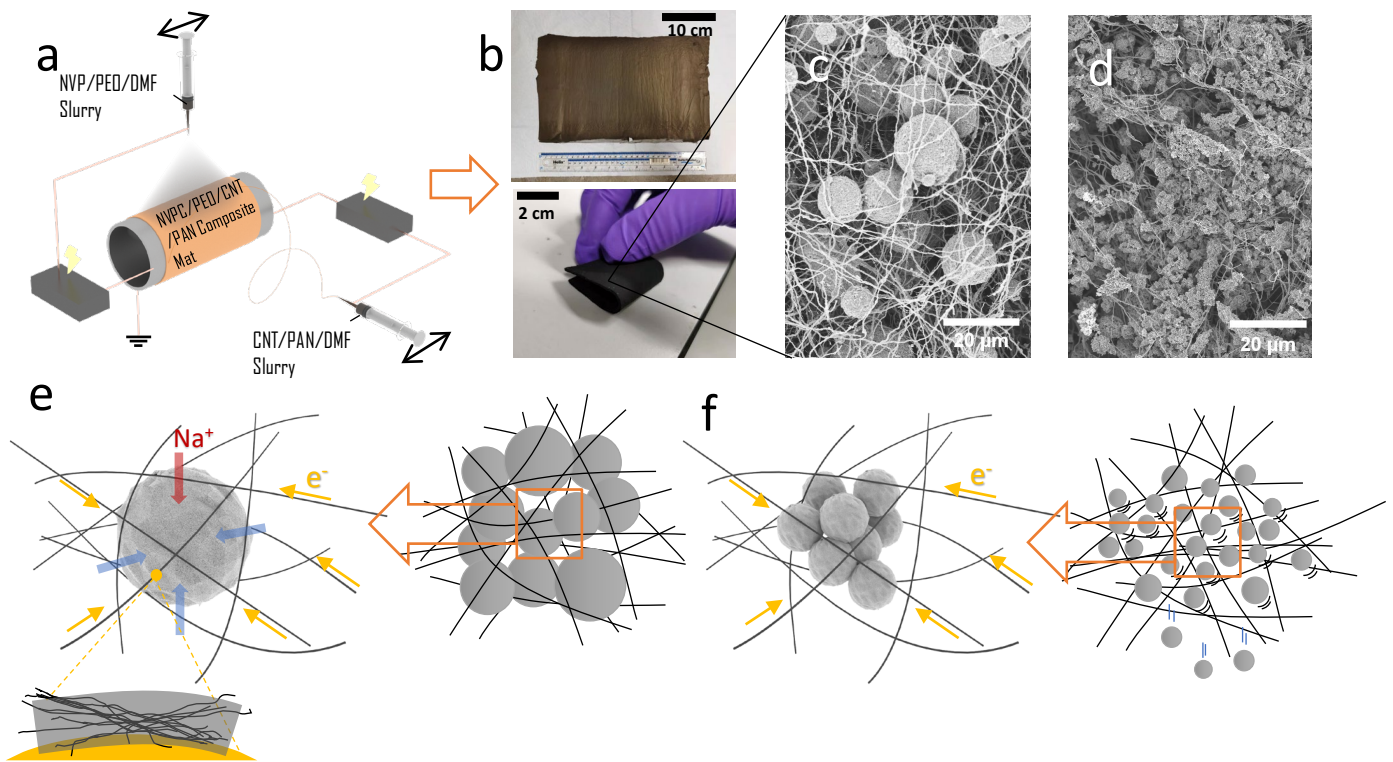

**Figure 1. Fabrication and 2D morphology of the co-esp NVPC/CNTF electrodes. a.** Schematic diagram of co-electrospinning-electrospraying fabrication set-up. **b.** Photographic pictures of 600 cm$^2$ as-spun NVPC/PAN electrode (above) and calcined 20 cm$^2$ co-esp NVPC/CNF electrode (below, containing weight ratio of CNT: CNF: NVPC of 1:1.5:97.5). SEM images of co-esp electrodes with **c.** pristine micron-sized NVPC particles and **d**. ball-milled nano-sized NVPC particles. The schematic diagrams of the NVPC/CNTF co-ESP electrode **e.** composed of pristine micron-sized NVPC particles; **f**. composed of ball-milled NVPC particles.

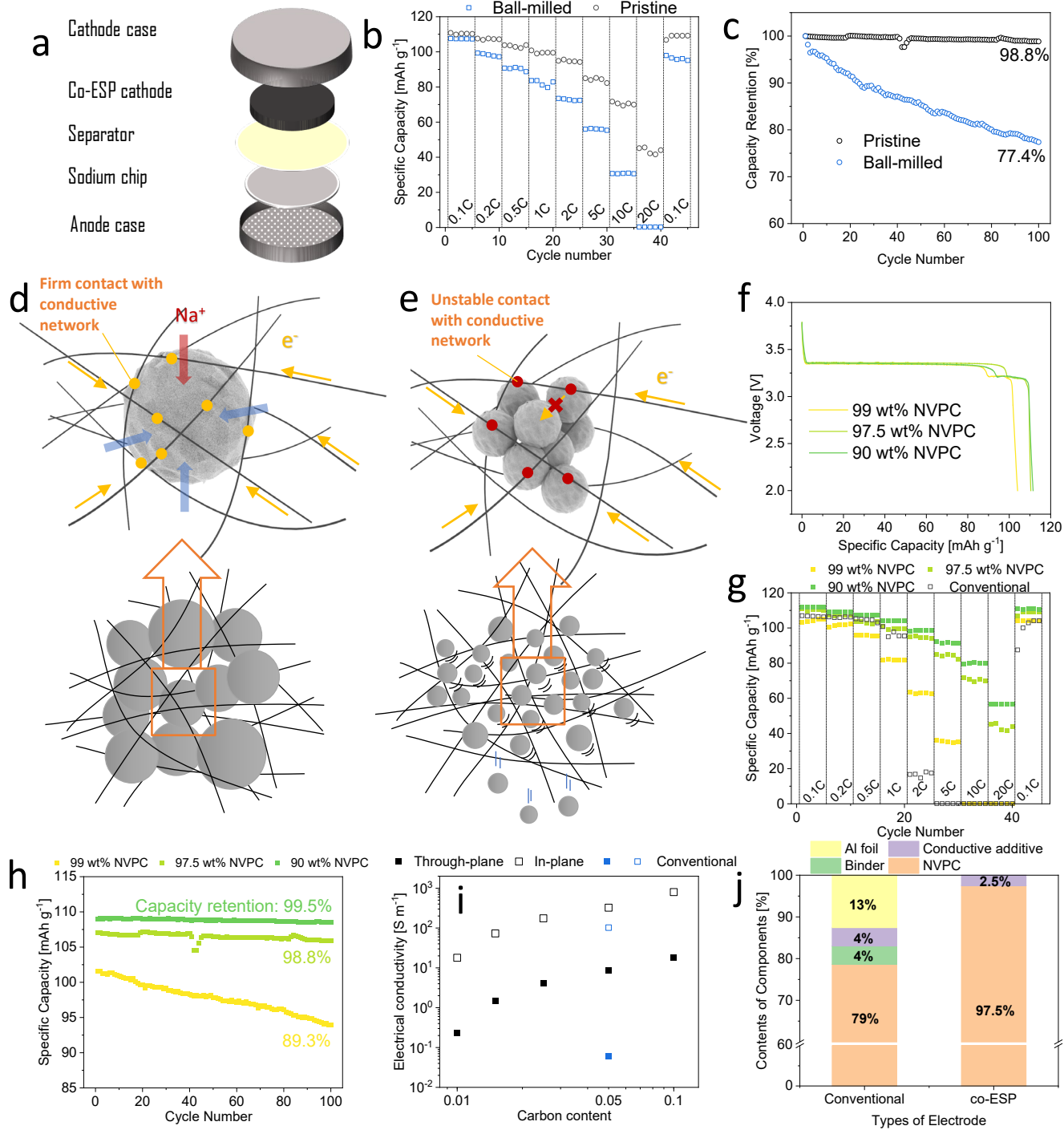

**Figure 2. The performance of co-ESP NVPC cathodes with different particle sizes and active content: a.** Schematic diagram of a sodium-ion battery half cell; Half cell performance of co-ESP NVPC cathode consist of pristine and ball-milled NVPC: **b.** Rate performance and **c.** 0.2C cycling stability; The schematic diagrams of the morphology and electron transportation path of NVPC/CNTF co-ESP electrode with **d.** pristine NVPC particles and **e.** ball-milled NVPC particles; Half cell performance of co-ESP NVPC with different active contents: **f.** The third discharge curve and **g.** rate performance and **h.** 0.2 C cycling stability; **i.** Electric conductivity of co-ESP NVPC electrode with different CNTF content component's; **j.** composition of co-ESP and conventional slurry-casted NVPC cathode with 25 mg cm$^{-2}$ areal loading

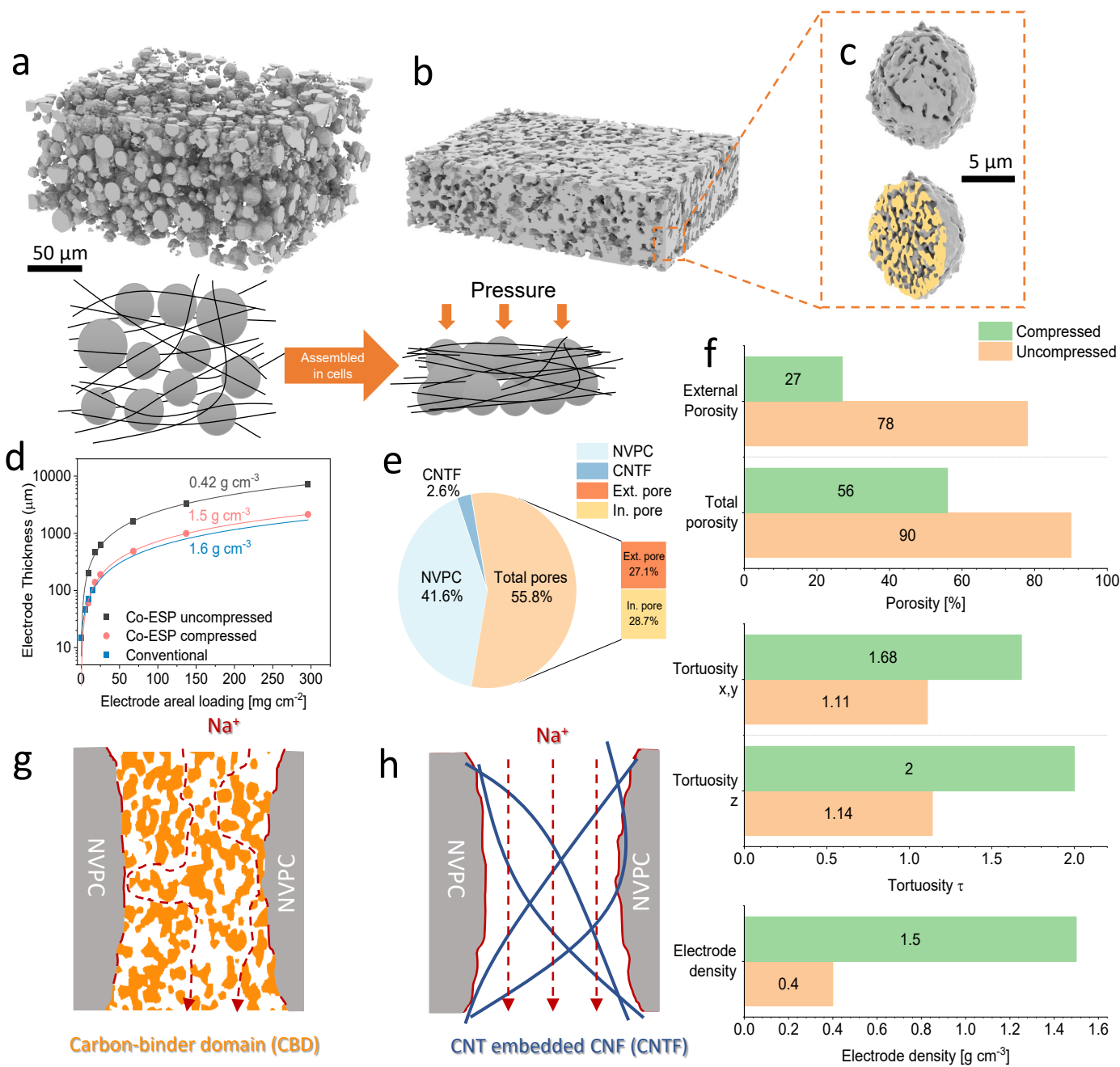

**Figure 3. Physical properties and 3D morphology of co-ESP NVPC electrodes:** The schematics and 3D reconstruction of NVPC/CNTF electrodes, reconstructed from micro-CT scans: **a.** Uncompressed; **b.** compressed; **c.** fine structure of a single NVPC particle (cross-section indicated in yellow); **d.** summary of structural parameters acquired from XCT; **e.** the volume ratio of different components in a compressed NVPC/CNTF electrode; **f.** the thickness of compressed and uncompressed NVPC/CNTF electrodes with different areal loadings and conventional electrodes (current collector included); The schematics of sodium ion transportation in the pores of **g.** conventional electrodes and **h.** co-ESP electrodes;

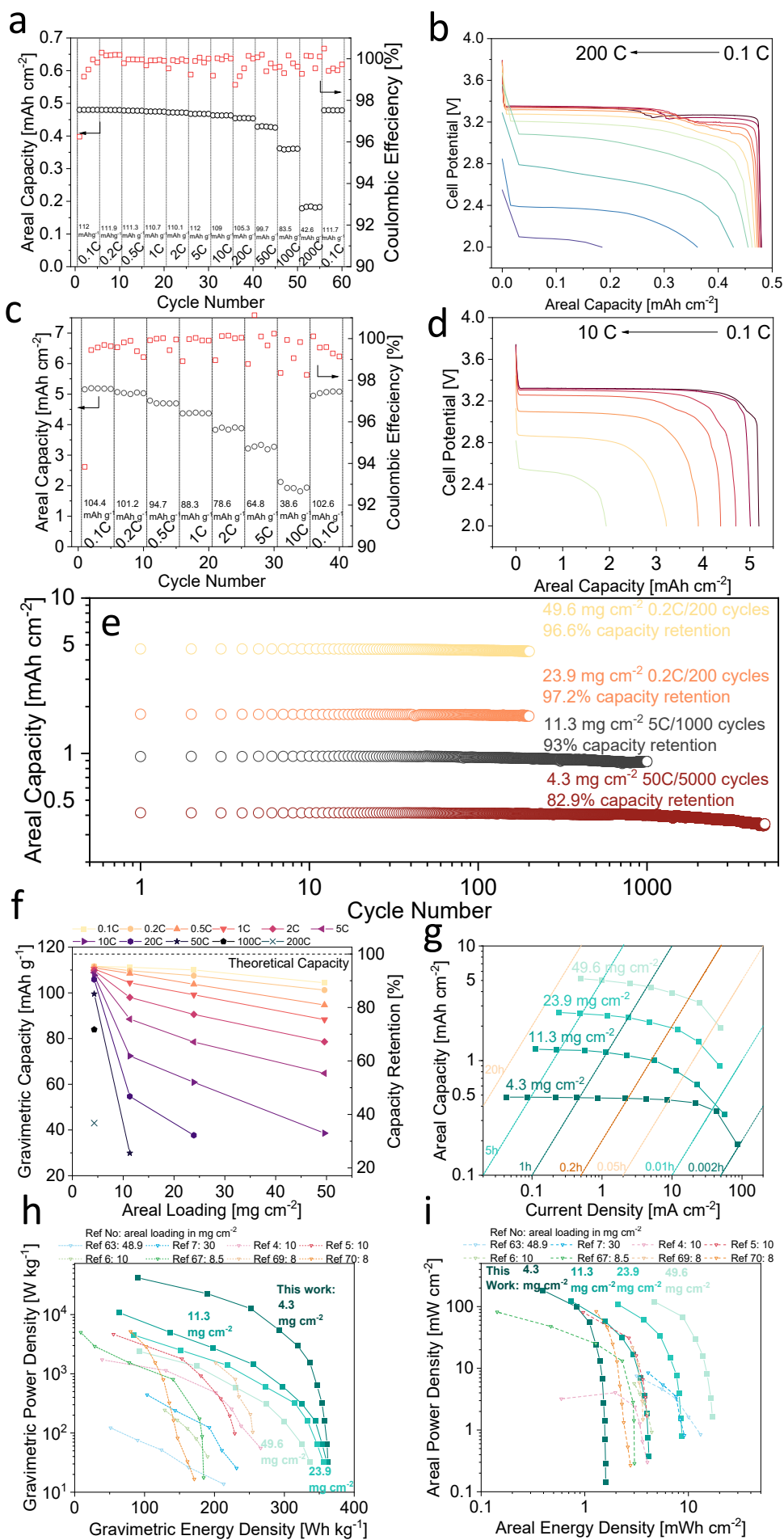

**Figure 4. The performance of co-ESP NVPC cathodes with different areal loading:** 4.3 mg cm$^{-2}$ cathode's **a.** rate performance and **b.** voltage profile; 49.6 mg cm$^{-2}$ cathode's **c.** rate performance and **d.** voltage profile; **e.** Cycling stability of different areal loading half cells; **f.** the change of specific discharge capacity with areal loading and cycling rate; **g.** The change of areal capacity with areal current; Ragone plots of **h.** gravimetric energy density versus power density and **i.** areal energy density versus areal power density, including the data acquired from previous sodium-ion battery half cells for comparison.

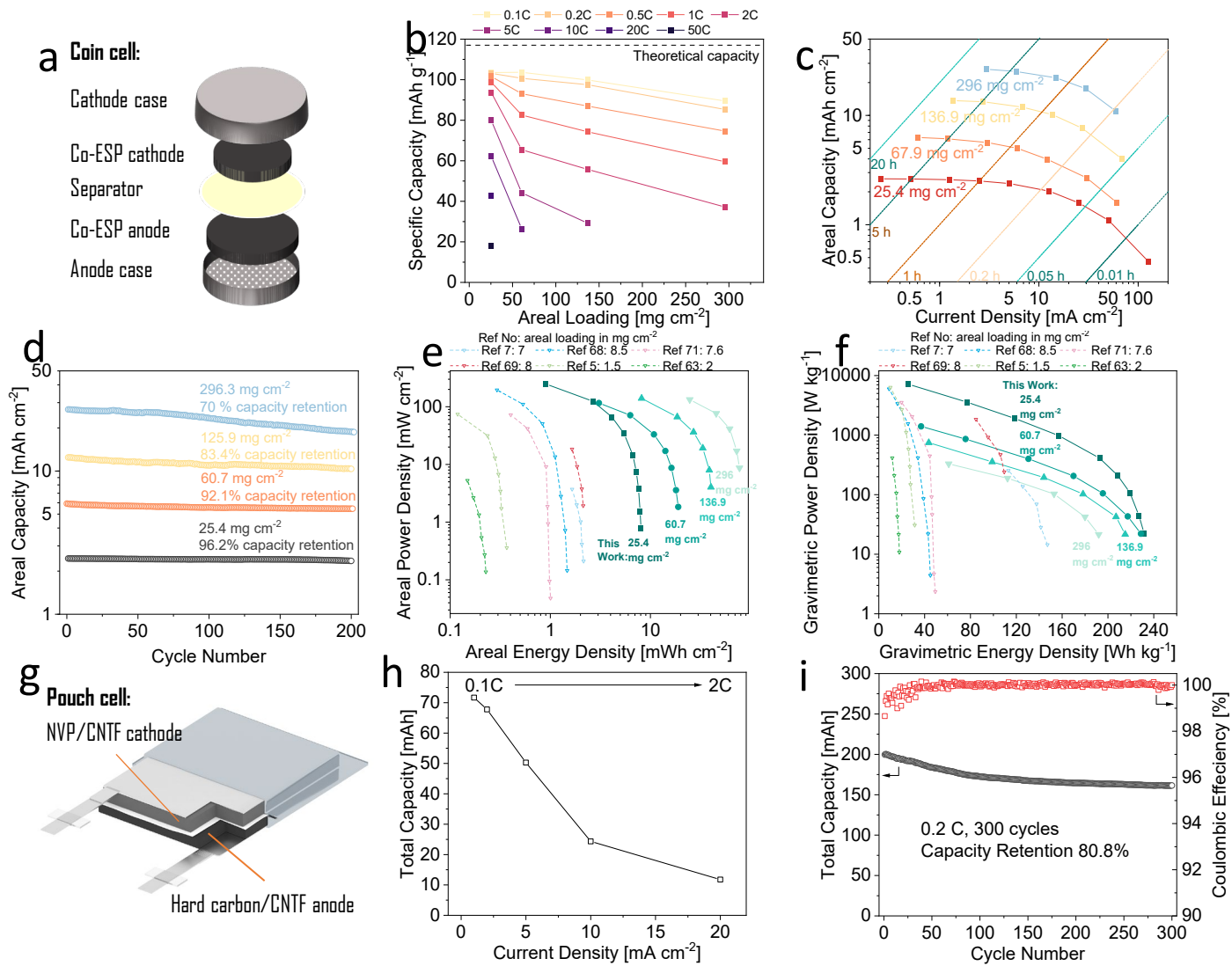

**Figure 5 The performance of sodium ion batteries full cells and pouch cells made of co-ESP NVPC cathodes and co-ESP HC anodes: a.** Schematic diagram of a sodium-ion battery full cell; **b.** the change of specific discharge capacity with areal loading and cycling rate; **c.** The change of areal capacity with areal current; **d.** Cycling stability of different areal loading full cells; Ragone plots of **e.** gravimetric energy density versus power density and **f.** areal energy density versus areal power density, including the data acquired from previous sodium-ion battery full cells for comparison; pouch cell performance with 100 mg cm$^{-2}$ cathode loading: **g.** Schematic diagram of a sodium-ion battery pouch cell; **h.** Rate performance and **i.** Cycling performance of 0.2 Ah, 100 mg cm$^{-2}$ cathode loading pouch cell.

# Supplementary Appendix 1: Figures

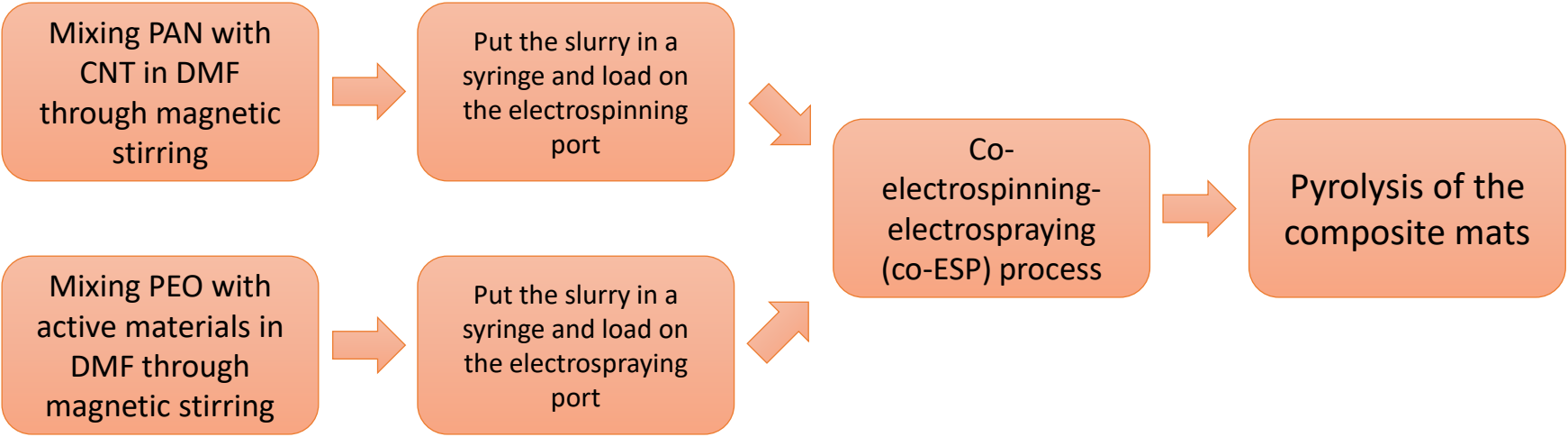

**Figure S1** Flow chart showing the process of fabricating co-ESP electrode

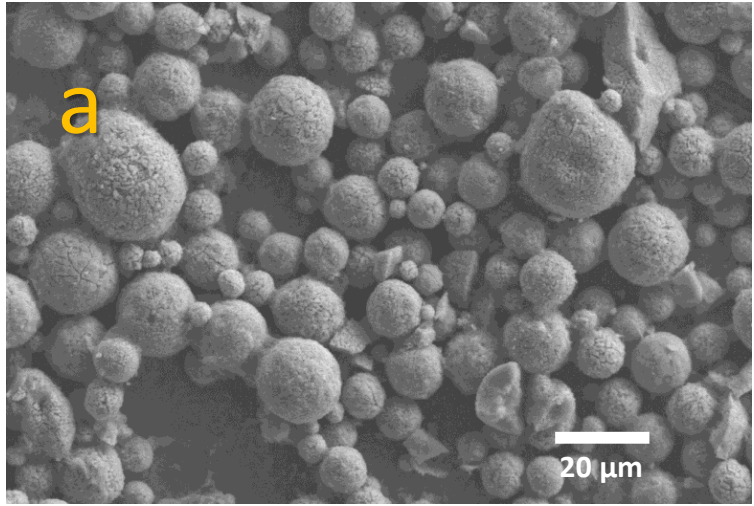 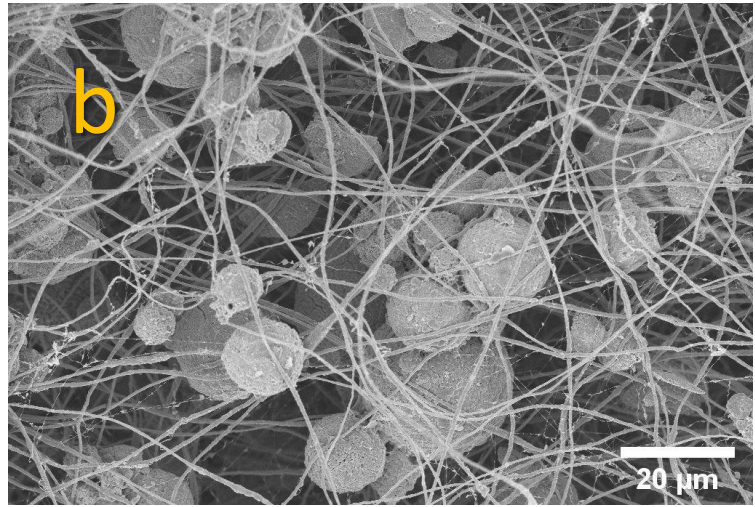 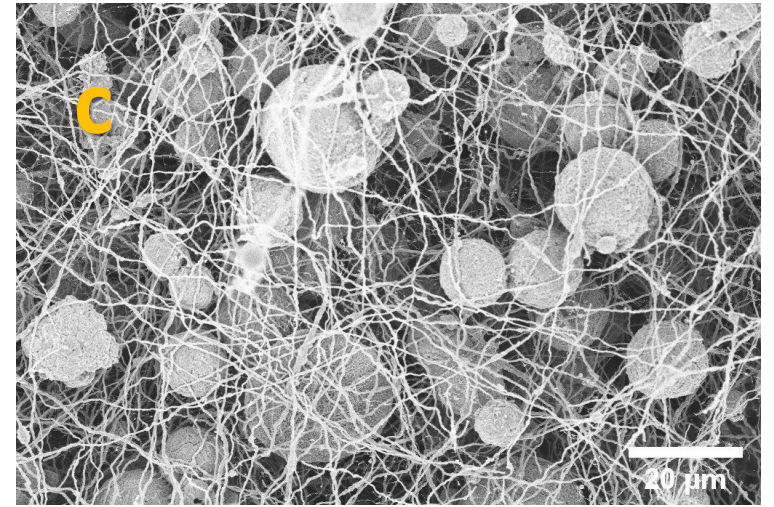

**Figure S2** Morphology of a. Pristine NVPC particles; b. NVPC-PEO/PAN-CNT composite mat, right after the co-ESP fabrication; c. NVPC/CNT-CNF electrode, after pyrolysis.

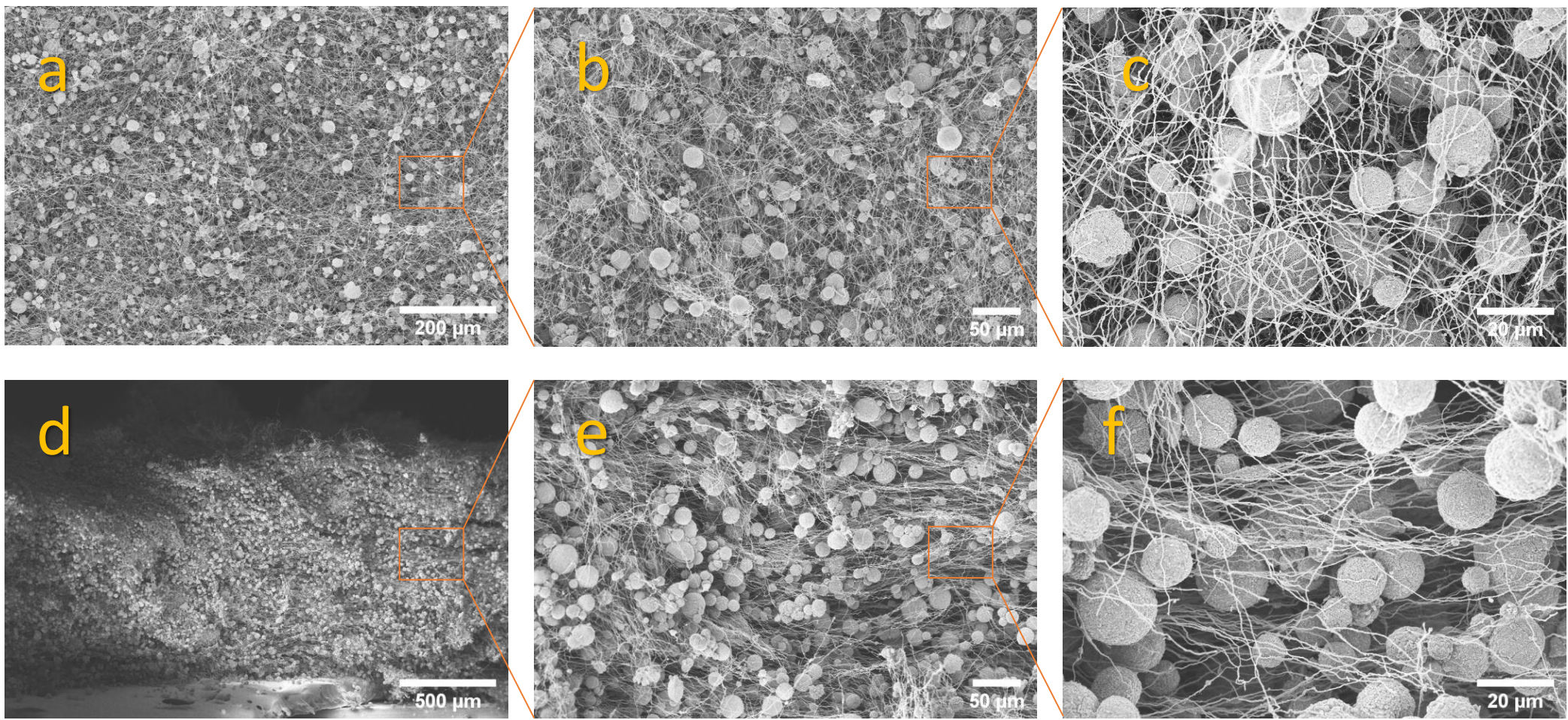

**Figure S3** Morphology of 97.5% NVPC/CNTF from **a-c.** top view and **d-f.** cross view of different magnifications

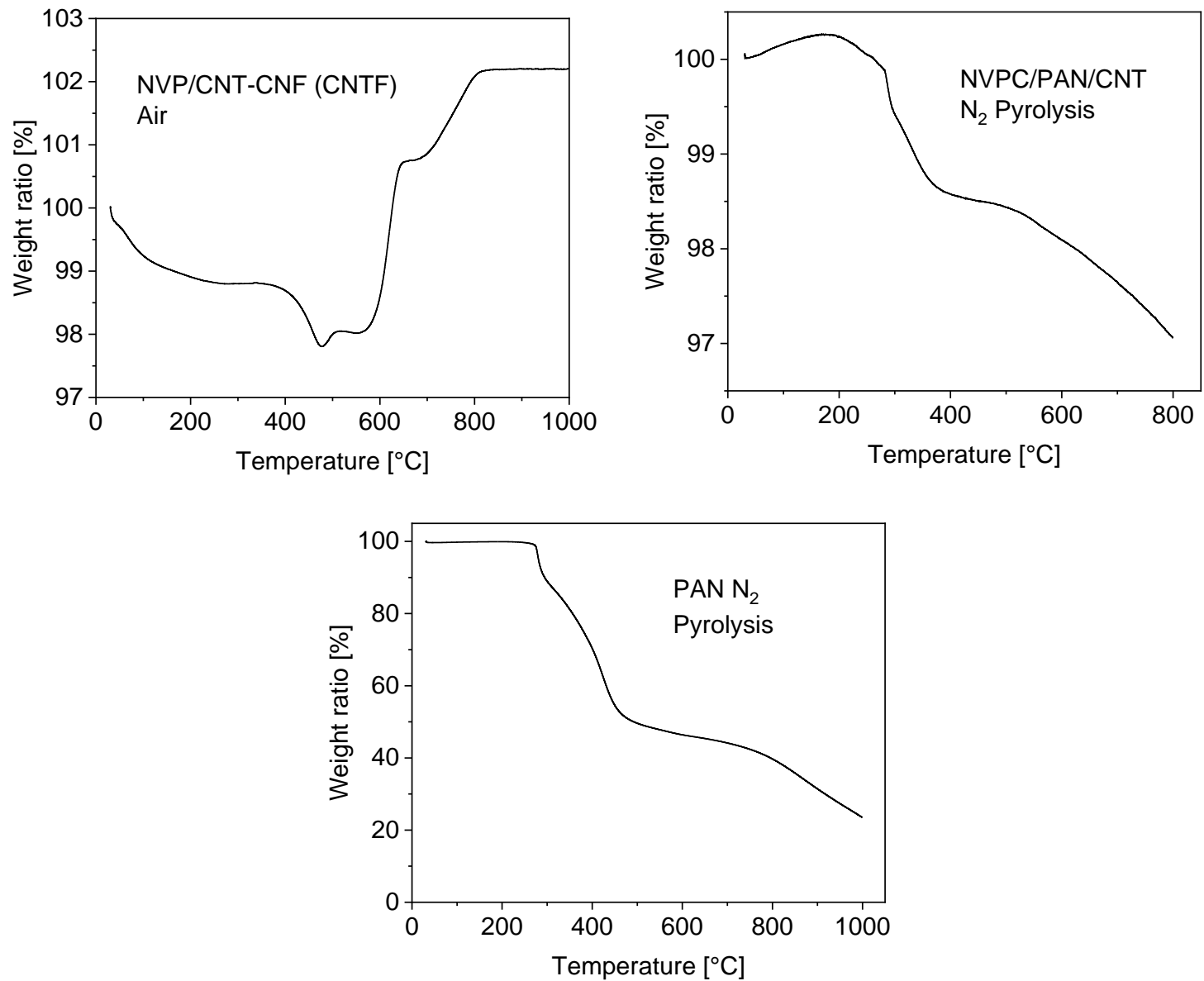

**Figure S4** TGA of NVPC/CNTF in air, NVPC/PAN-CNT right after co-ESP fabrication in N$_2$ and pure PAN fibres in N$_2$

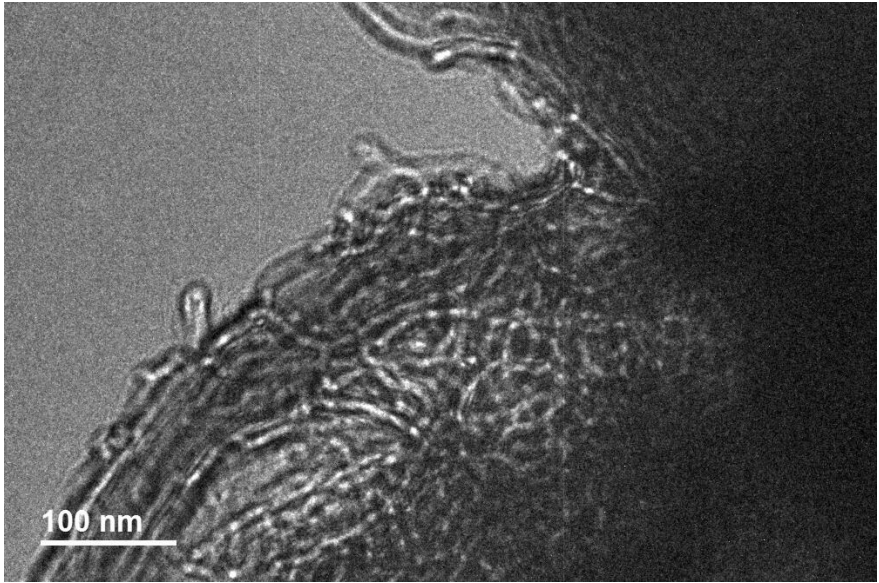 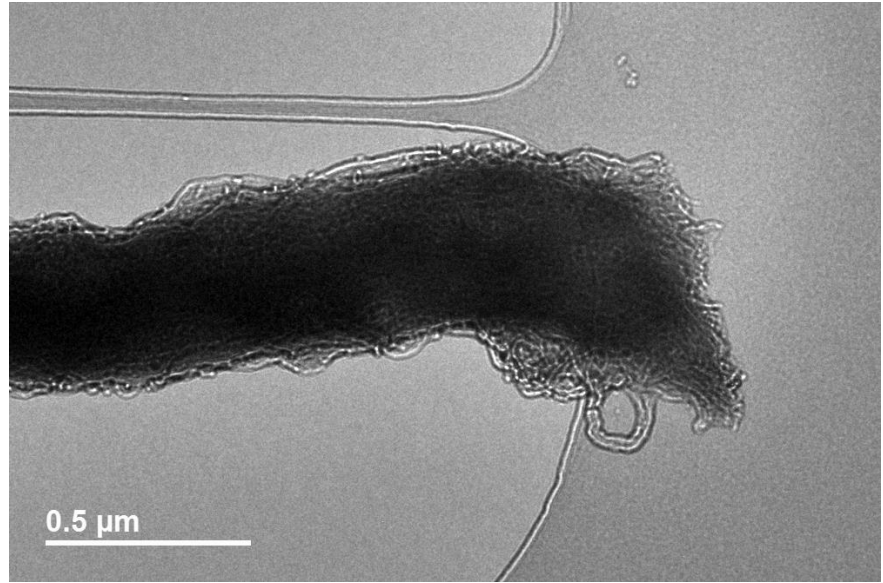

**Figure S5** TEM of carbon nanotube embedded carbon fibre (CNTF)

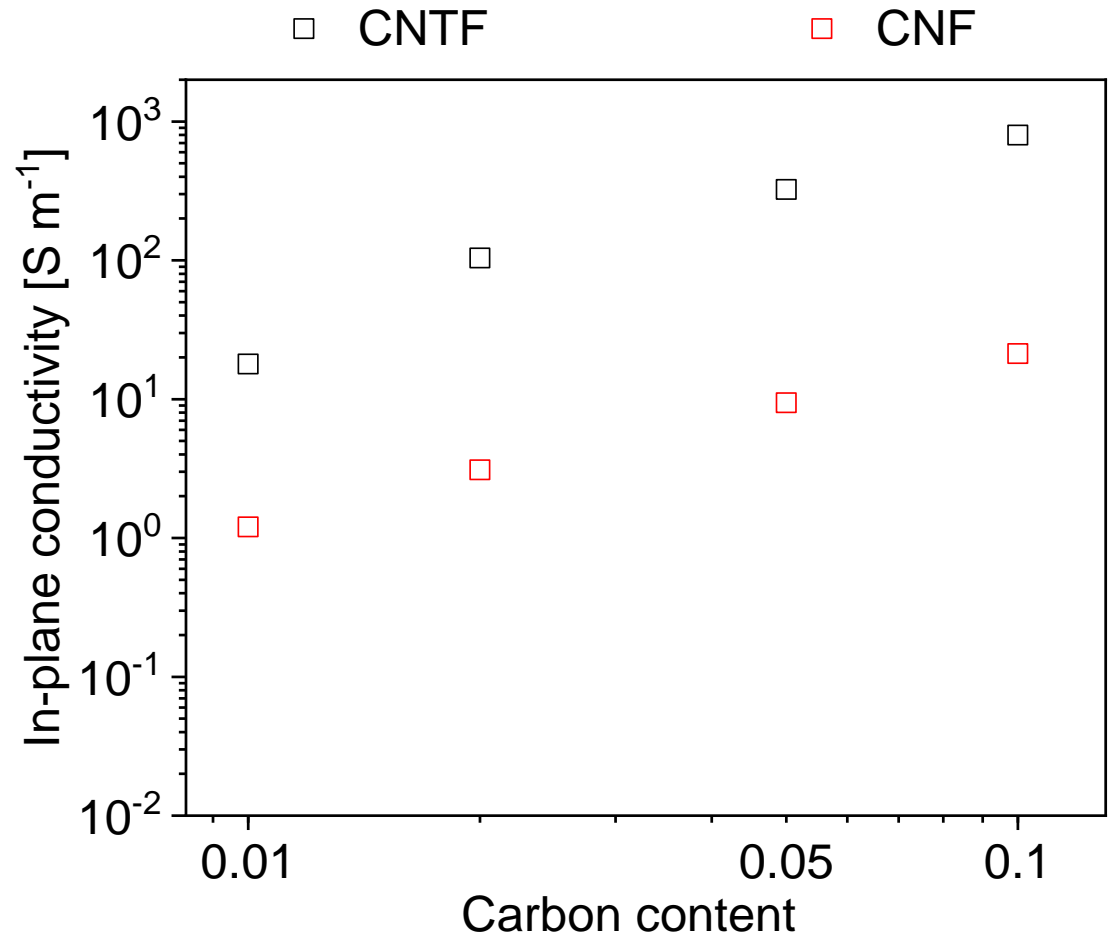

**Figure S6** In-plane conductivity of PAN-derived CNF network and CNT embedded CNF (CNTF) network

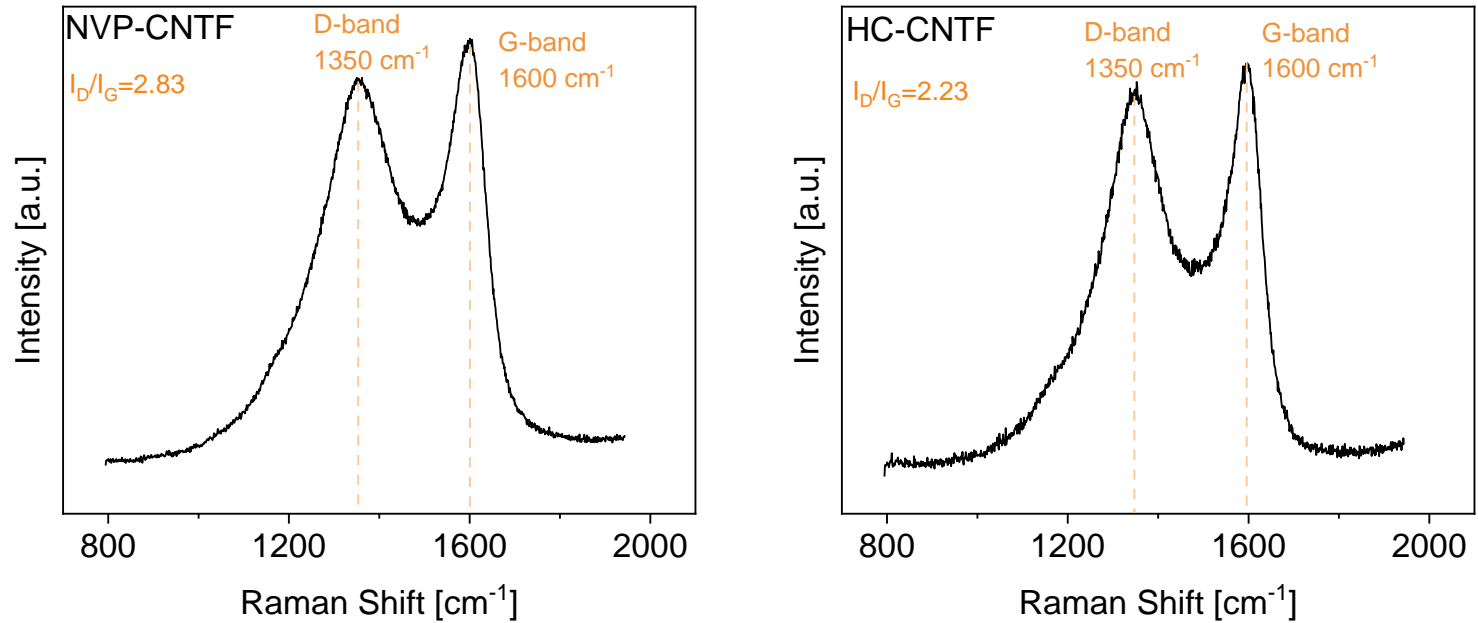

**Figure S7** Raman spectrum of co-ESP NVPC/CNTF electrodes and HC/CNTF electrodes

The $I_D/I_G$ value is inversely related to the degree of graphitisation of the carbon sample (Ferrari and Robertson 2000). The $I_D/I_G$ values of NVPC-CNTF and HC-CNTF are 2.83 and 2.23, respectively. The lower $I_D/I_G$ of the HC-CNTF is caused by its higher pyrolysis temperature (1100 vs. 850 °C). Both samples have lower $I_D/I_G$ than previously reported PAN-derived carbon fibres pyrolyzed at same temperatures (Wang, Serrano et al. 2003). This is mainly due to the presence of CNT in the CNTF fibres(Bokobza and Zhang 2012).

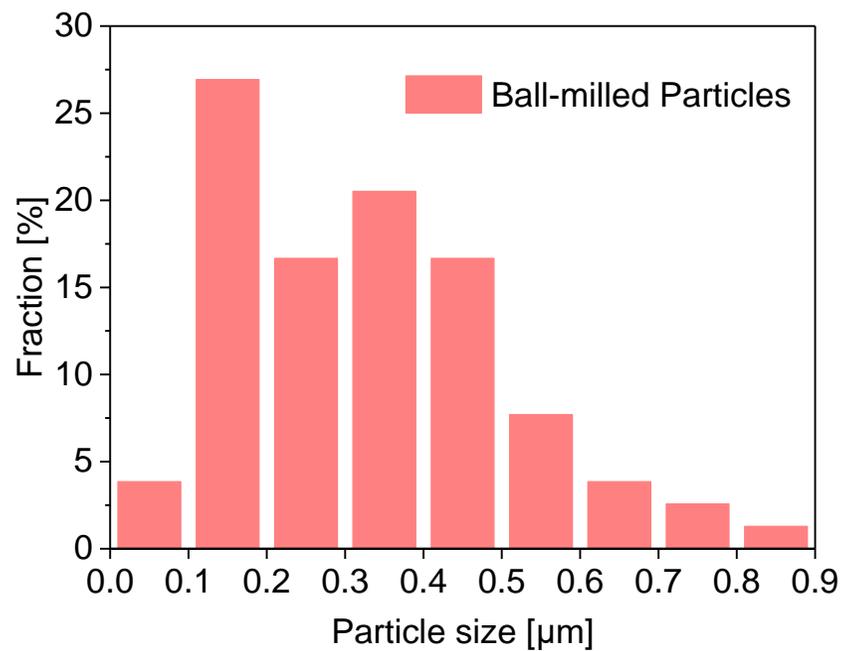 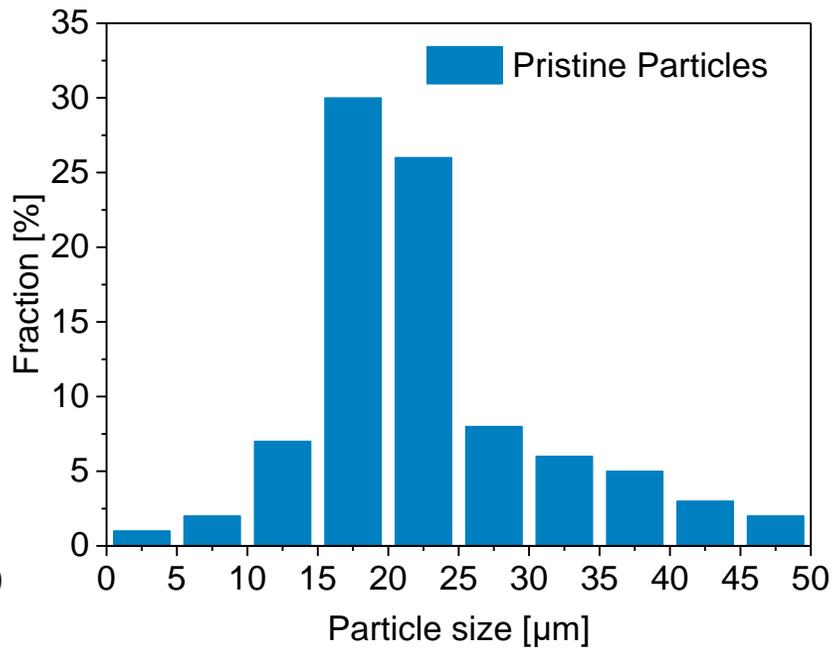

**Figure S8** Particle size distribution of NVPC and ball-milled NVPC particles

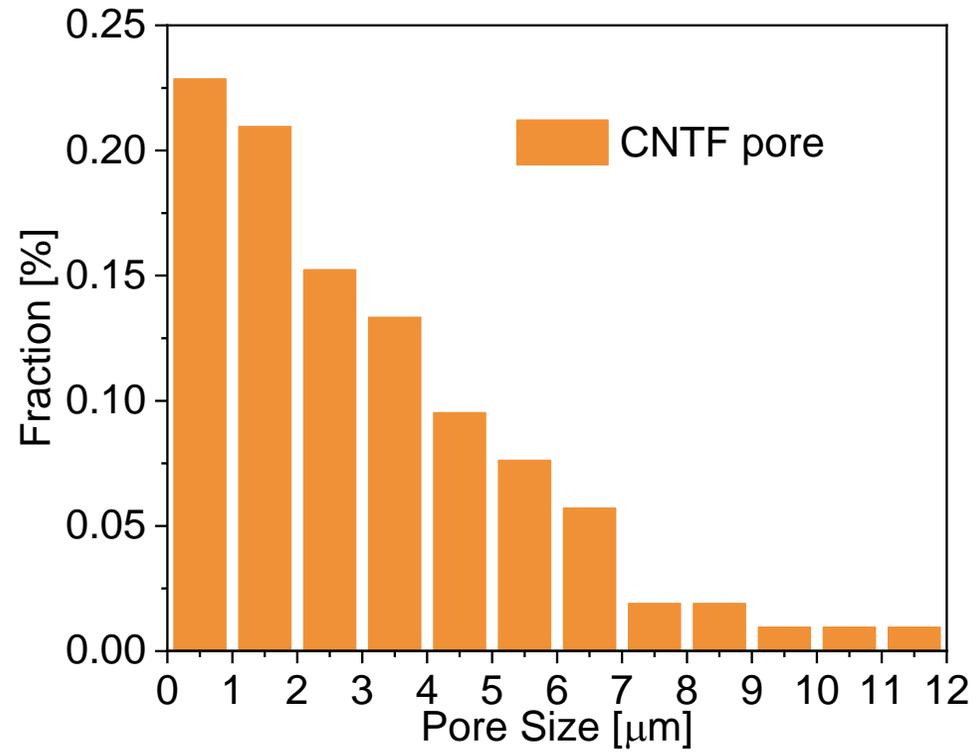

**Figure S9** Pore size distribution of CNTF network

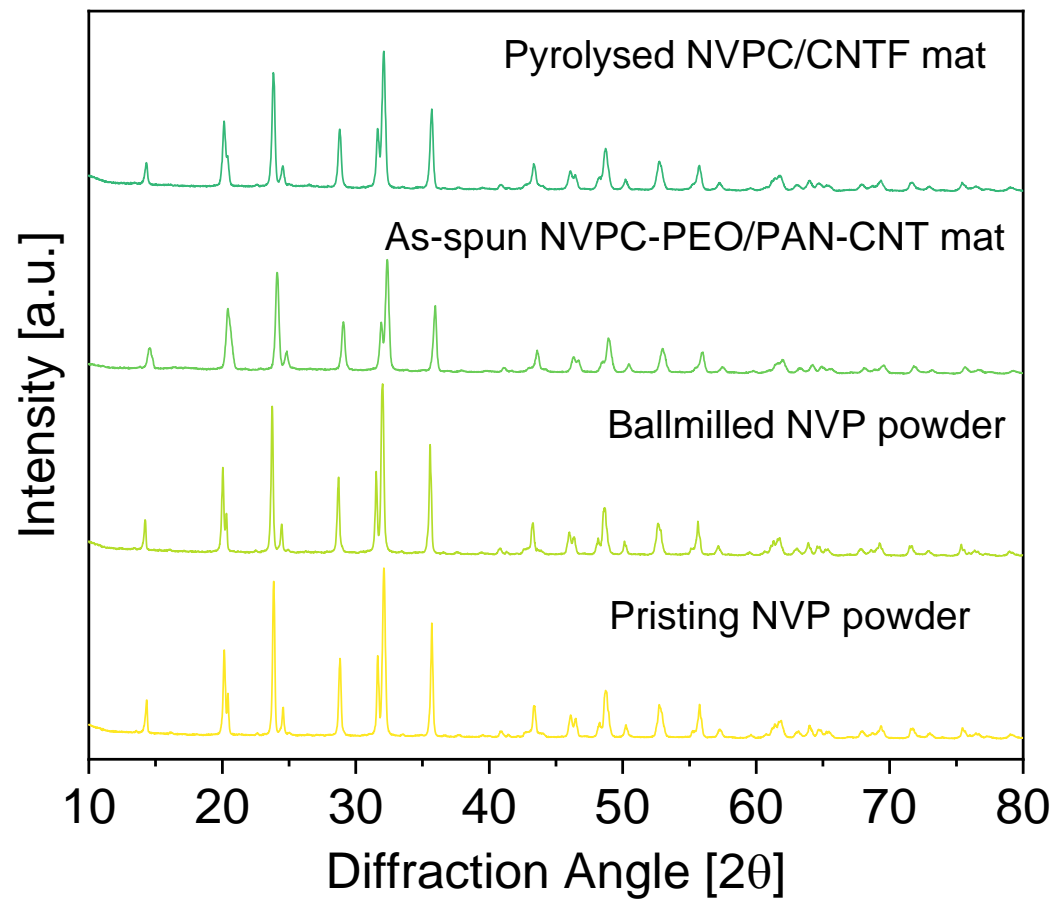

**Figure S10** The X-ray diffraction (XRD) pattern of pristine and ball-milled NVPC particles and NVPC/CNTF electrode at different stages of preparation

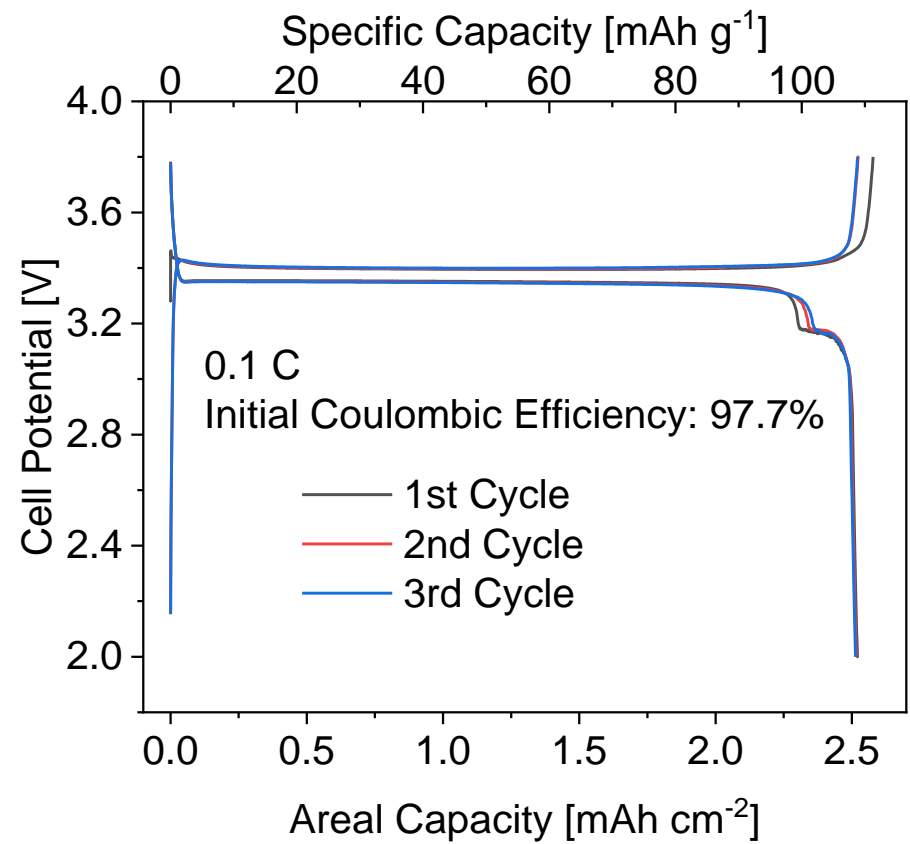 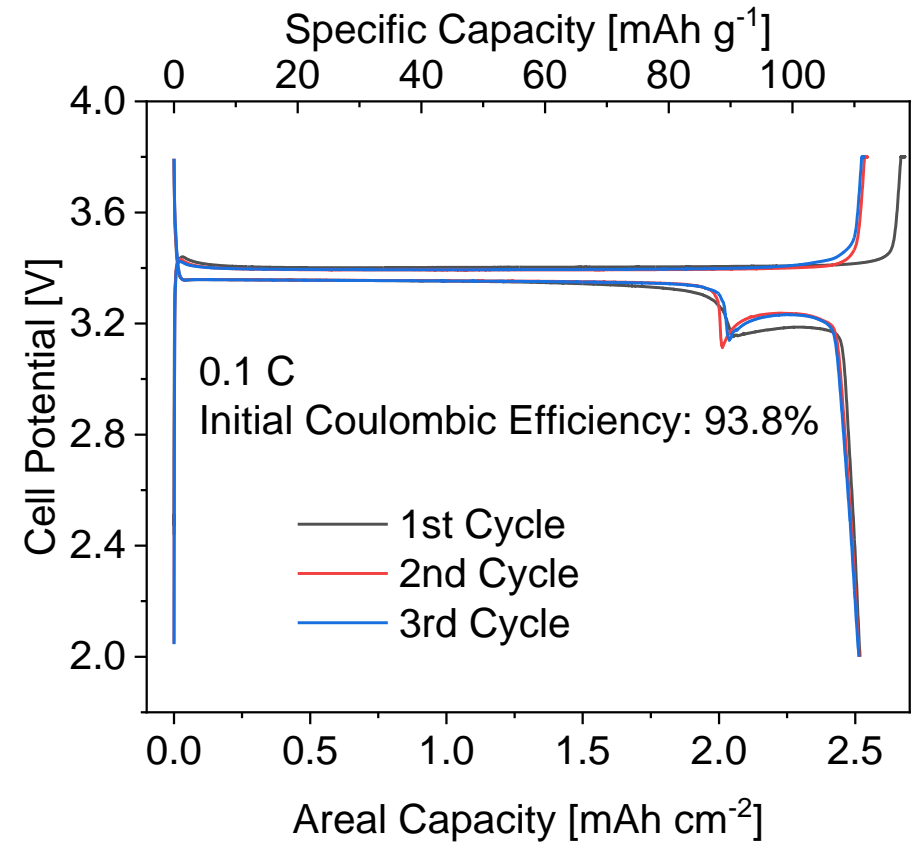

**Figure S11** First three cycles of **a)** pristine and **b)** ball-milled NVPC/CNTF half cells

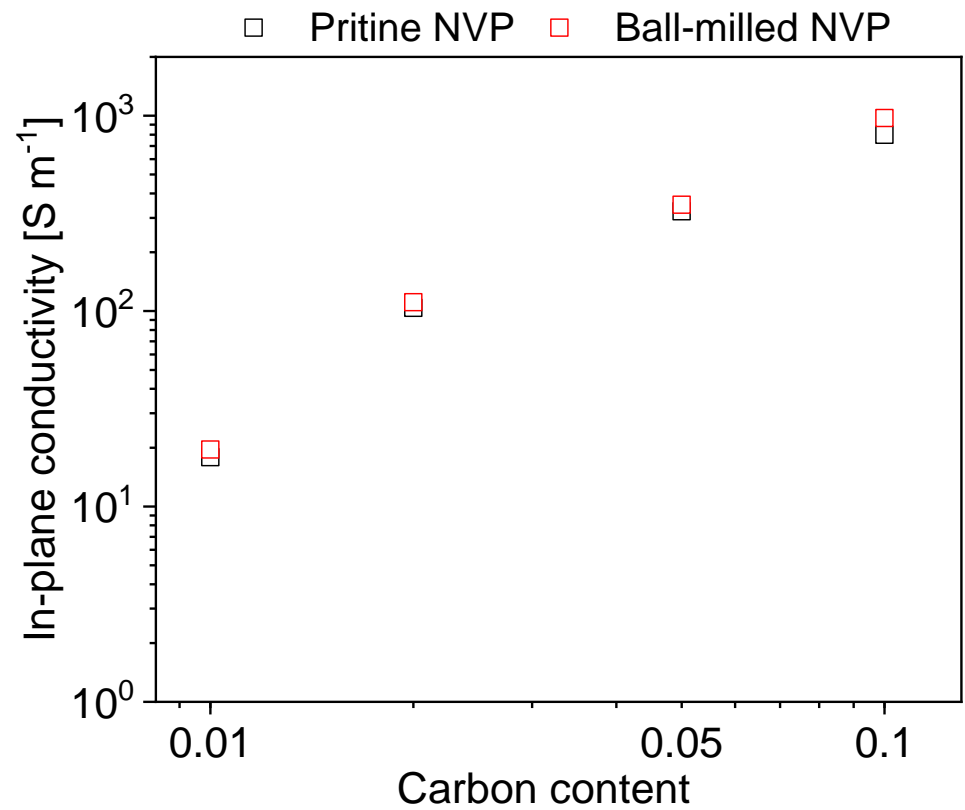

**Figure S12** In-plane conductivity of co-ESP electrodes made by pristine NVPC and ball-milled NVPC

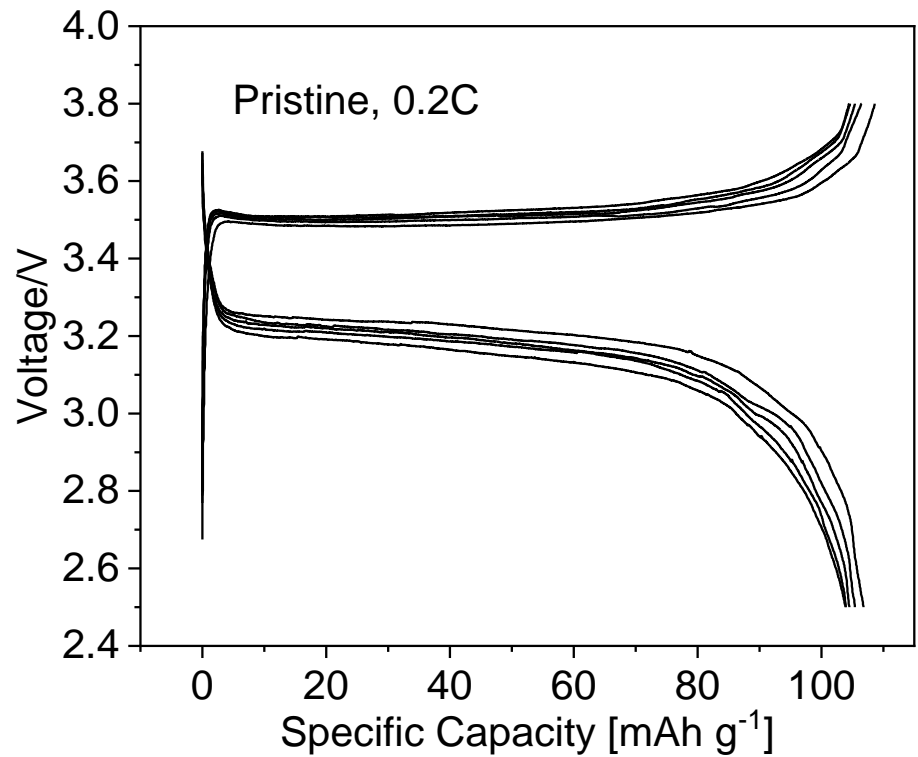 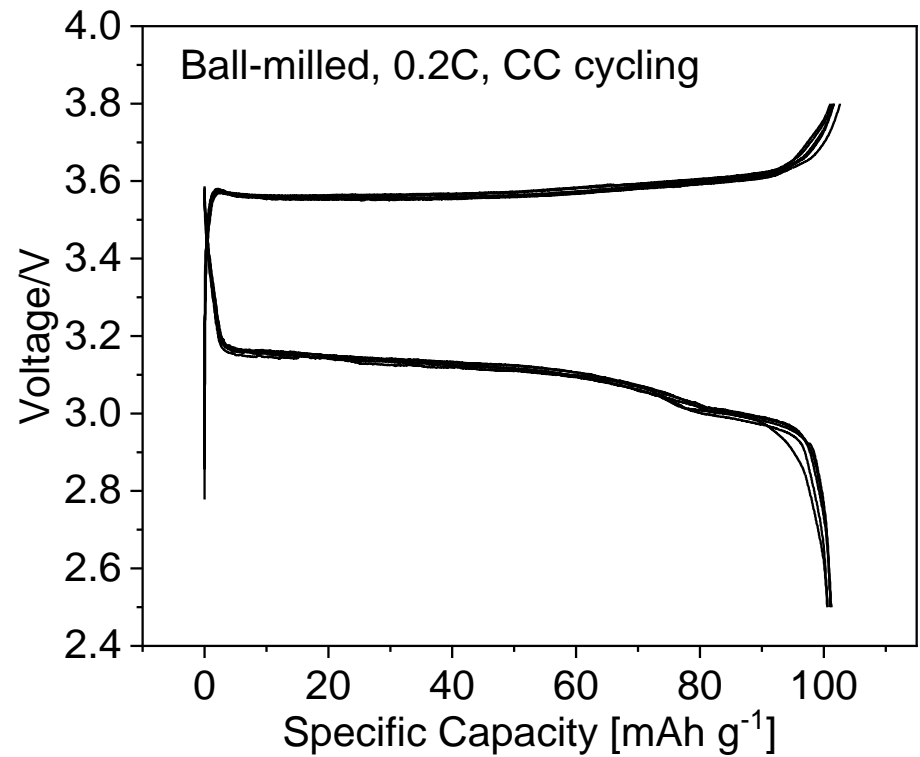

**Figure S13** 0.2C voltage profile of conventional slurry-casted electrode with pristine and ball-milled NVPC particles, CC cycling

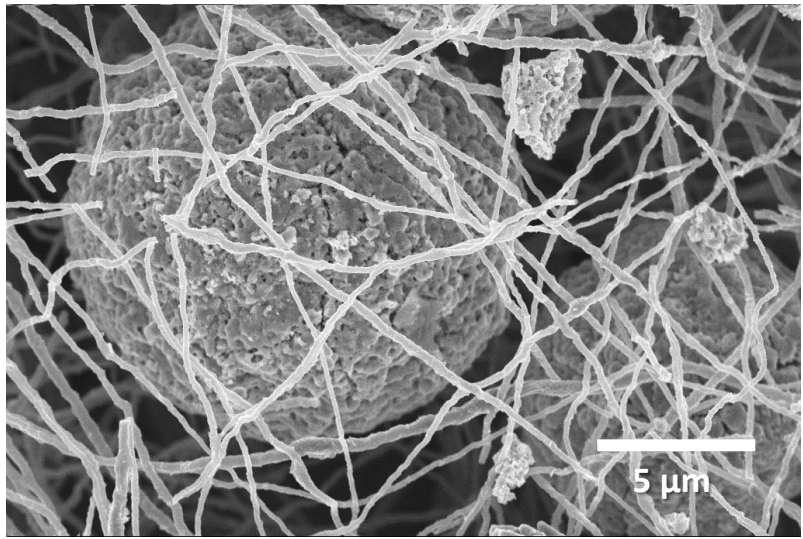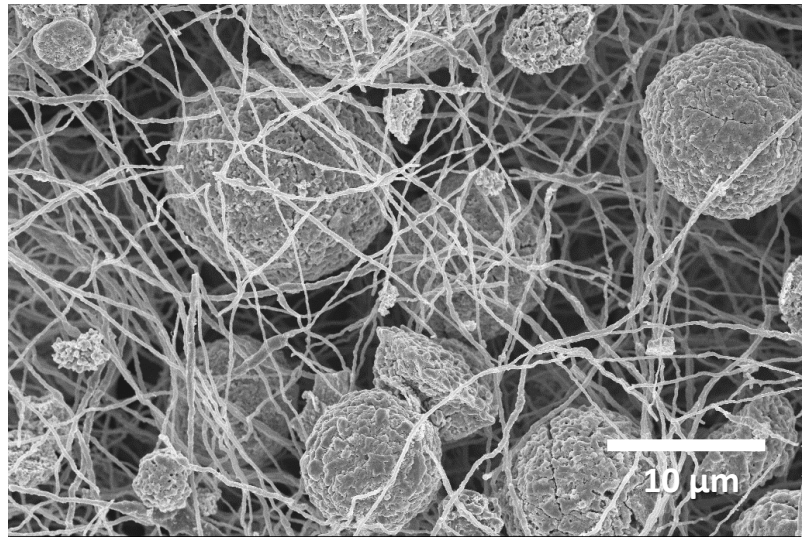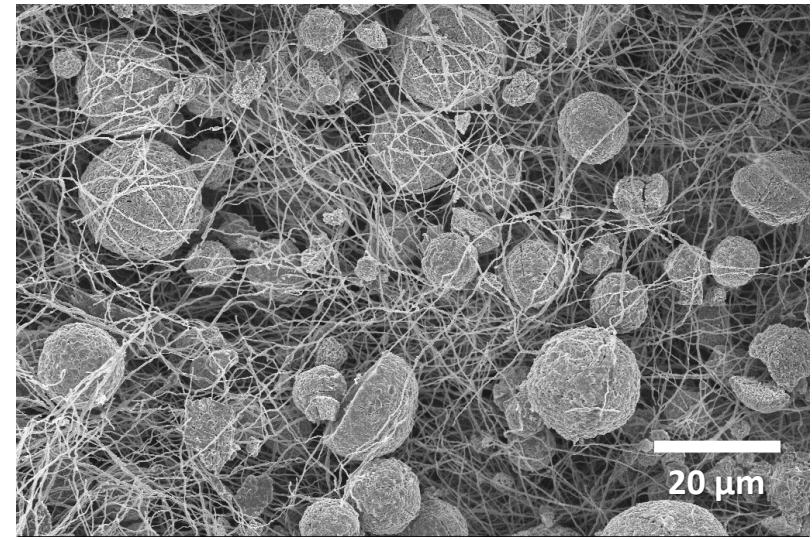
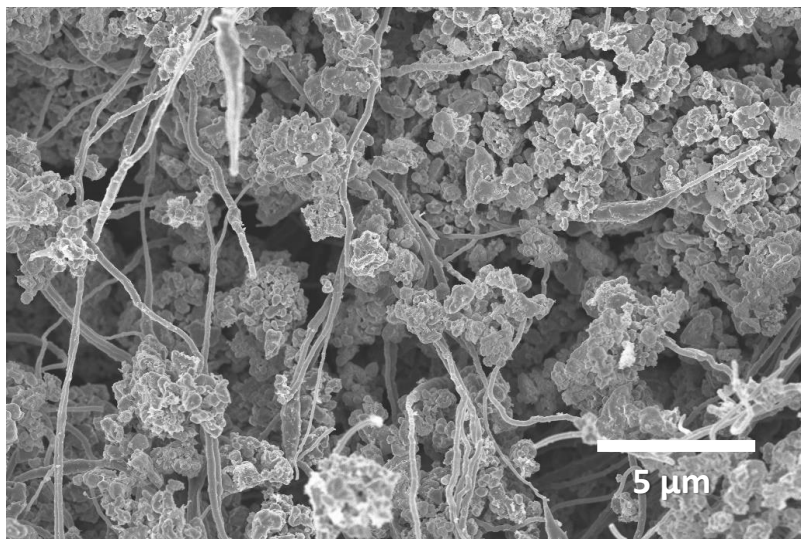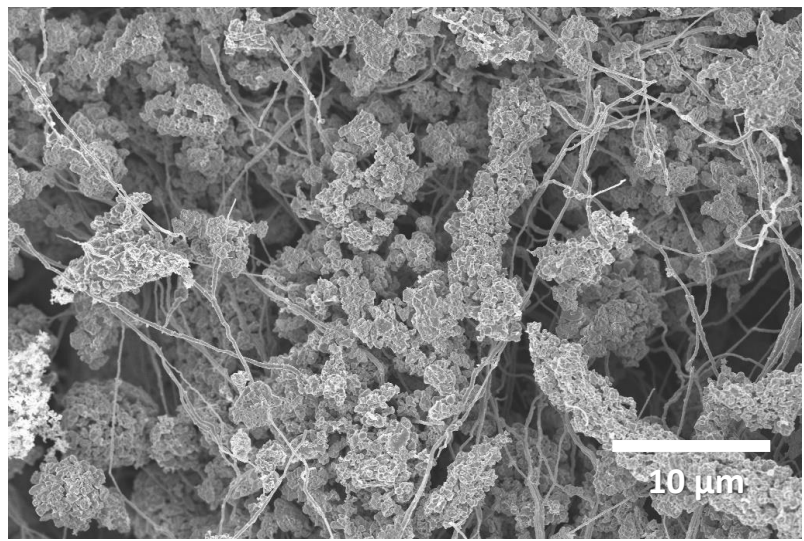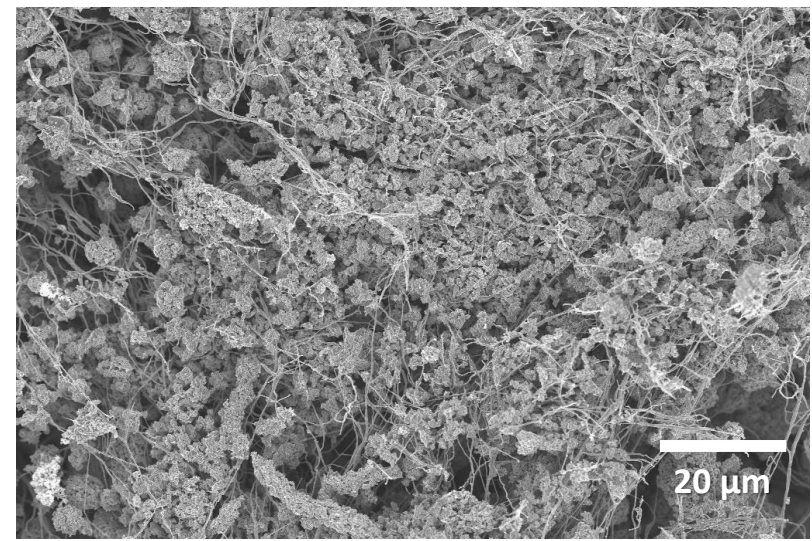

**Figure S14** Photo/SEM images of co-ESP NVPC electrodes from **top**: pristine and **bottom**: ball-milled after 100 cycles (97.5wt% active content)

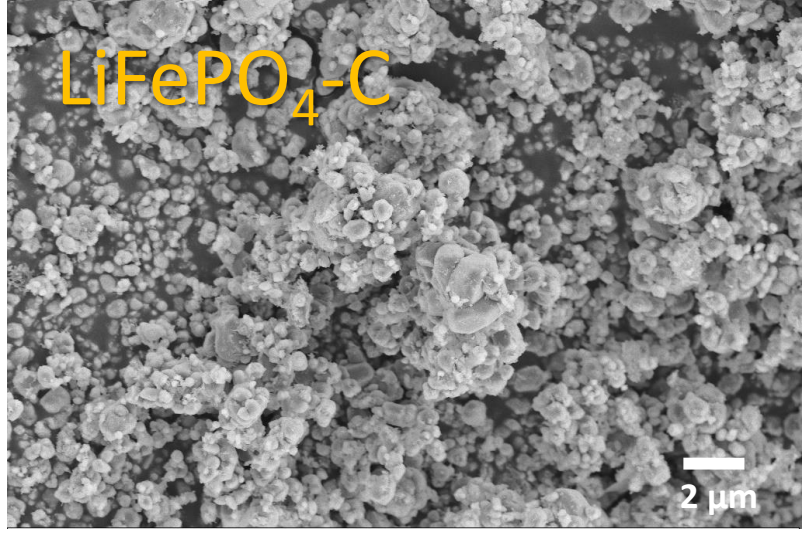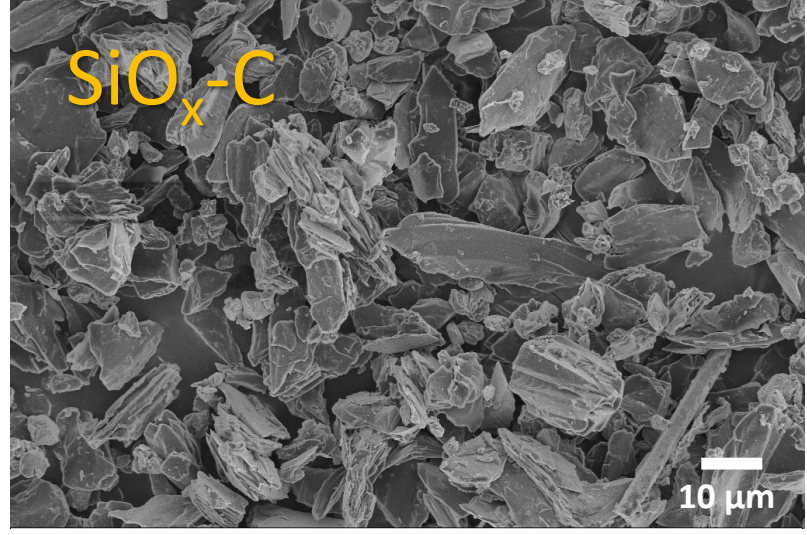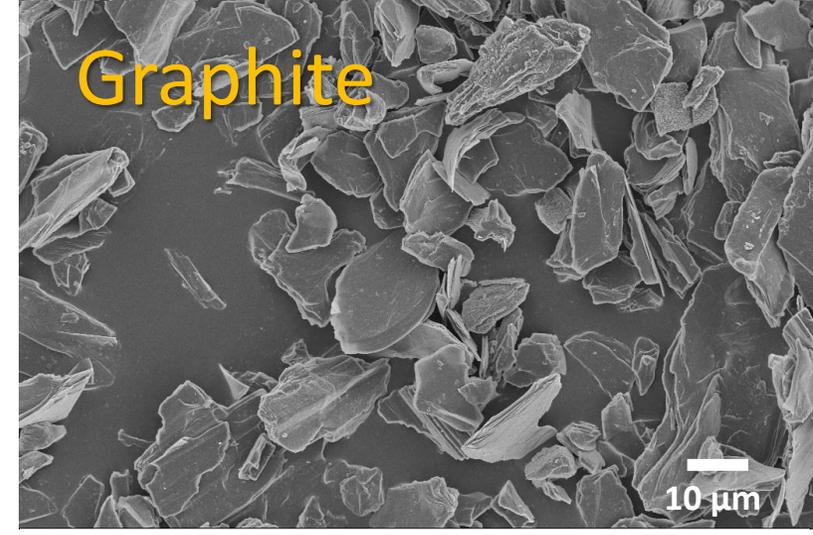
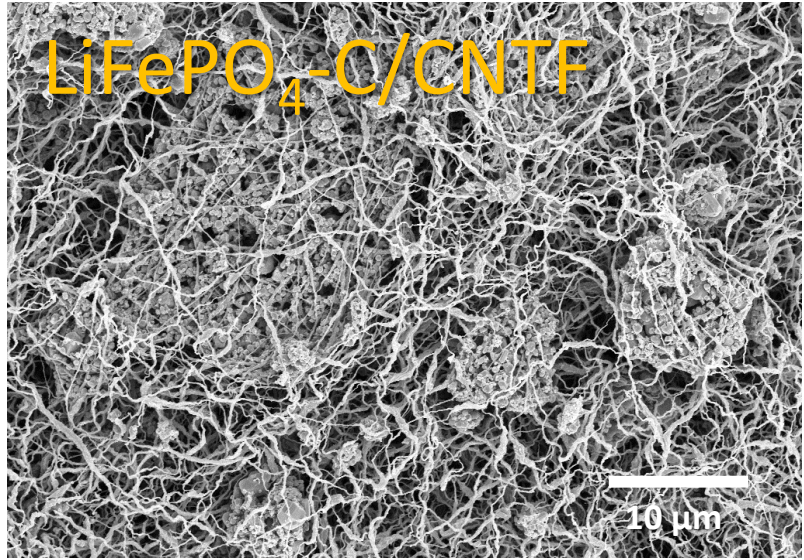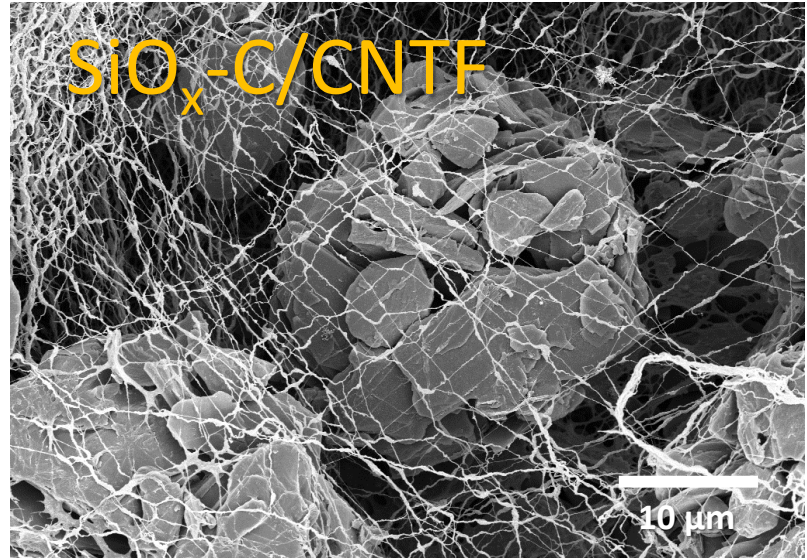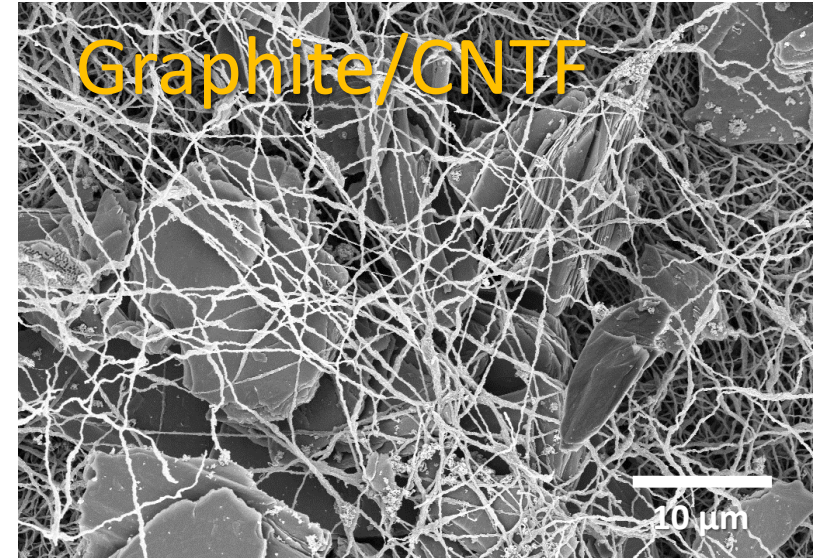

**Figure S15** Morphologies of LiFePO$_4$-C , SiO$_x$-C and graphite electrode produced by co-ESP method

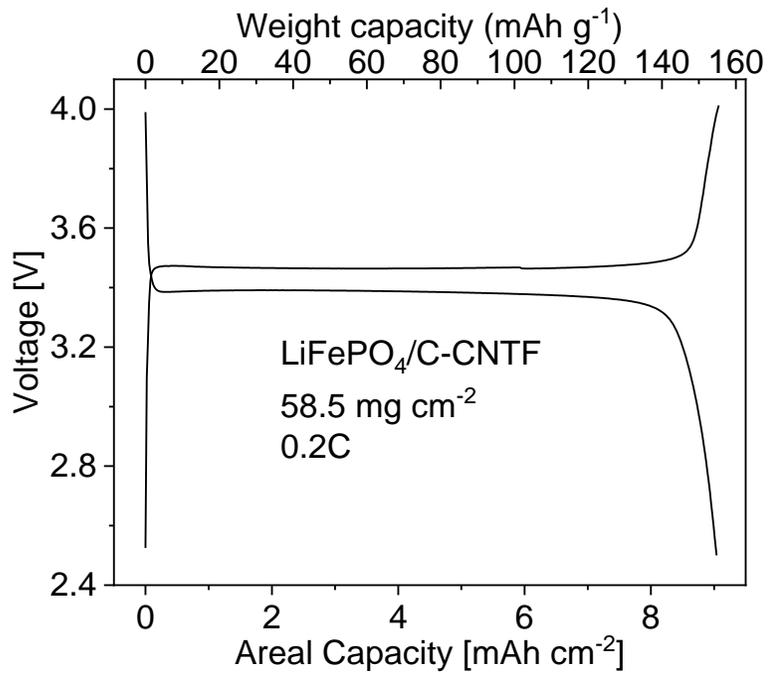 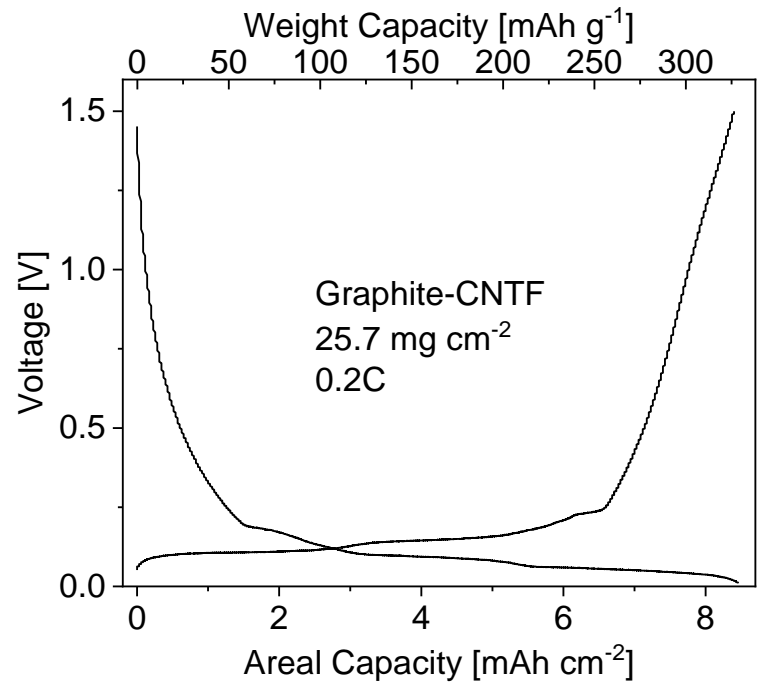 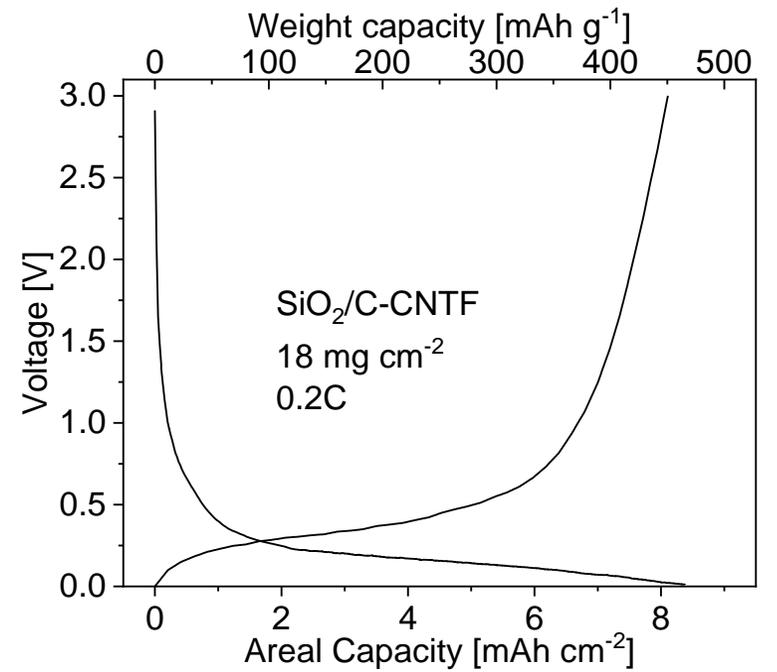

**Figure S16** Voltage profiles of high areal loading LiFePO$_4$-C, graphite, and SiO$_x$-C electrode produced by co-ESP method

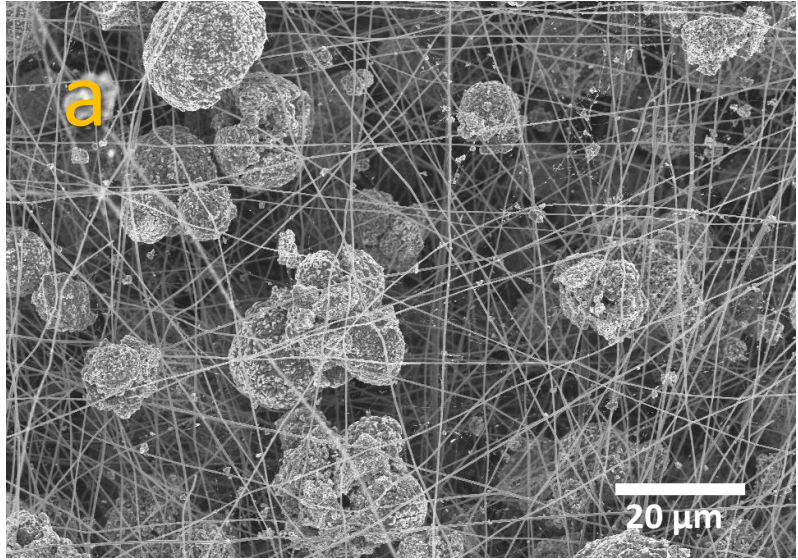 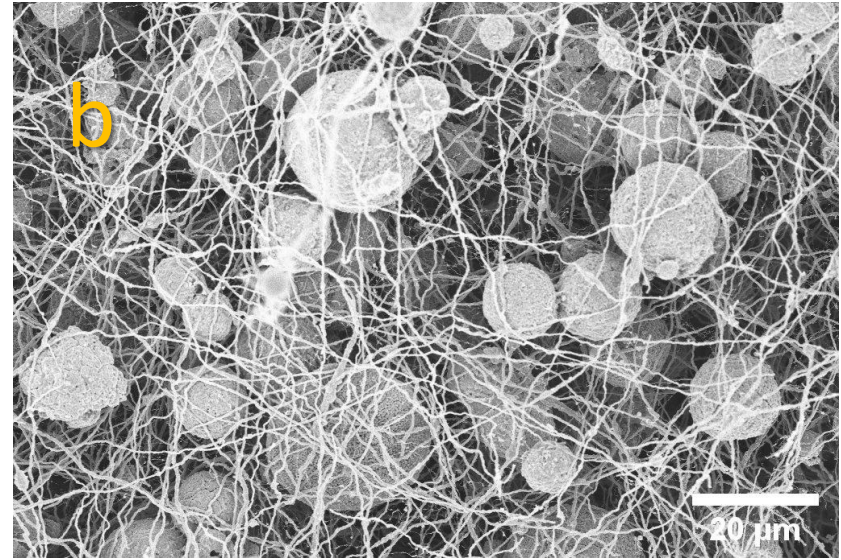
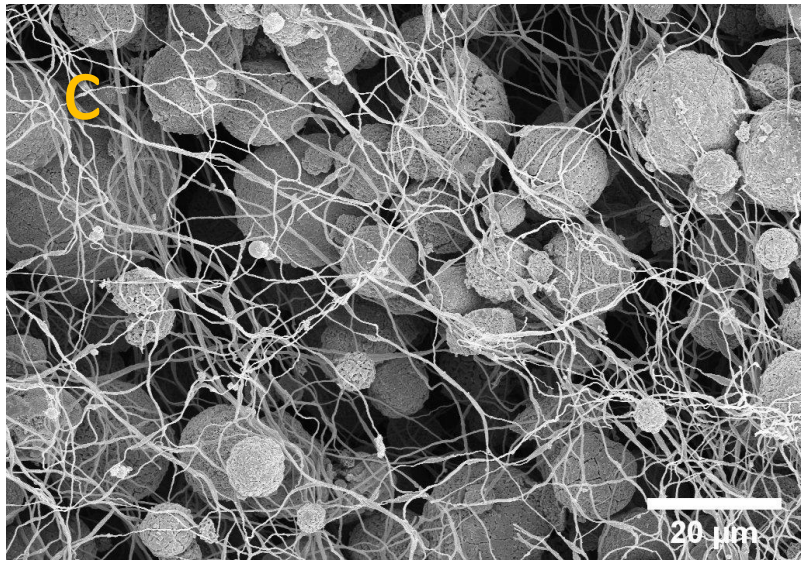 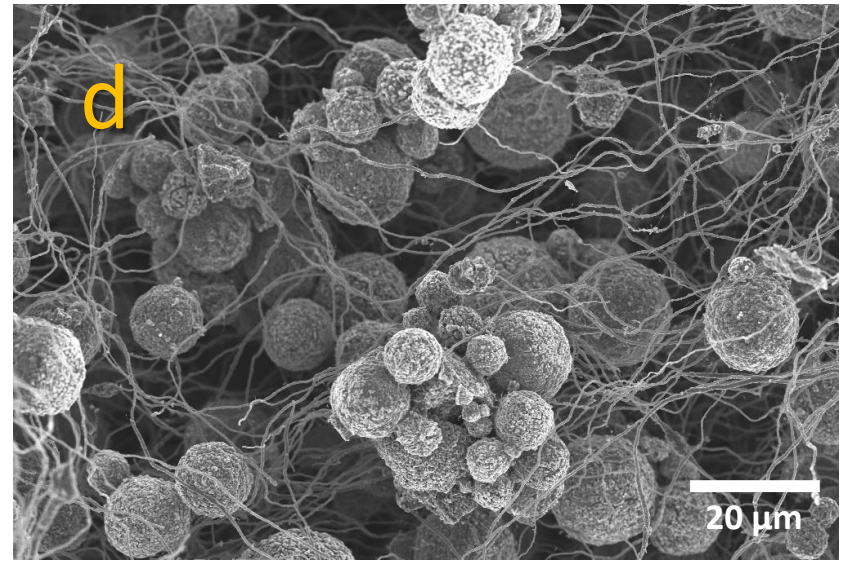

**Figure S17** Morphology of NVPC/CNTF with a. 90 wt%; b. 97.5 wt%; c. 98 wt%; d. 99 wt% NVPC content

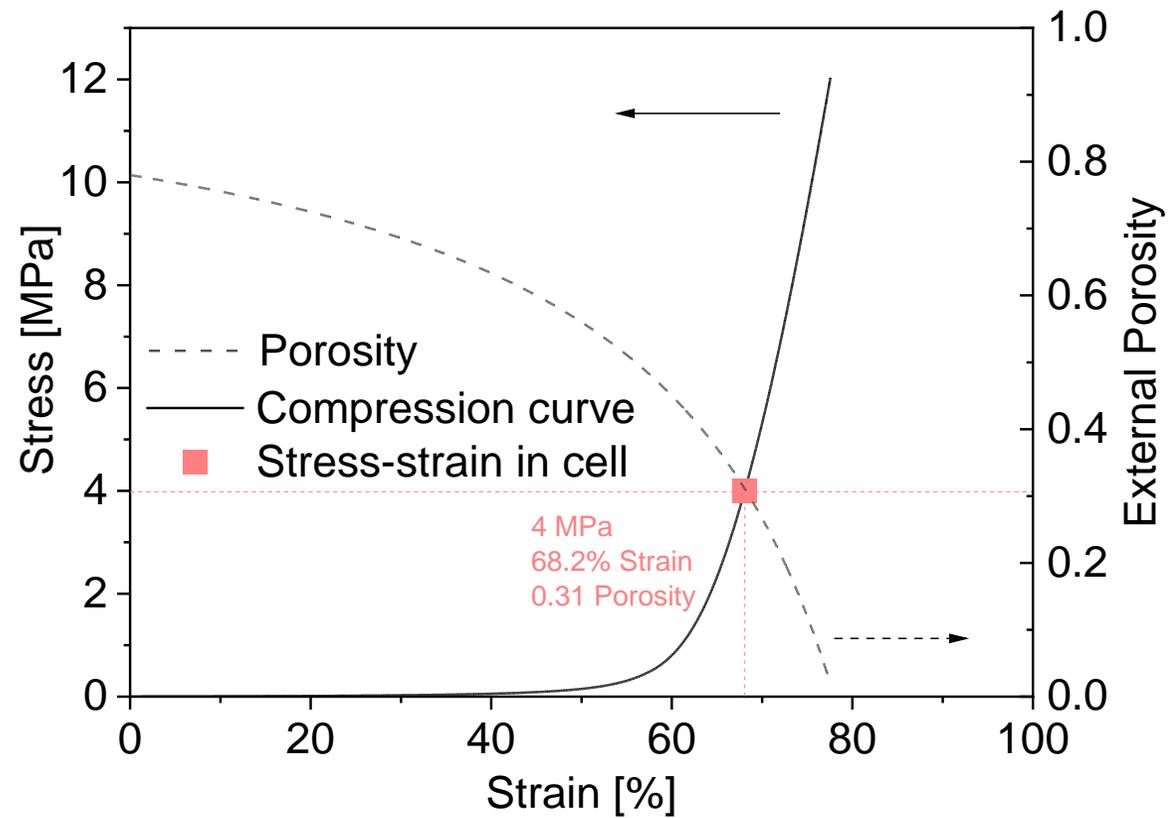

**Figure S18** The relation of stress, strain and external porosity of the NVPC/CNTF electrodes

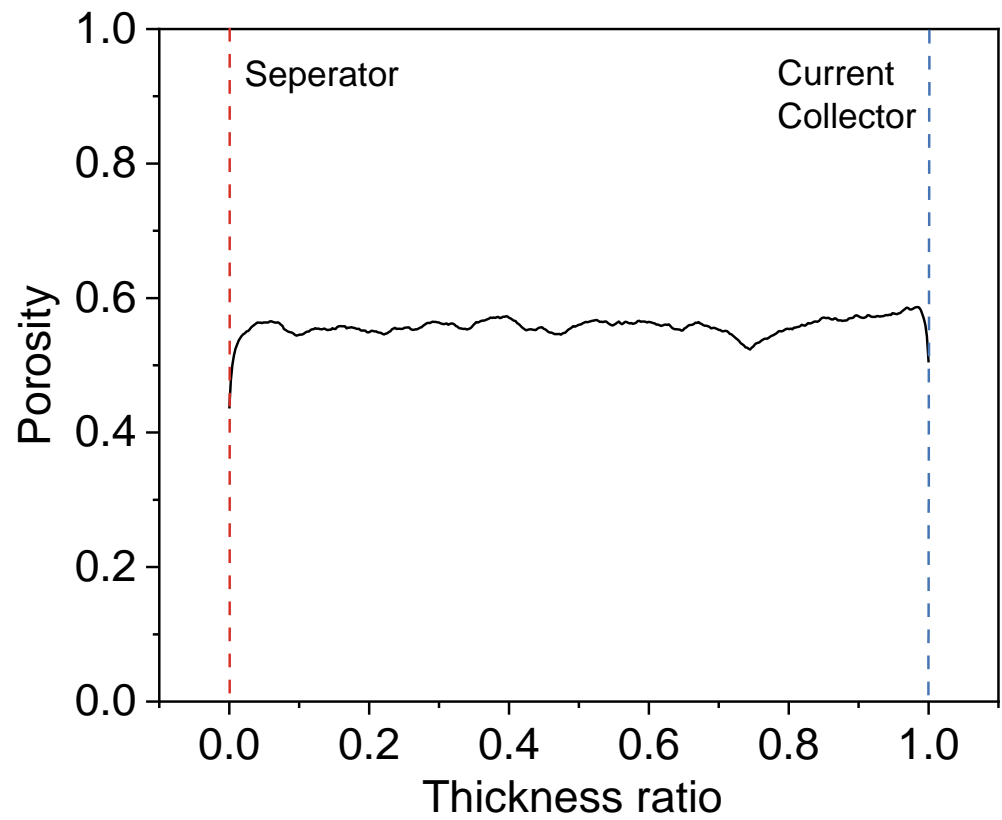

**Figure S19** Uniformity of porosity across thickness

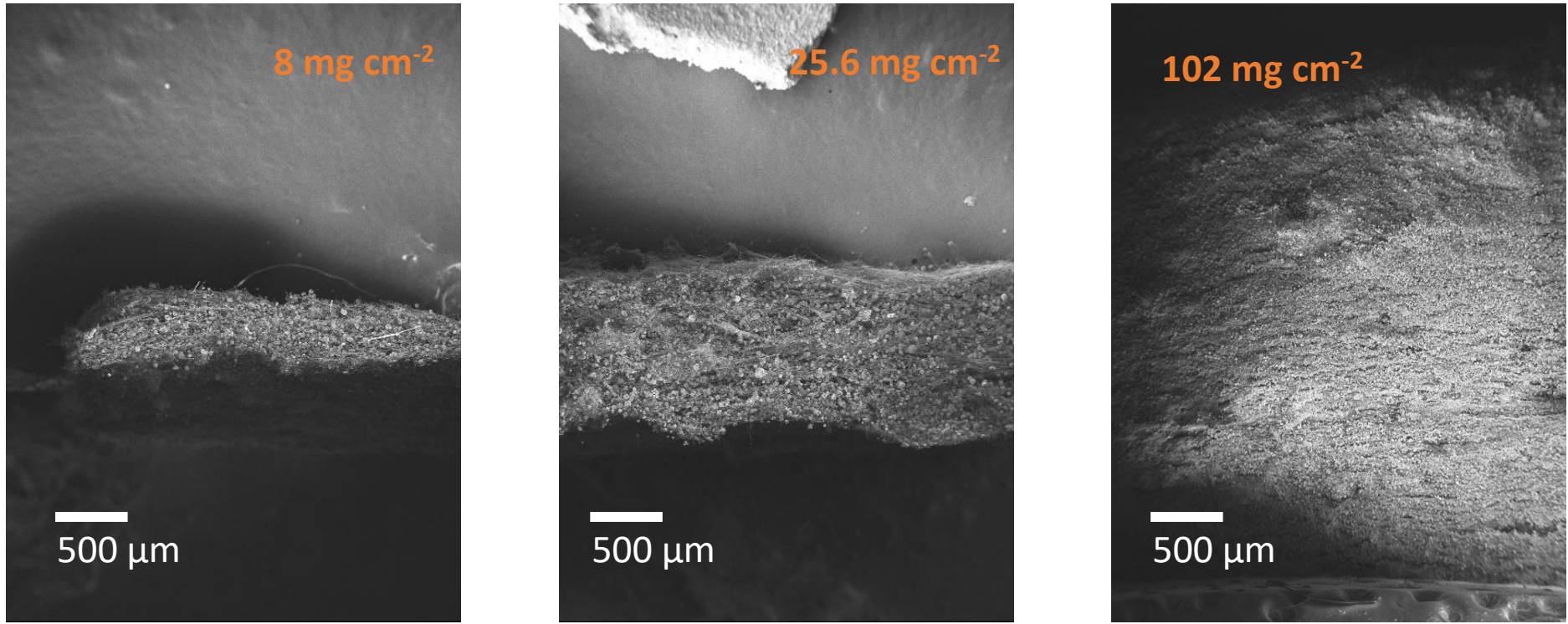

**Figure S20** Cross-section morphologies of co-ESP NVPC/CNTF electrodes of different areal loadings, uncompressed. Left to right: 8 mg cm$^{-2}$, 25.6 mg cm$^{-2}$, 102 mg cm$^{-2}$

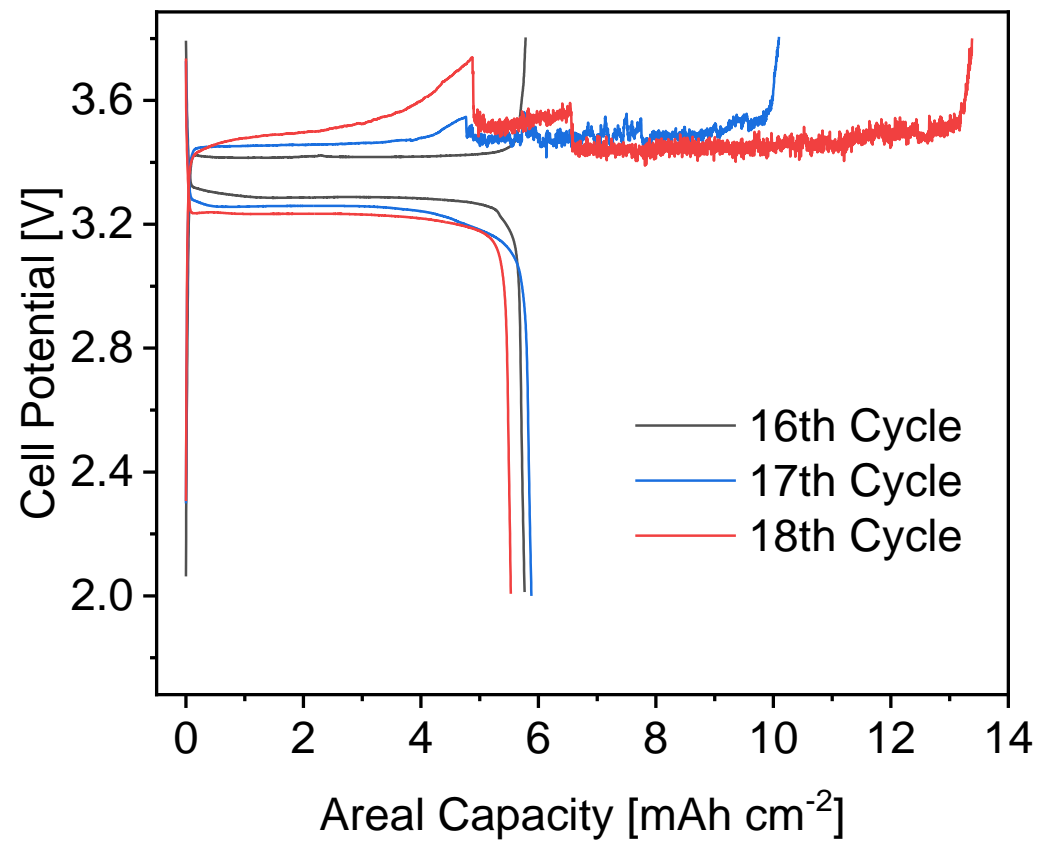

**Figure S21** 16-18th cycles of NVPC/CNTF half cells with FEC 56.7 mg cm$^{-2}$, the overcharging started from the 17th cycle in this cell even FEC is added in the electrolyte.

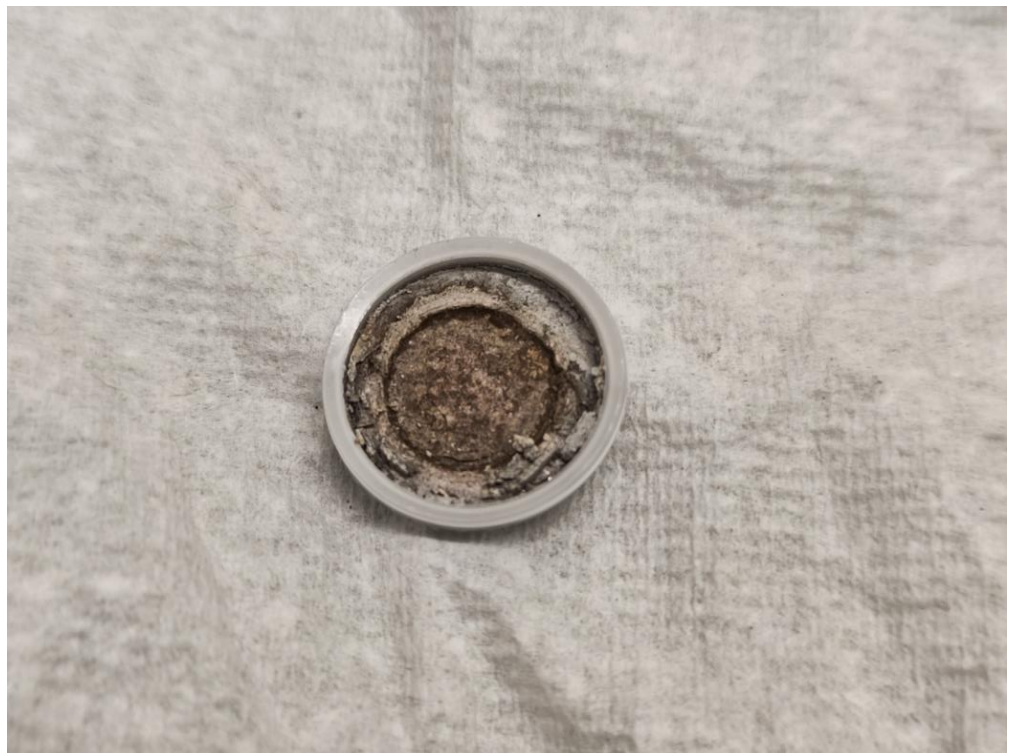

**Figure S22** Sodium anode after overcharging in a NVPC/CNTF half-cell with 60.7 mg cm$^{-2}$ areal loading, 20 cycles in total.

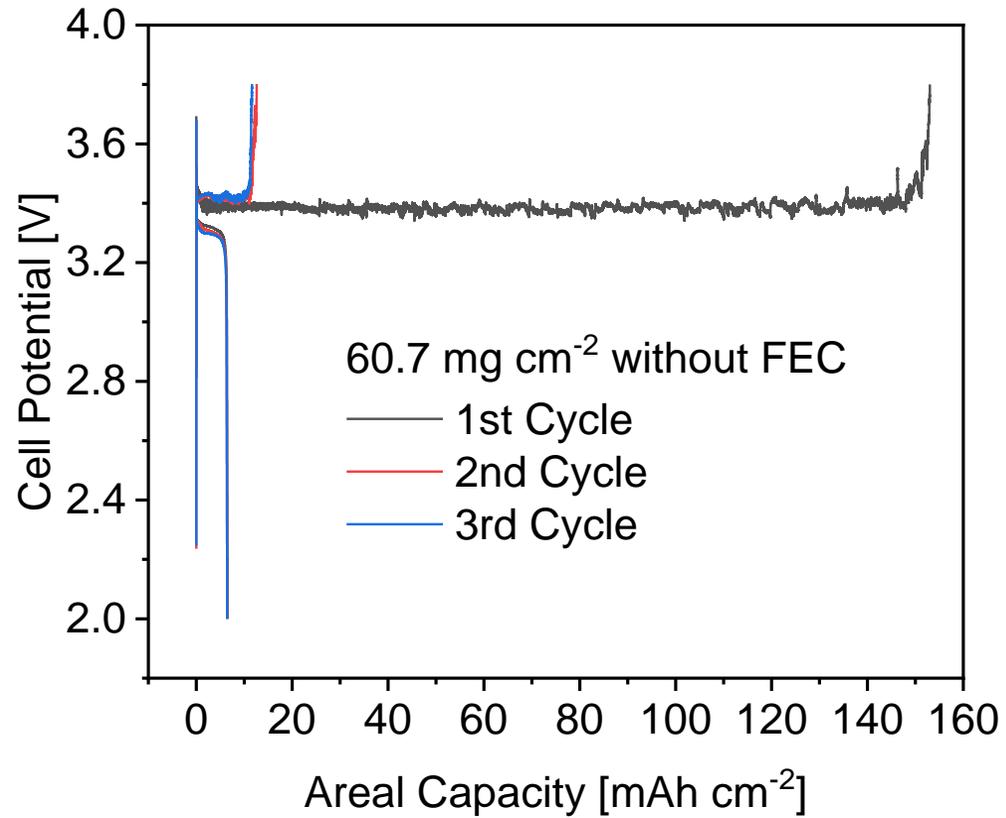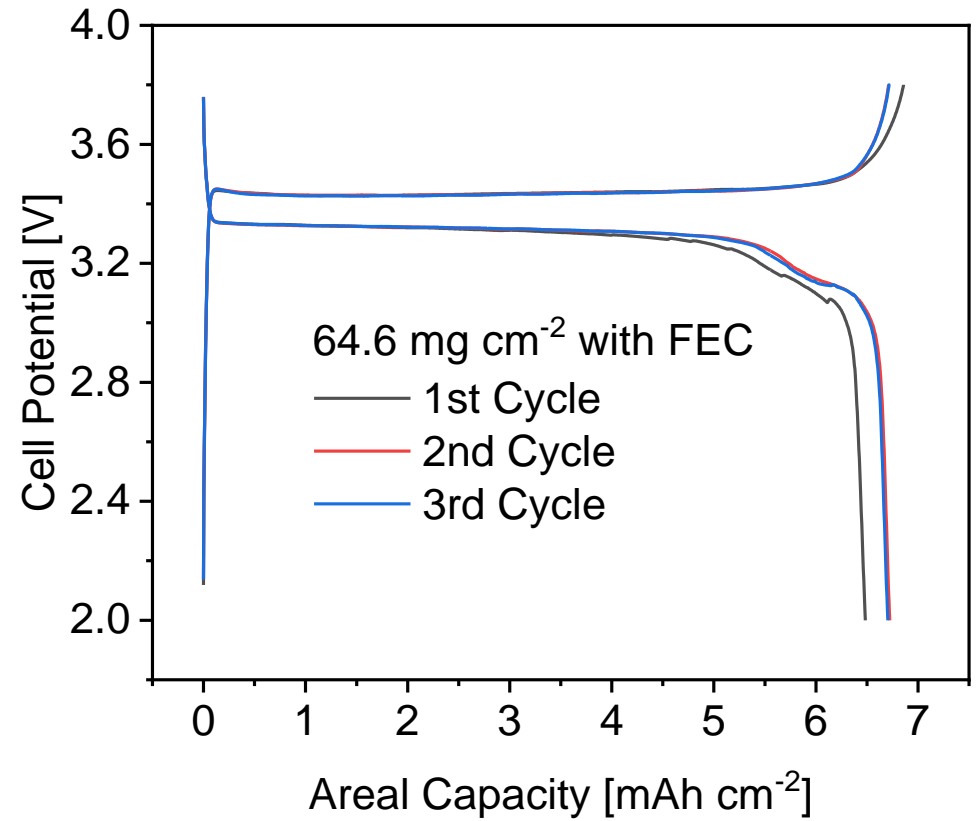

**Figure S23** First three cycles of NVP/CNT-CNF half cells without FEC 60.7 mg cm$^{-2}$ or with FEC 64.6 mg cm$^{-2}$

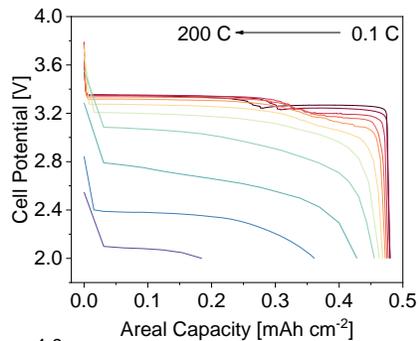
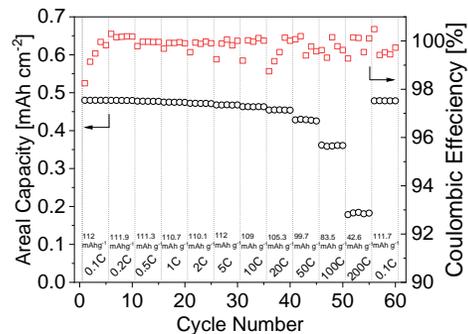
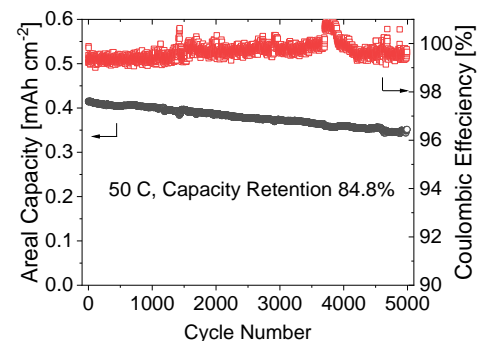

4.3 mg cm$^{-2}$

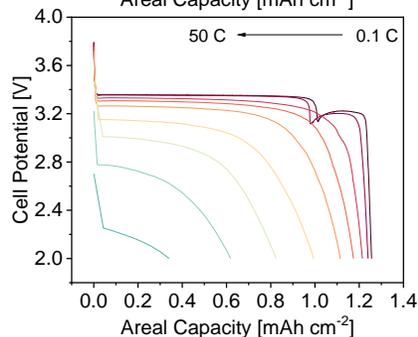
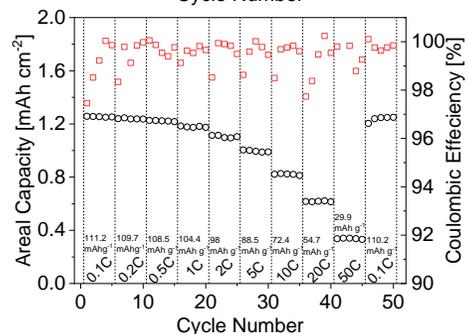
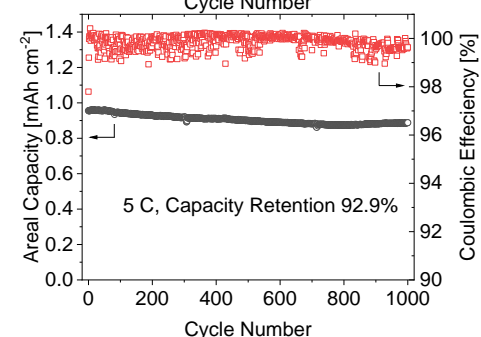

11.3 mg cm$^{-2}$

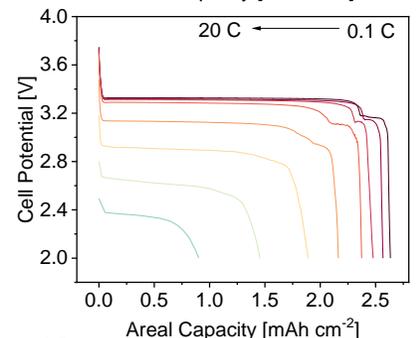
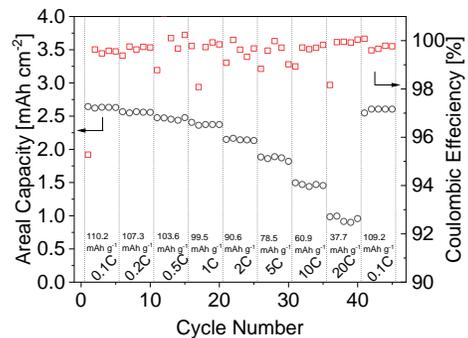
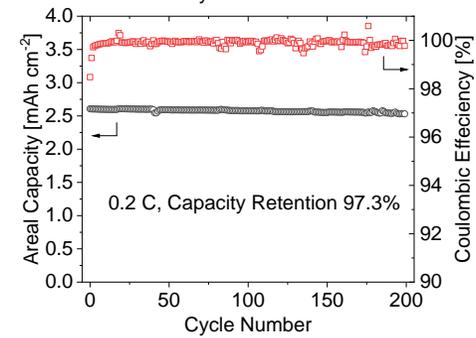

23.9 mg cm$^{-2}$

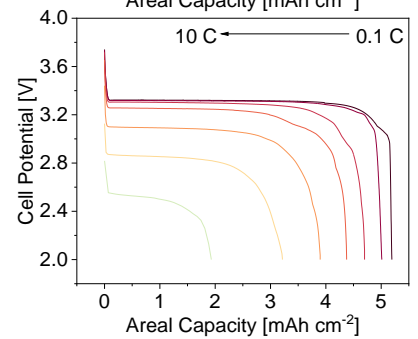
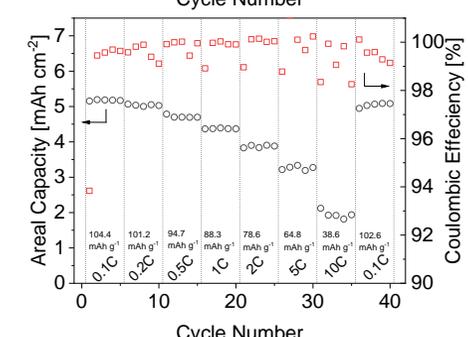
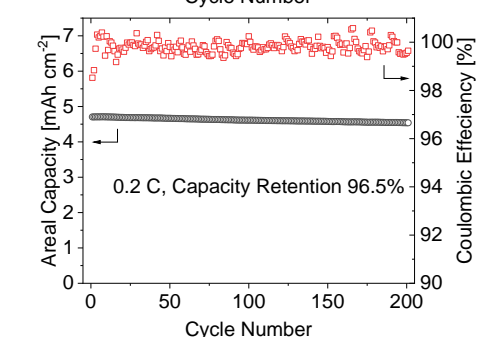

49.6 mg cm$^{-2}$

**Figure S24** Cycling data of NVP/CNTF half cells of different areal loading

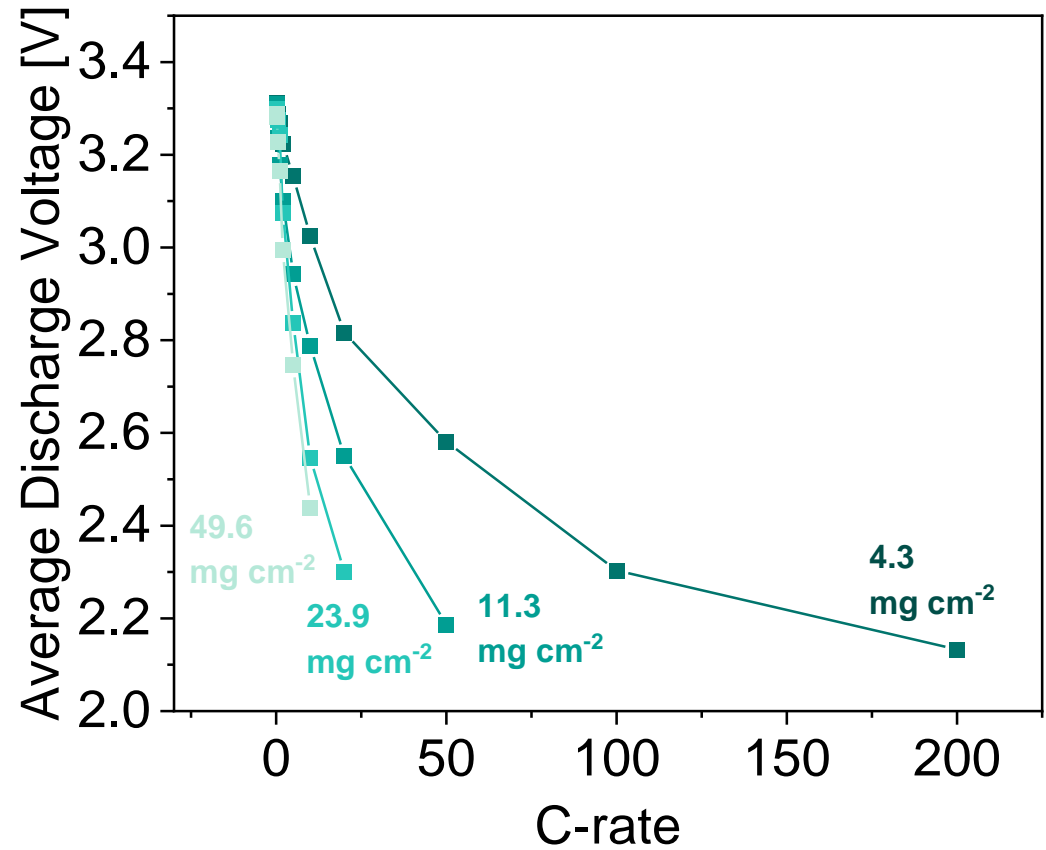

**Figure S25** Average discharge voltage of co-ESP NVPC half-cells of different areal loading and C-rate

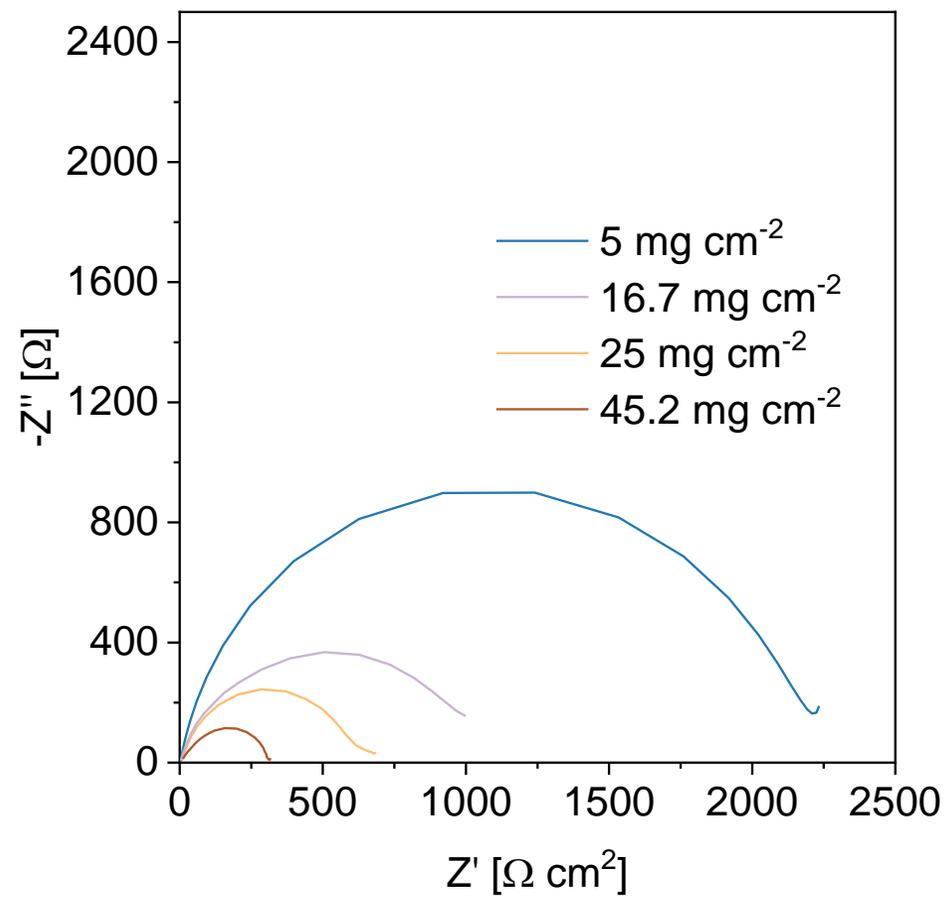

**Figure S26** Nyquist of NVP/CNTF half cell of different loading, 0% state of charge

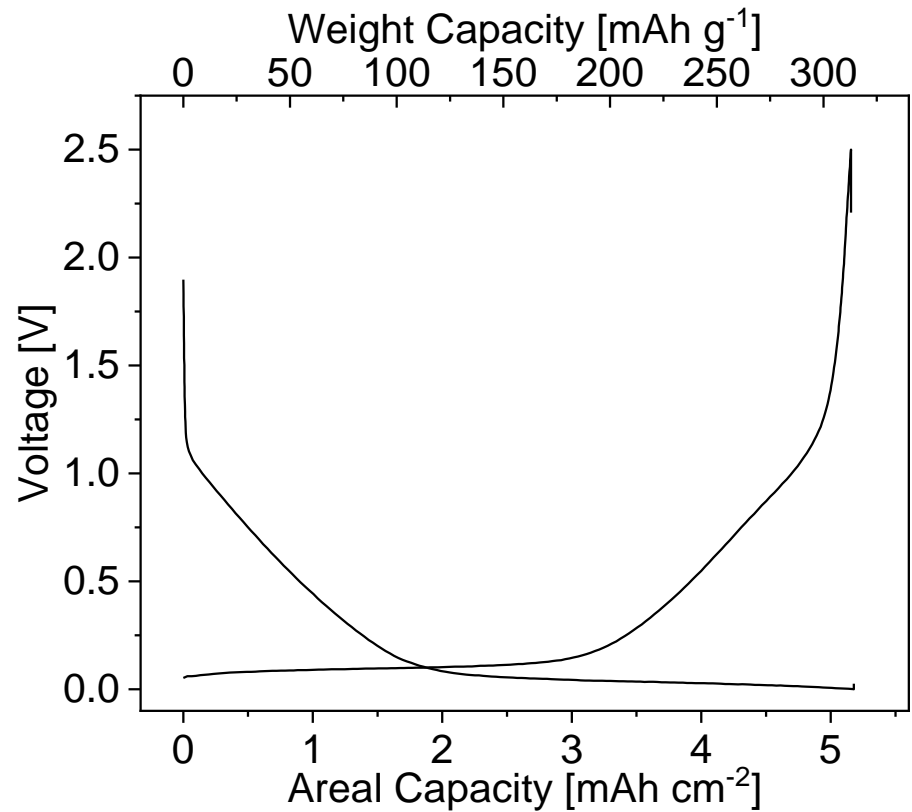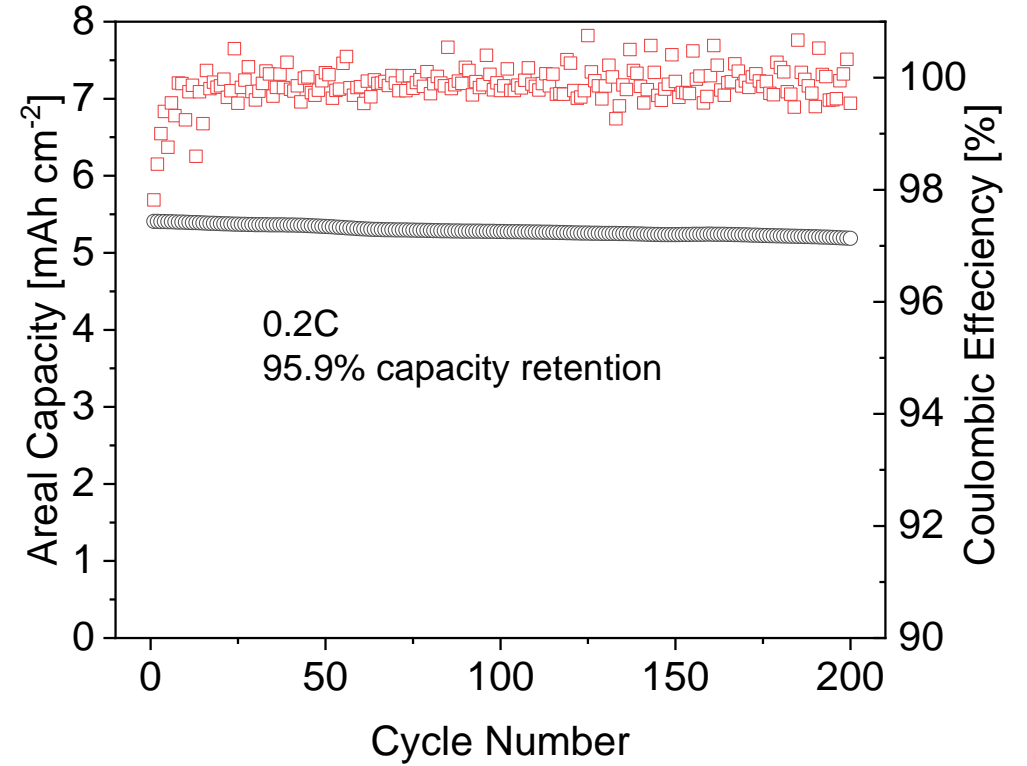

**Figure S27** Voltage profile (left) and cycling stability (right) of glucose-derived hard carbon half cells, areal loading 16.5 mg cm⁻², 0.2C

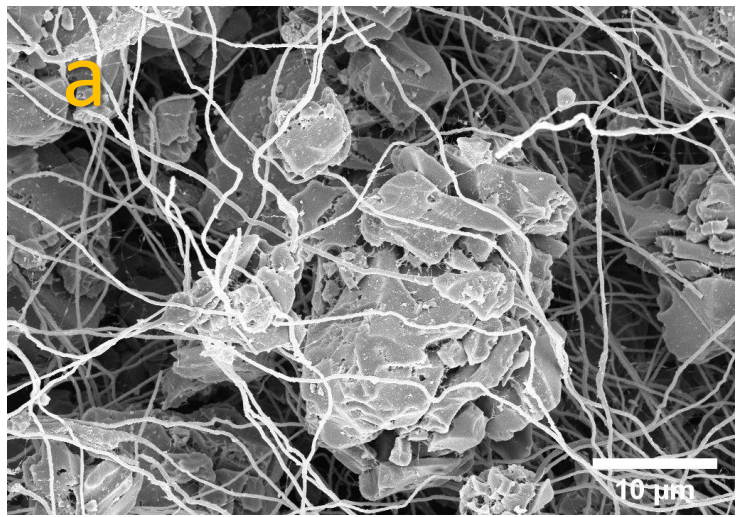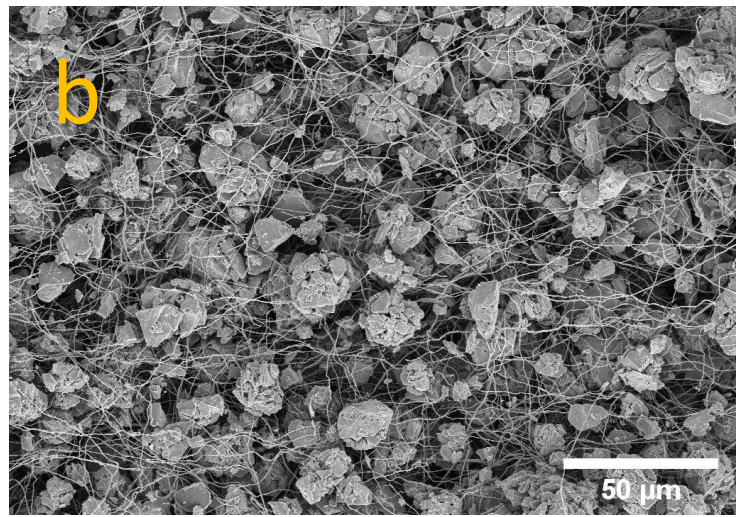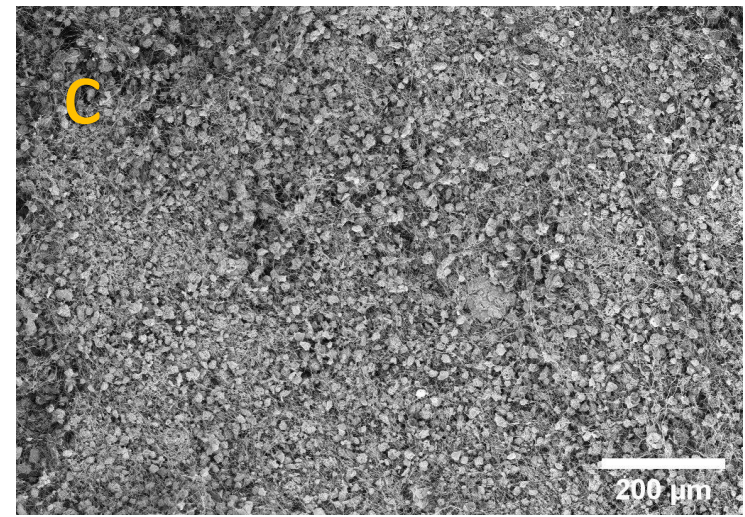

**Figure S28** Morphology of co-ESP hard carbon/CNTF electrodes

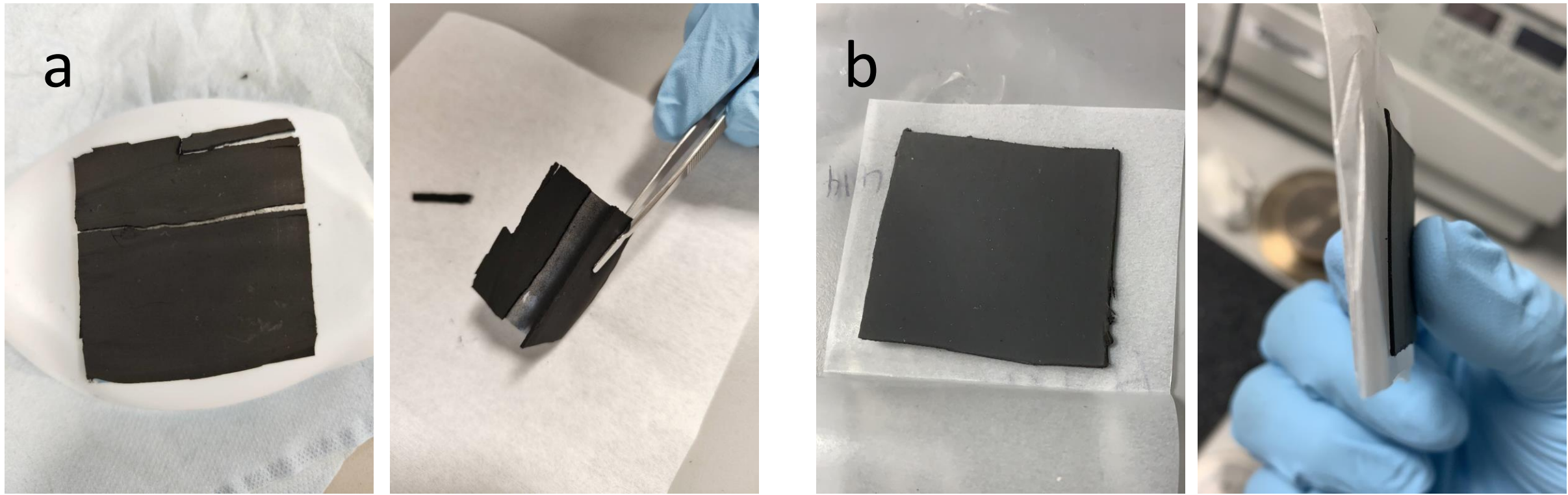

**Figure S29** Photos of **a.** cracked and delaminated conventional NVPC electrodes with 80 mg cm$^{-2}$ areal loading; **b.** compressed 100 mg cm$^{-2}$ co-ESP NVPC/CNTF pouch cell electrode

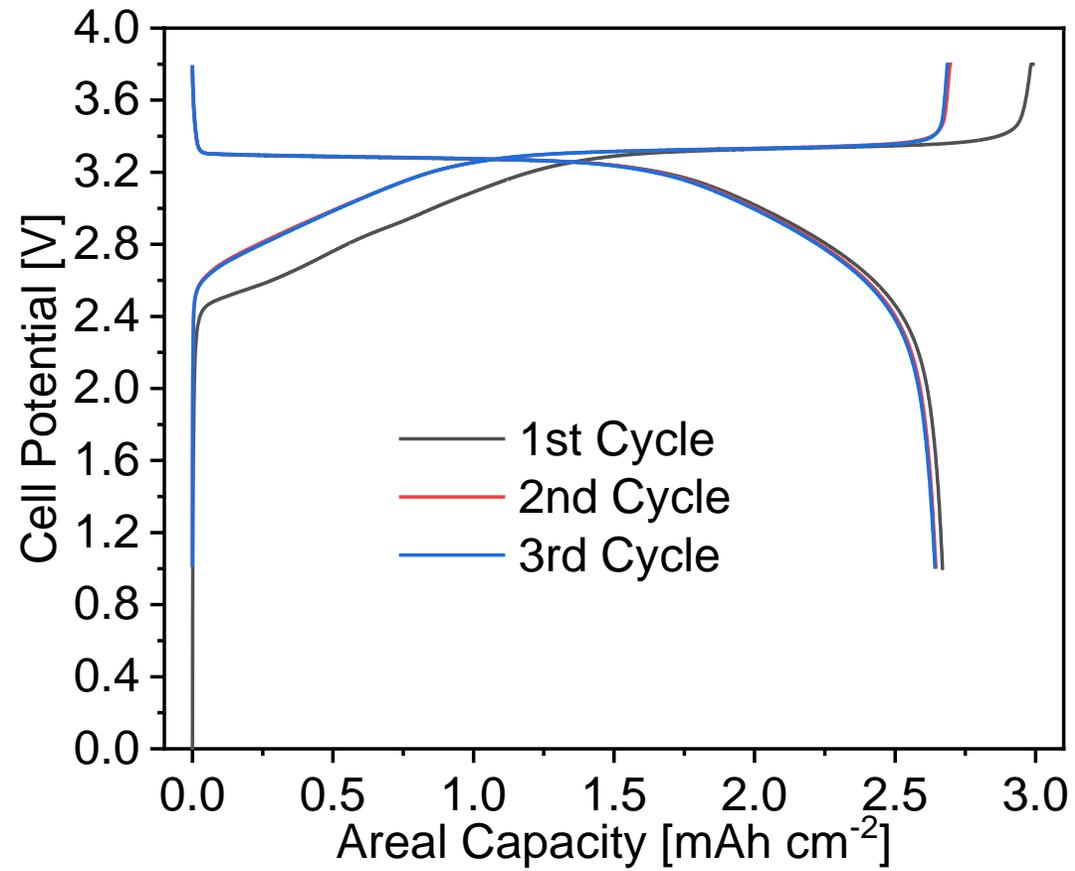

**Figure S30** First three cycles of co-ESP NVPC/CNTF full cells with 25.4 mg cm$^{-2}$ areal loading

25.4 mg cm$^{-2}$
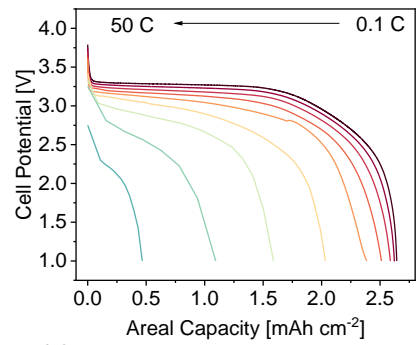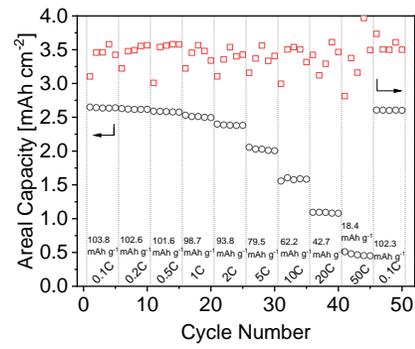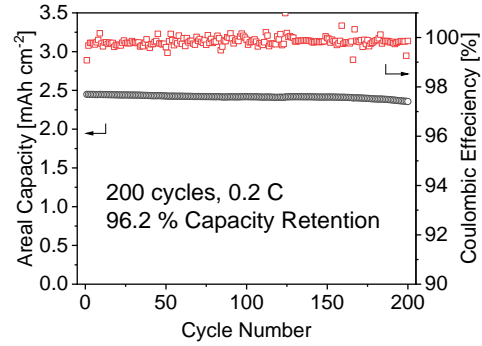

60.7 mg cm$^{-2}$
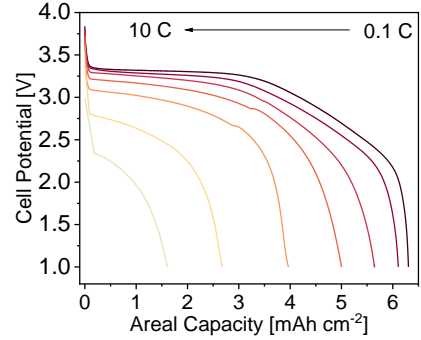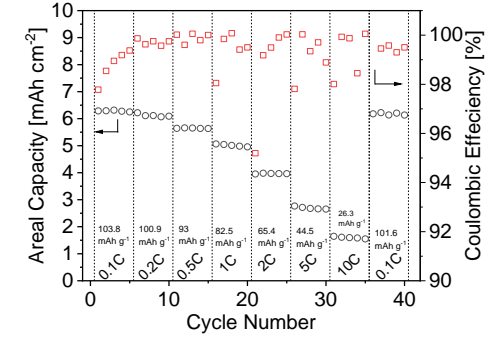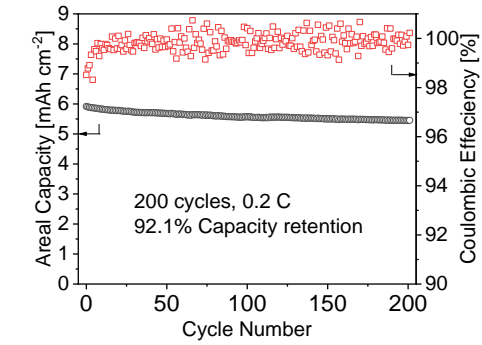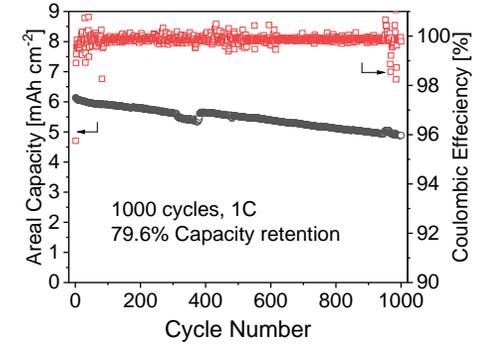

136.9 mg cm$^{-2}$
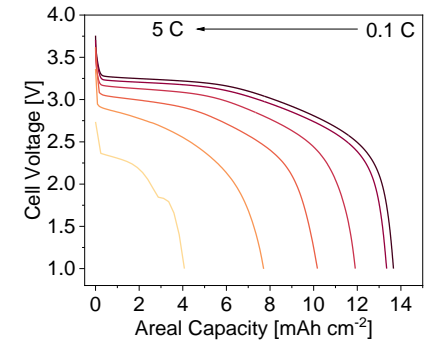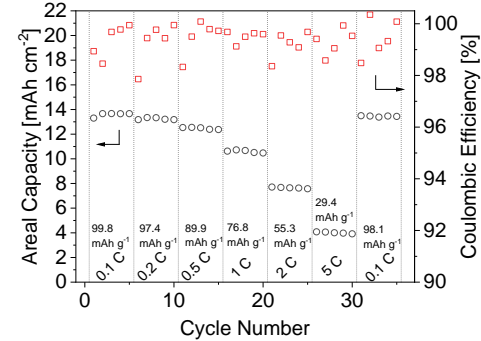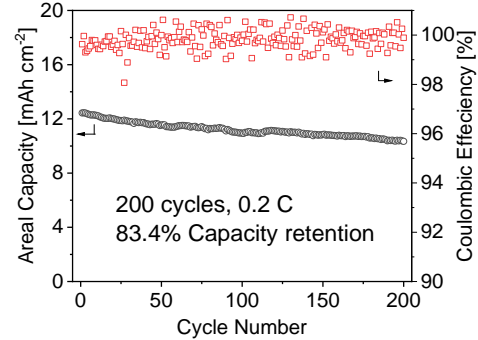

296 mg cm$^{-2}$
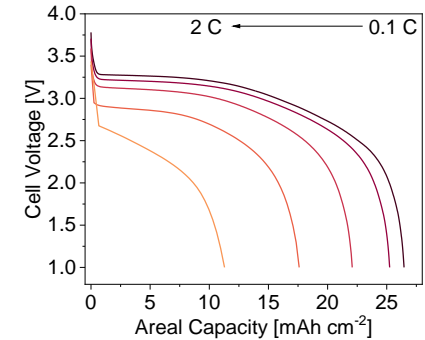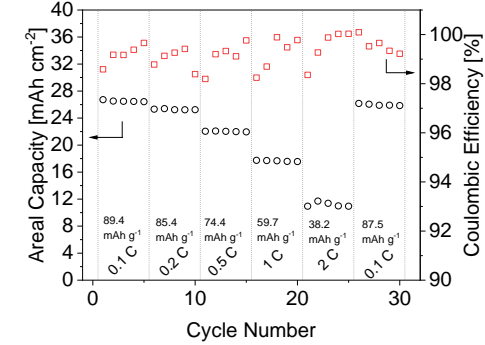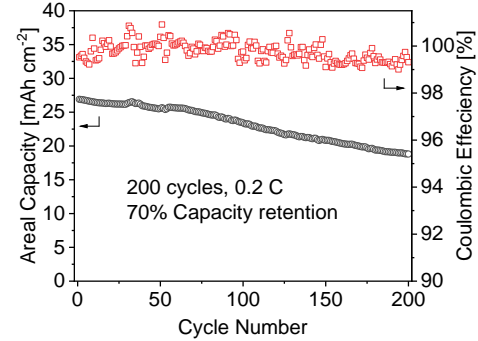

**Figure S31** Detailed electrochemical data of full cells made of co-ESP NVPC/CNTF cathodes and HC/CNTF anodes of different areal loading

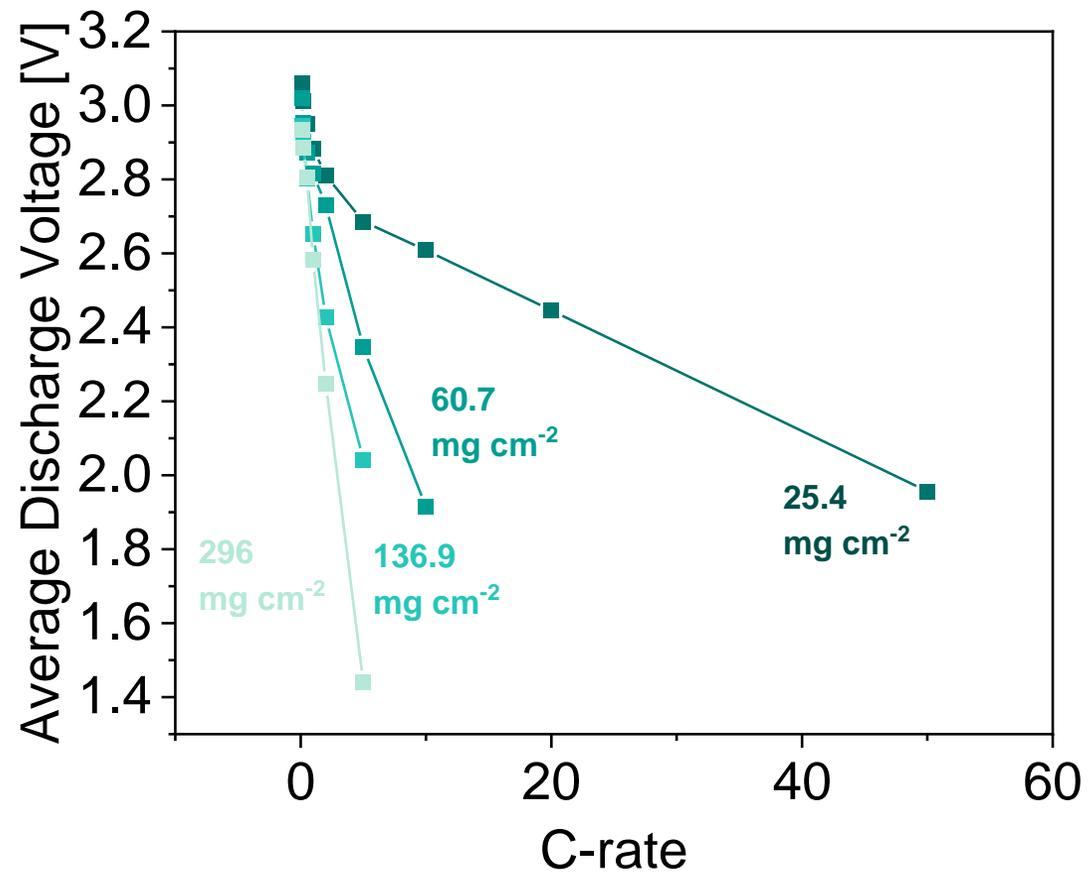

**Figure S32** Average discharge voltage of co-ESP full-cells of different areal loading and C-rate

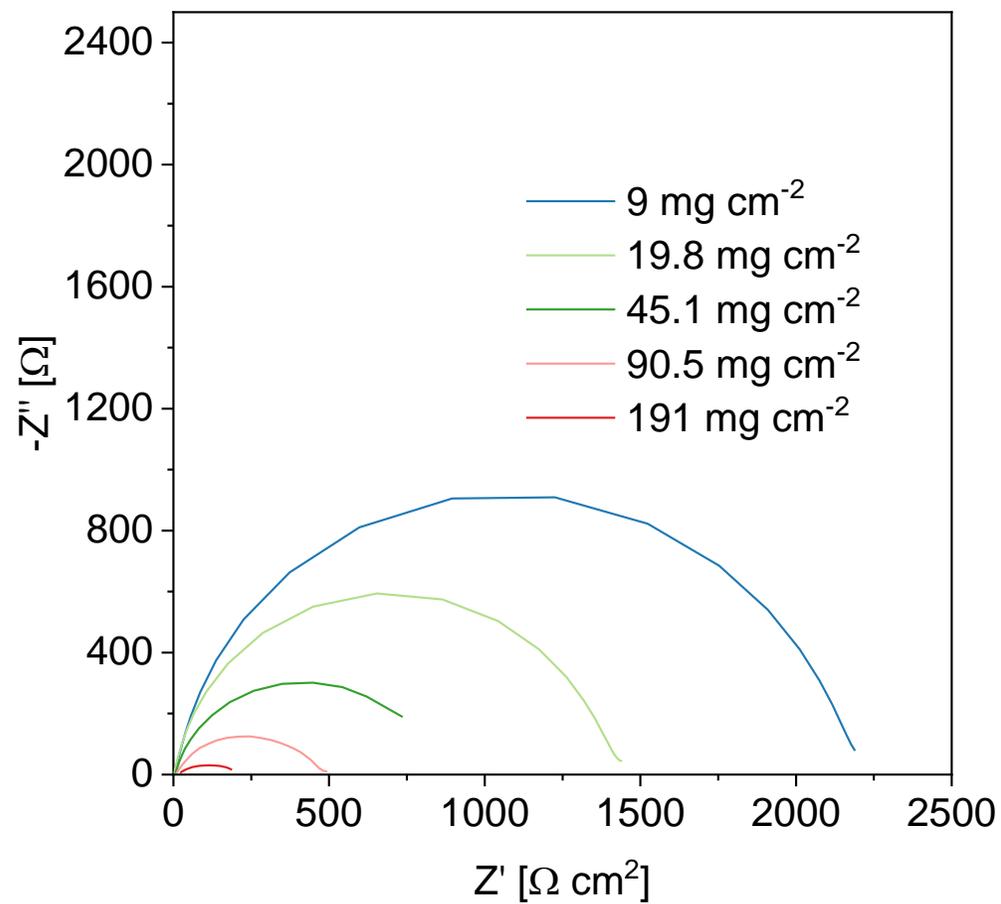 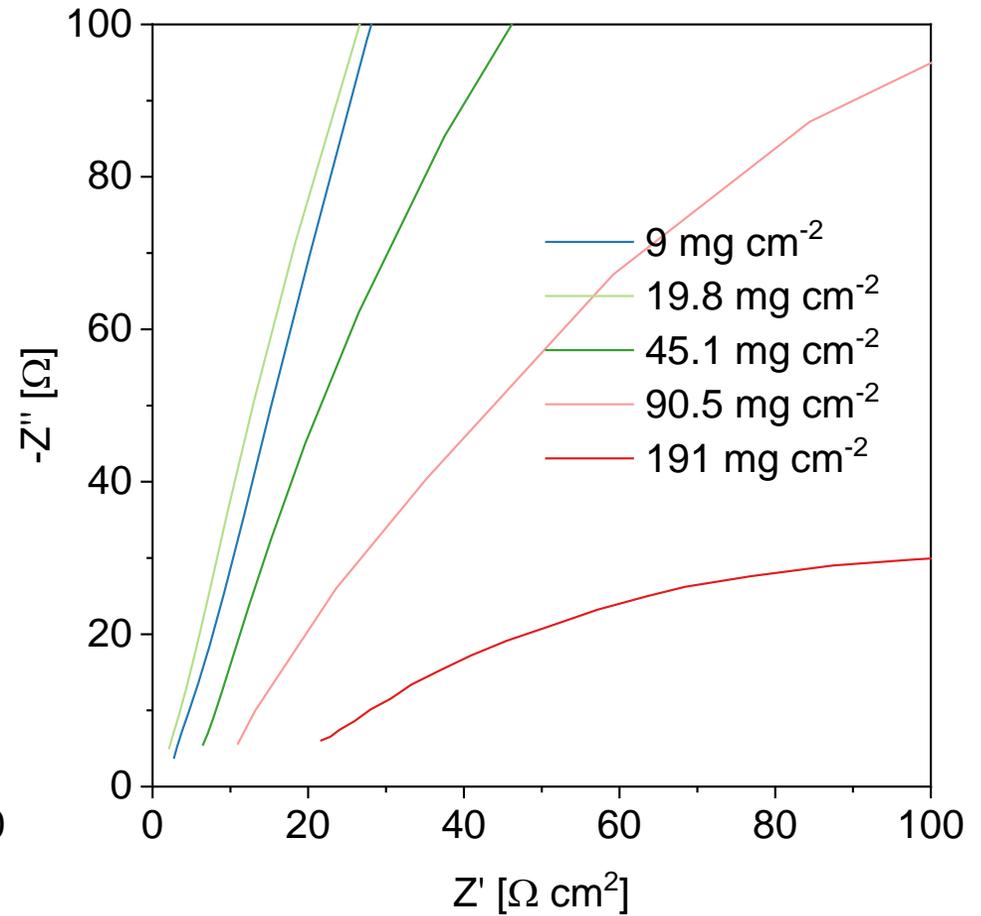

**Figure S33** EIS of full cells made of co-ESP NVPC/CNTF cathodes and HC/CNTF anodes, of different cathode loading

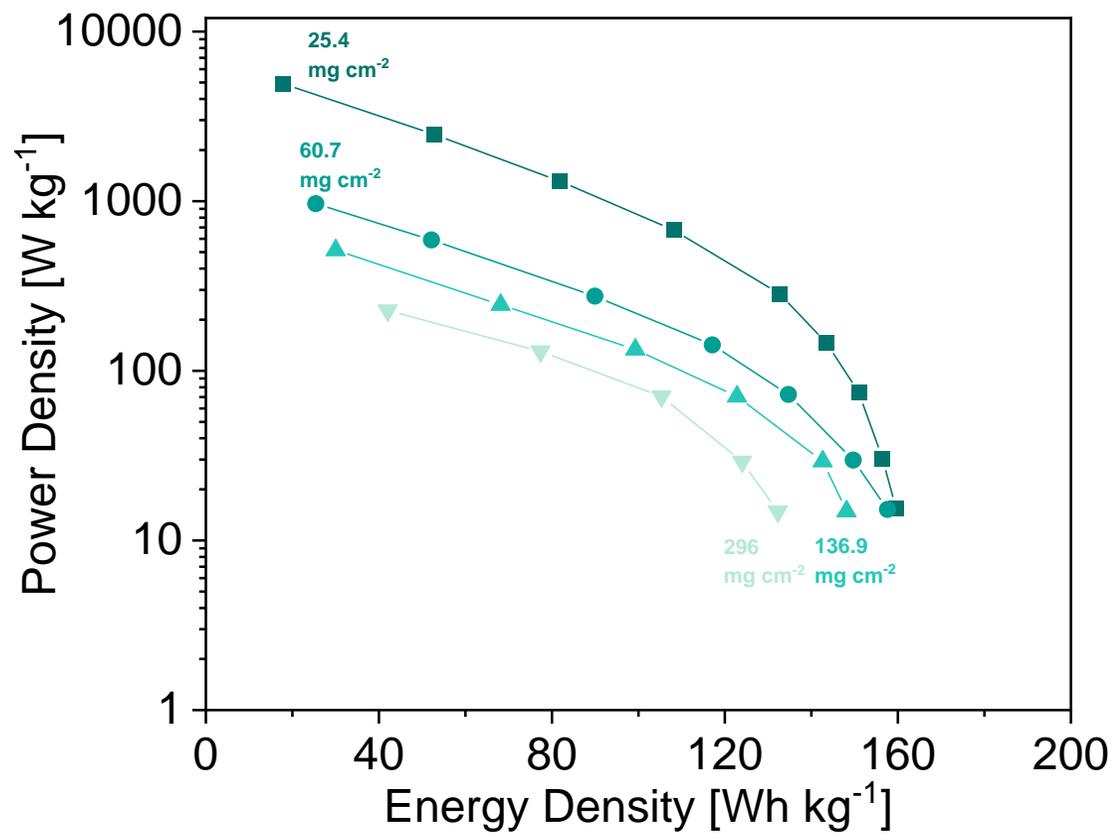

**Figure S34** Gravimetric energy/power density of co-ESP full cells considering weight of electrolyte and separator

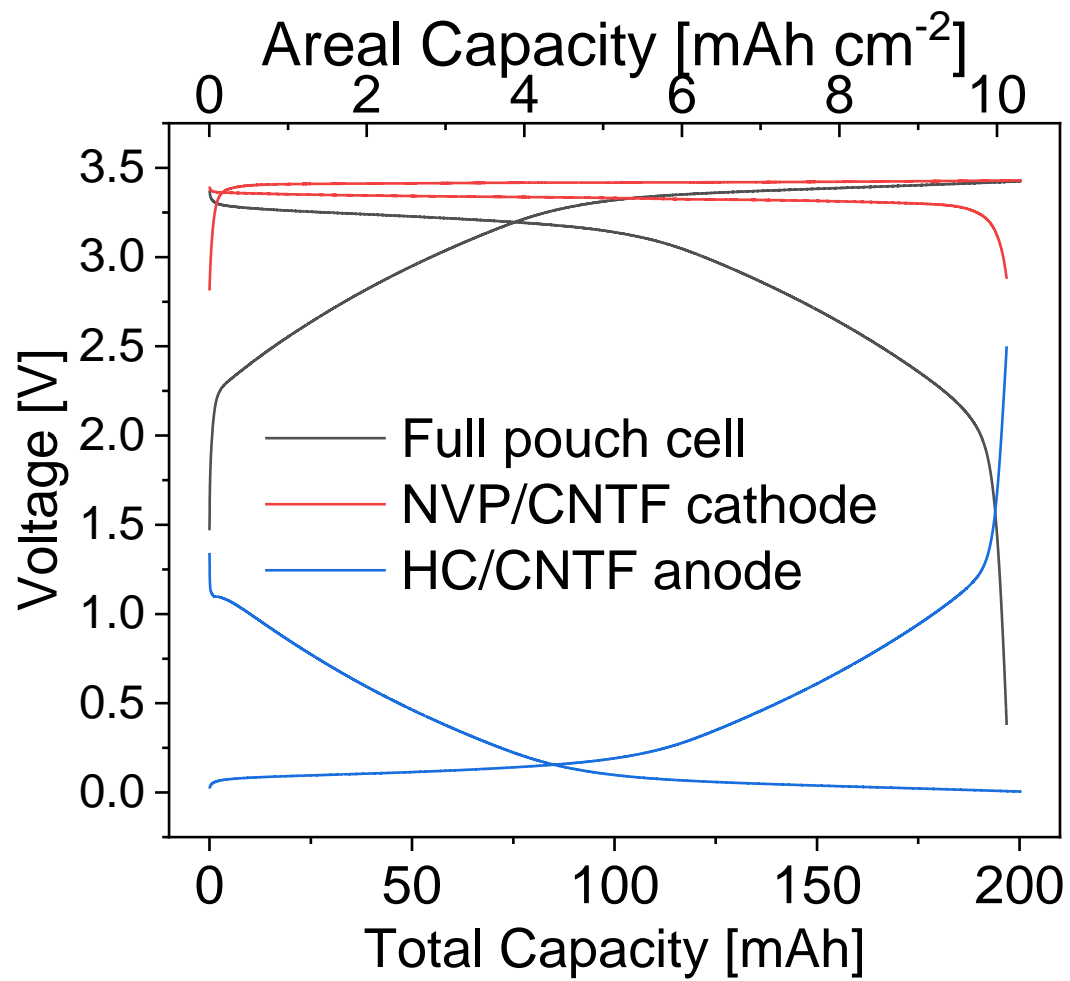 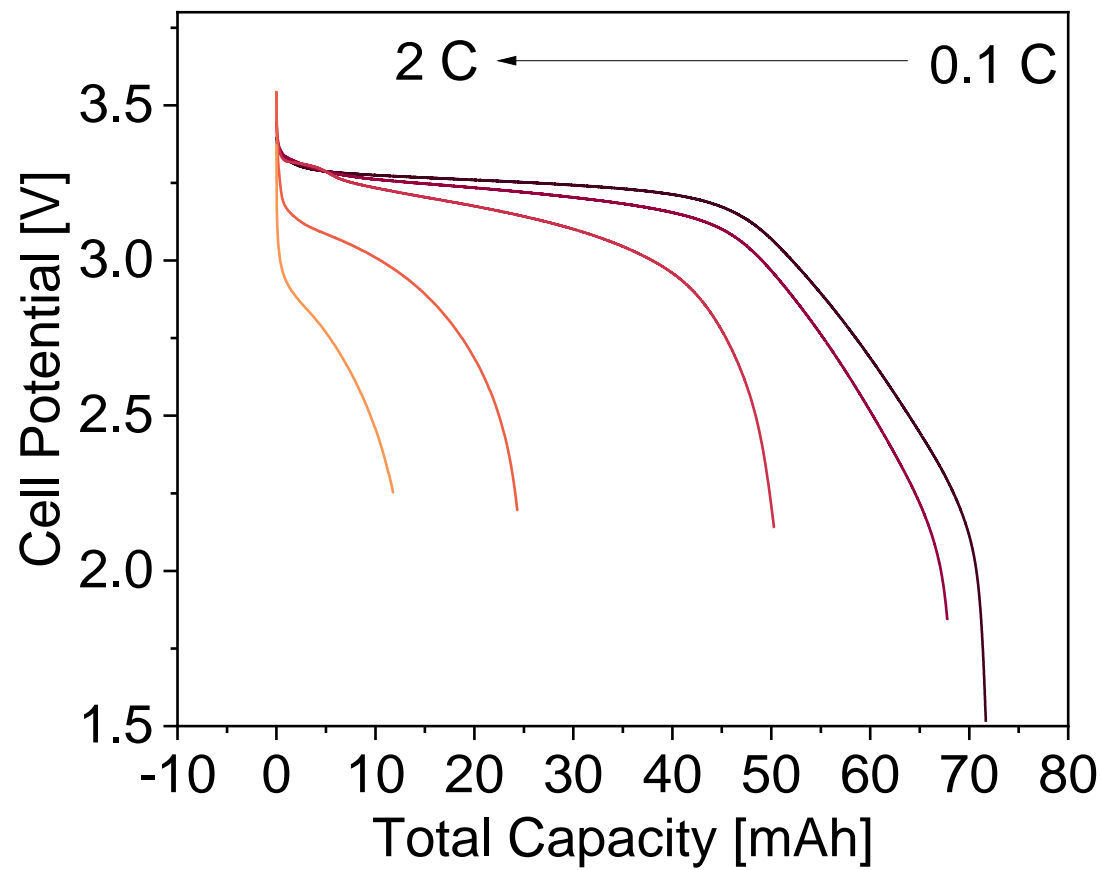

**Figure S35 Left:** Voltage profiles of a single-layer, 200 mAh pouch cell with 100 mg cm$^{-2}$ areal loading, 0.2 C, with a three-electrodes set-up; **Right**: voltage profiles of a 70 mAh pouch cell at different rates. CC-CV cycles. The discharge stopped when the anode's voltage reach 2.5V, or the cathode's voltage reach 2.5V, whichever happened first.

# Supplementary Appendix 2: Tables

**Table S1** Weight of different components for electrospinning and electrospraying, for electrodes with different active content, per 1000 mg NVP cathode

| Active content | Electrospraying recipe | | | Electrospinning recipe | | |
|---|---|---|---|---|---|---|
| | NVPC/mg | PEO/mg | DMF/mL | PAN/mg | CNT/mg | DMF/mL |
| 90 wt% | $10^3$ | 200 | 1 | 200 | 40 | 4 |
| 95 wt% | $10^3$ | 200 | 1 | 100 | 20 | 2 |
| 97.5 wt% | $10^3$ | 200 | 1 | 50 | 10 | 1 |
| 98 wt% | $10^3$ | 200 | 1 | 40 | 8 | 0.8 |
| 99 wt% | $10^3$ | 200 | 1 | 20 | 4 | 0.4 |

## Table S2 Microstructural parameters extracted from micron and nano-XCT

|  | Synchrotron Micro-XCT uncompressed[1] | Synchrotron Micro-XCT compressed | Nano-XCT uncompressed | Calculated compressed electrode parameters[2] |
|---|---|---|---|---|
| Total Porosity [vol%] | 90 | 49.3 | 93.1 | 55.8 |
| Porosity inside particles [vol%] | N/A | N/A | 4.2 | 28.7 |
| Porosity outside particles [vol%] | 90 | 49.3 | 88.9 | 27.1 |
| NVPC volume fraction [vol%] | 10 | 50.7 | 6.5 | 41.6 |
| CNTF volume fraction [vol%] | N/A | N/A | 0.4 | 2.6 |
| Pore through-plane tortuosity | 1.15 | 2.00 | 1.33 | 2.00 |
| Pore in-plane tortuosity (average x/y direction) | 1.09 | 1.68 | 1.12 | 1.68 |

1. The voxel size of micron-CT was 350 nm, nano-CT 47 nm
2. Because of the large field of view of micro-CT, the following parameters were acquired from micro-CT results: porosity outside particles, NVPC volume fraction, through-plane and in-plane tortuosities. Micro-CT was unable to resolve the CNTF fibres and the intra-particle pores. Therefore, the CNTF volume fraction, porosity inside particles were acquired from nano-CT results.

**Table S3** co-ESP composition 97.5 wt% NVP/CNF fabric electrode on a 300 cm$^2$ substrate

| Aimed areal loading [mg cm$^{-2}$] | Electrospraying recipe | | | Electrospinning recipe | | |
| --- | --- | --- | --- | --- | --- | --- |
| | NVP/mg | PEO/mg | DMF/mL | PAN/mg | CNT/mg | DMF/mL |
| 5 | 1.5×10$^3$ | 300 | 1.5 | 75 | 15 | 1.5 |
| 20 | 6×10$^3$ | 1.2×10$^3$ | 6 | 300 | 60 | 6 |
| 100 | 3×10$^4$ | 6×10$^3$ | 30 | 1.5×10$^3$ | 300 | 30 |

## Table S4 Performance comparison with published sodium-ion half batteries work

| Electrode fabrication method | Cathode composition[1] | Max half cell areal loading/capacity[3] (areal loading: rate, areal capacity)[2] | Rate Capacity and retention (discharge rate, specific capacity in mAh g$^{-1}$, capacity retention)[3] | Half cell cycling (number of cycles, rate, capacity retention) | Max areal Energy Density (mWh cm$^{-2}$)/Power Density (mW cm$^{-2}$)[4] | Max gravimetric energy density (Wh kg$^{-1}$)/power density (W kg$^{-1}$)[4] | Ref |
|---|---|---|---|---|---|---|---|
| **Co-ESP** | **97.5% Commercial NVPC/CNTF** | **49.6 mg cm$^{-2}$: 0.1C, 5.2 mAh cm$^{-2}$; 10C, 1.9 mAh cm$^{-2}$** | **0.1C, 106.7, 91.2%; 10C, 49.1, 42%** | **200 cycles, 0.2C, 93.1%** | **20.5/145.4** | **336.4/2390.7** | **This work** |
| | | **4.3 mg cm$^{-2}$: 0.1C, 0.48 mAh cm$^{-2}$; 200C, 0.18 mAh cm$^{-2}$** | **0.1C, 111.6, 95.4%; 200C, 41.9, 35.8%** | **5000 cycles, 50C, 84.5%** | **1.59/183.4** | **362.2/41805** | |
| Slurry-casting | 79.5% HNVP/CB/PVDF/Al foil | 48.9 mg cm$^{-2}$: 0.01C, 3.8 mAh cm$^{-2}$; 0.5C, 1.1 mAh cm$^{-2}$ | 0.05C, 77, 70.1%; 0.5C, 23, 20.9% | 48.9 mg cm$^{-2}$ no cycling data | 13.1/7.52 | 213.1/122.3 | Ref 61 |
| Phase inversion | 76% NVP/carbon | 30 mg cm$^{-2}$: 0.1C, 2.8 mAh cm$^{-2}$; 2C, 1.4 mAh cm$^{-2}$ | 0.1C, 91.7, 78.4%; 2C, 46, 39.3% | 30 mg cm$^{-2}$ no cycling data | 9.15/8.46 | 231.7/440.8 | Ref 7 |
| Spray-drying# | 66.2% NVP/CB/PVDF/Al foil | 10 mg cm$^{-2}$: 0.2C, 1.2 mAh cm$^{-2}$; 10C, 0.27 mAh cm$^{-2}$ | 0.2C, 120, 102.5%; 10C, 29, 24.8% | 100 cycles, 1C, 100% | 4.02/3.2 | 266.1/1725.8 | Ref 4 |
| Slurry-casting | 57.6% NVP@C@CNT/CB/PVDF/Al foil | 10 mg cm$^{-2}$: 0.5C, 1.2 mAh cm$^{-2}$; 30C, 0.36 mAh cm$^{-2}$ | 0.5C, 117, 100%; 30C, 36, 30.8% | 450 cycles, 15C, 95.2% | 3.98/80.4 | 229.1/4631 | Ref 5 |
| Hydrothermal | 43% NCO/Ni foam | 10 mg cm$^{-2}$: 0.33C, 1.66 mAh cm$^{-2}$; 2.1C, 1 mAh cm$^{-2}$, | 0.33C, 166.3, 98.5%; 2.1C, 100.5, 60.4% | 100 cycles, 1C, 100% | 4.42/5.67 | 190.2/243.8 | Ref 6 |
| Sol-gel | 51.9% NVP/CNF/C65/PVDF/Al foil | 8.5 mg cm$^{-2}$: 0.1C, 0.89 mAh cm$^{-2}$; 2C, 0.84 mAh cm$^{-2}$ | 0.1C, 105, 89.7%; 2C, 99.6, 85.1% | 200 cycles, 2C, 100% | 3.01/81.6 | 184.5/4992 | Ref 66 |
| Slurry-casting | 56% NVPF/GO/CB/PVDF/Al foil | 8 mg cm$^{-2}$: 0.4C, 0.98 mAh cm$^{-2}$; 6.3C, 0.85 mAh cm$^{-2}$ | 0.4C, 123, 96.9%; 6.3C, 106, 83% | 200 cycles, 3 C, 96% | 3.64/21.8 | 254.8/1523.2 | Ref 67 |
| Slurry-casting | 49.6% NVP/CNF/C | 8 mg cm$^{-2}$: 0.1C, 0.82 mAh cm$^{-2}$; 40C, 0.5 mAh cm$^{-2}$ | 0.1C, 103, 88%; 40C, 62, 53% | 500 cycles, 1C, 95.9% | 2.76/83.2 | 171.1/5158.4 | Ref 68 |
| Precipitation& | 72.6% NVP/CNF/C | 7.6 mg cm$^{-2}$: 0.05C, 0.87 mAh cm$^{-2}$; 100C, 0.5 mAh cm$^{-2}$ | 0.05C, 114, 97.4%; 100C, 66, 56.4% | 700 cycles, 1C, 89.7% | 2.85/ NA | 270/ NA | Ref 69 |

\# The reported capacity exceeded the theoretical value of the cathode material, the source did not specify the reason.
& The work did not give the voltage profile at higher C-rate.
1. Calculated from the highest areal loading reported in the publication that has rate and cycling data. Estimated 15 μm-thick aluminium current collector weight: 4 mg cm$^{-2}$, 10 μm-thick copper current collector: 9 mg cm$^{-2}$
2. Showing the areal capacity from the lowest and highest cycling rate.
3. The specific capacity of the highest areal loading reported in the publication, capacity retention at different rate; NMC811's theoretical capacity is 200 mAh g$^{-1}$; LCO's theoretical capacity is 170 mAh g$^{-1}$.
4. The energy/power densities are re-calculated considering the weight and volume of all electrode components, including active material, current collector, binder, conductive additive, based on the data reported in the literature. The method of calculation is shown in Supplementary Appendix 4.

## Table S5 Performance comparison with published lithium-ion half-cell work

| Fabrication method | Cathode composition[1] | Max half cell areal loading/capacity[3] (areal loading: rate, areal capacity)[2] | Rate Capacity and retention (discharge rate, specific capacity in mAh g$^{-1}$, capacity retention)[3] | Half cell cycling (number of cycles, rate, capacity retention) | Max areal Energy Density (mWh cm$^{-2}$)/Power Density (mW cm$^{-2}$)[4] | Max gravimetric energy density (Wh kg$^{-1}$)/power density (W kg$^{-1}$)[4] | Ref |
|---|---|---|---|---|---|---|---|
| **Co-ESP** | **97.5% Commercial NVPC/CNTF** | **49.6 mg cm$^{-2}$: 0.1C, 5.2 mAh cm$^{-2}$; 10C, 1.9 mAh cm$^{-2}$** | **0.1C, 106.7, 91.2%; 10C, 49.1, 42%** | **200 cycles, 0.2C, 93.1%** | **20.5/145.4** | **336.4/2390.7** | This work |
| | | **4.3 mg cm$^{-2}$: 0.1C, 0.48 mAh cm$^{-2}$; 200C, 0.18 mAh cm$^{-2}$** | **0.1C, 111.6, 95.4%; 200C, 41.9, 35.8%** | **5000 cycles, 50C, 84.5%** | **1.59/183.4** | **362.2/41805** | |
| Mechanical pressing | 79% LCO/KB/Carbon Cloth | 71 mg cm$^{-2}$: 0.08C, 10 mAh cm$^{-2}$; 0.8C, 8.7 mAh cm$^{-2}$ | 0.08C, 137, 85.6%; 1.6C, 97, 60.6% | 40 cycles, 0.15C, 91% | 37.7/50.4 | 419.9/709.9 | [ref] |
| Slurry-casting& | 80% LCO/cellulose/CNT | 86 mg cm$^{-2}$: 0.1C, 12.1 mAh cm$^{-2}$ | 0.1 C, 141, 88.1%; no rate | 20 cycles, 0.04C, 90% | 46/5.23 | 427.7/48.6 | [ref] |
| Impregnation | 100% LCO | 206 mg cm$^{-2}$: 0.05C, 24.7 mAh cm$^{-2}$; 1C, 7.79 mAh cm$^{-2}$ | 0.05 C, 120, 75%; 1C, 37.8, 23.6% | 27 cycles, 0.5C, 87% | 93.9/95.6 | 455.6/464 | [ref] |
| Slurry-casting | 90.2% NMC811/biopolymer/CNT/Al foil | 47.7 mg cm$^{-2}$: 0.1C, 8.84 mAh cm$^{-2}$; 1C, 3 mAh cm$^{-2}$ | 0.05C, 185, 92.5%; 1C, 62.9, 31.4% | 50 cycles, 0.2C, 92.5% | 29.9/28.9 | 625.8/604.3 | [ref] |
| Slurry-casting | 75% LFP/kb/CNF/PVDF | 108 mg cm$^{-2}$: 0.05C, 17 mAh cm$^{-2}$; 0.5C, 10 mAh cm$^{-2}$ | 0.05C, 157.4, 92.6%; 0.5C, 92.6. 54.5% | 40 cycles, 0.2C, 94.3% | 57.5/26.8 | 399/190 | [ref] |
| Slurry-casting | 71% LFP/CB/PVDF/Al foil | 128 mg cm$^{-2}$: 0.025C, 19.8 mAh cm$^{-2}$; 1C, 0.81 mAh cm$^{-2}$ | 0.025C, 155, 91.2%; 1C, 6.34, 3.7% | 150 cycles, 0.1C, 73% | 65.3/56 | 358.1/310.6 | [ref] |
| Slurry-casting& | 95.7% NMC811/CNT/Al foil | 99 mg cm$^{-2}$: 0.05C, 17.8 mAh cm$^{-2}$; 0.25C, 7.2 mAh cm$^{-2}$ | 0.05C, 180, 90%; 0.25C, 72.7, 36.4% | 10 cycles, 0.1C, 96% | 65.9/ NA | 631.2/ NA | [ref] |
| Slurry-casting& | 80% LFP/CNT/EVA | 49 mg cm$^{-2}$: 0.2C, 7.5 mAh cm$^{-2}$; 2C, 3 mAh cm$^{-2}$ | 0.2C, 153.1, 90%; 2C, 61.2, 36% | N/A | 24.4/NA | 416.3/NA | [ref] |
| Slurry-casting | 67.3% LFP/CB/PVDF | 70 mg cm$^{-2}$: 0.05C, 10.9 mAh cm$^{-2}$; 1C, 1.5 mAh cm$^{-2}$ | 0.05C, 155.7, 91.6%; 1C, 21.4, 12.6% | 200 cycles, 0.1C, 70% | 36/16.4 | 346.1/157.7 | [ref] |
| Extrusion | 93.9 % LFP/carbon/Al foil | 90 mg cm$^{-2}$: 1/24C, 12.8 mAh cm$^{-2}$; 1/12C, 11.1 mAh cm$^{-2}$ | 1/24C, 142.2, 83.6%; 1/12C, 123.3, 72.5% | 7 cycles, 1/12C, 86.6% | 42.2/4.2 | 440.3/43.8 | [ref] |
| Hot-pressing& | 76.5% NMC712/CNT/PVDF/Al foil | 70 mg cm$^{-2}$: 0.1C, 13.2 mAh cm$^{-2}$; 0.5C, 6.8 mAh cm$^{-2}$ | 0.1C, 188.6, 94.3%; 0.5C, 97.1, 48.6% | 30 cycles, 0.1C, 96% | 48.8/NA | 533.3/NA | [ref] |

& The work did not give the voltage profile at higher C-rate.

1. Calculated from the highest areal loading reported in the publication that has rate and cycling data. Estimated 15 μm-thick aluminium current collector weight: 4 mg cm$^{-2}$, 10 μm-thick copper current collector: 9 mg cm$^{-2}$
2. Showing the areal capacity from the lowest and highest cycling rate.
3. The specific capacity of the highest areal loading reported in the publication, capacity retention at different rate; NMC811's theoretical capacity is 200 mAh g$^{-1}$; LCO's theoretical capacity is 170 mAh g$^{-1}$.
4. The energy/power densities are re-calculated considering the weight and volume of all electrode components, including active material, current collector, binder, conductive additive, based on the data reported in the literature. The method of calculation is shown in Supplementary Appendix 4.

**Table S6** Fabrication composition of co-ESP hard carbon

| Aimed areal loading [mg cm$^{-2}$] | Electrospraying recipe | | | Electrospinning recipe | | |
|---|---|---|---|---|---|---|
| | NVP/mg | PEO/mg | DMF/mL | PAN/mg | CNT/mg | DMF/mL |
| 5 | 1.5×10$^3$ | 600 | 3 | 160 | 32 | 3.2 |
| 10 | 3×10$^3$ | 1.2×10$^3$ | 6 | 320 | 64 | 6.4 |
| 25 | 7.5×10$^3$ | 3×10$^3$ | 15 | 800 | 160 | 16 |
| 40 | 1.2×10$^4$ | 4.8×10$^3$ | 24 | 1.28×10$^3$ | 256 | 25.6 |

**Table S7** Specification of sodium-ion battery coin cell using co-ESP electrodes, 25.4 mg cm$^{-2}$ cathode loading

| | Separator | Cathode | Anode |
|---|---|---|---|
| Active materials used | Polypropylene | NVPC/CNTF | Hard carbon/CNTF |
| Active materials loading [wt%] | N/A | 97.5 % | 98 % |
| Conductive additive [wt%] | | 2.5% | 2 % |
| Binder [wt%] | | | |
| Current collector [wt%] | | | |
| Areal loading [mg cm$^{-2}$] | | 25.4 | 3.9 |
| Compressed thickness [μm] | 25 | 136.1 | 92.1 |
| Area [cm²] | 2.3 | 0.78 | 1.77 |
| Total Weight [mg] | 3 | 20.3 | 7 |
| Electrolyte weight [mg]¹ | 11.6 | | |
| Energy density gravimetric [Wh kg$^{-1}$]² | 156.4 | Cell areal capacity 0.2 C [mAh cm$^{-2}$] | 2.63 |

1. The electrolyte weight is calculated by: $W_{electrolyte} = V_{cell} \varphi \rho_{electrolyte}$, $V_{electrode}$ is the volume of the electrodes and separator, $\varphi$ is the porosity of the electrodes and separator, $\rho_{electrolyte}$ is the density of the electrolyte.
2. Including the weight of electrode components, separator, and electrolyte

**Table S8** Specification of sodium-ion battery coin cell using co-ESP electrodes, 296 mg cm$^{-2}$ cathode loading

| | Separator | Cathode | Anode |
|---|---|---|---|
| Active materials used | Polypropylene | NVPC/CNTF | Hard carbon/CNTF |
| Active materials loading [wt%] | N/A | 97.5 % | 98 % |
| Conductive additive [wt%] | | 2.5% | 2 % |
| Binder [wt%] | | | |
| Current collector [wt%] | | | |
| Areal loading [mg cm$^{-2}$] | | 296 | 46 |
| Compressed thickness [µm] | 25 | 1586 | 1086 |
| Area [cm²] | 2.3 | 0.78 | 1.77 |
| Total Weight [mg] | 3 | 236.8 | 83 |
| Electrolyte weight [mg][1] | 135 | | |
| Energy density gravimetric [Wh kg$^{-1}$][2] | 124 | Cell areal capacity 0.2 C [mAh cm$^{-2}$] | 26.4 |

1. The electrolyte weight is calculated by: $W_{electrolyte} = V_{cell} \varphi \rho_{electrolyte}$, $V_{electrode}$ is the volume of the electrodes and separator, $\varphi$ is the porosity of the electrodes and separator, $\rho_{electrolyte}$ is the density of the electrolyte.
2. Including the weight of electrode components, separator, and electrolyte

**Table S9** Specification of sodium-ion battery coin cell using conventional electrodes

|  | Separator | Cathode | Anode |
|---|---|---|---|
| Active materials used | Polypropylene | NVPC | Hard carbon |
| Active materials loading [wt%] | N/A | 69.8 % | 29.3 % |
| Conductive additive [wt%] |  | 3.1 % | 0 % |
| Binder [wt%] |  | 4.6 % | 3.3 % |
| Current collector [wt%][1] |  | 22.5 % | 67.4 % |
| Areal loading [mg cm$^{-2}$] |  | 16 | 3.9 |
| Thickness [μm] | 25 | 110 | 62 |
| Area [cm$^2$] | 2.3 | 1.1 | 1.9 |
| Weight [mg] | 3 | 25.2 | 25.3 |
| Electrolyte weight [mg][2] | 10.1 |  |  |
| Energy density gravimetric [Wh kg$^{-1}$][3] | 89.5 | Cell areal capacity 0.2 C [mAh cm$^{-2}$] | 1.7 |

1. Estimated 15 μm-thick aluminium current collector weight: 4 mg cm$^{-2}$, 10 μm-thick copper current collector: 9 mg cm$^{-2}$.
2. The electrolyte weight is calculated by: $W_{electrolyte} = V_{cell} \varphi \rho_{electrolyte}$, $V_{electrode}$ is the volume of the electrodes and separator, $\varphi$ is the porosity of the electrodes and separator, $\rho_{electrolyte}$ is the density of the electrolyte, 1.26 g mL$^{-1}$.
3. Including the weight of electrode components, separator, and electrolyte

## Table S10 Performance comparison with published sodium-ion full batteries work

| Fabrication method | Cathode composition[1] | Max cathode areal loading/capacity (areal loading: rate, areal capacity)[2] | Rate Capacity and retention (discharge rate, specific capacity in mAh g$^{-1}$, capacity retention)[3] | Anode composition[1] | Full cell cycling (number of cycles, rate, capacity retention) | Max areal Energy Density (Wh cm$^{-2}$)/Power Density (W cm$^{-2}$)[4] | Max gravimetric Energy Density (Wh kg$^{-1}$)/Power Density (W kg$^{-1}$)[4] | Ref |
|---|---|---|---|---|---|---|---|---|
| **Co-ESP** | 97.5% Commercial NVP/CNT/CNF, no binder, no current collector | 296 mg cm$^{-2}$: 0.1C, 26.5 mAh cm$^{-2}$; 1C, 11 mAh cm$^{-2}$ | 0.1C, 89.4, 76.4%; 2C, 46.5, 39.7% | 98% Hard carbon-CNF/CNT, no binder, no current collector | 200 cycles, 0.2C, 84.7% | 77.7/133.1 | 191.9/328.8 | This work |
|  |  | 25.4 mg cm$^{-2}$: 0.1C, 2.64 mAh cm$^{-2}$; 50C, 0.47 mAh cm$^{-2}$ | 0.1C, 103.8, 88.7%; 50C, 42.7, 36.5% |  | 200 cycles, 0.2C, 96.2% | 8.05/248.4 | 231.6/7152.6 |  |
| Phase inversion | 76% NVP/carbon, no binder, no current collector | 7 mg cm$^{-2}$: 0.1C, 0.7 mAh cm$^{-2}$; 2C, 0.6 mAh cm$^{-2}$ | 0.1C, 102, 87.2%; 2C, 89, 76.1% | Hard carbon, composition un specified | 100 cycles, 1C, 99% | 2.14/3.71 | 147.2/254.1 | Ref 7 |
| Sol-gel | 51.9% NVP/CNF/C65/PVDF/Al foil | 8.5 mg cm$^{-2}$: 0.1C, 0.85 mAh cm$^{-2}$; 2C, 0.76 mAh cm$^{-2}$ | 0.1C, 99.7, 85.2%; 2C, 90, 76.9% | 51.9% NVP/CNF/C65/PVDF/Al foil | 3000 cycles, 2C, 60.1% | 1.47/195.5 | 45/5968.5 | Ref 66 |
| Precipitation | 72.6% NVP/CNF/C, no binder, no current collector | 7.6 mg cm$^{-2}$: 0.05C, 0.84 mAh cm$^{-2}$; 100C, 0.41 mAh cm$^{-2}$ | 0.05C, 111, 94.7%; 100C, 54.5, 46.6% | 76.5% NTP/CNF/C, no binder, current collector | 1000 cycles, 1C, 91% 4000 cycles, 20C, 74.5% | 1/72.2 | 49.3/3538.7 | Ref 69 |
| Slurry-casting | 56% NVPF/GO/CB/PVDF/Al foil | 8 mg cm$^{-2}$: 0.78C, 0.88 mAh cm$^{-2}$; 6.3C, 0.74 mAh cm$^{-2}$ | 0.78C, 110, 85.9%; 6.3C, 92.5, 72.3% | 18.7% SnP/GO/CB/PA/Al foil | 200 cycles, 0.78C, 62% | 2.13/18.4 | 109.5/1840 | Ref 67 |
| Slurry-casting | 25.6% NVP@C@CNT/CB/PVDF/Al foil | 1.5 mg cm$^{-2}$: 0.05C, 0.77 mAh cm$^{-2}$; 25C, 0.45 mAh cm$^{-2}$ | 0.05C, 102, 87.2%; 25C, 60, 51.2% | 25.7% mesocarbon/CB/CMC-SBR/Al foil | 5000 cycles, 5C, 72.7% | 0.37/74.2 | 31.3/6366 | Ref 5 |
| Slurry-casting | 31.5% HNVP/CB/PVDF/Al foil | 2 mg cm$^{-2}$: 0.5C, 0.18 mAh cm$^{-2}$; 20C, 0.12 mAh cm$^{-2}$ | 0.5C, 90.8, 82.7%; 20C, 61.8, 56.3% | 31.5% NTP/CB/PVDF/Alfoil | 700 cycles, 2C, 88.8% | 0.23/5.27 | 17.9/414.9 | Ref 61 |

1. Calculated from the highest areal loading reported in the publication that has rate and cycling data. Estimated 15 μm-thick aluminium current collector weight: 4 mg cm$^{-2}$, 10 μm-thick copper current collector: 9 mg cm$^{-2}$
2. Showing the areal capacity from the lowest and highest cycling rate.
3. The specific capacity of the highest areal loading reported in the publication, capacity retention at different rate; NMC811's theoretical capacity is 200 mAh g$^{-1}$; LCO's theoretical capacity is 170 mAh g$^{-1}$.
4. The energy/power densities are re-calculated considering the weight and volume of all electrode components, including active material, current collector, binder, conductive additive, based on the data reported in the literature. The method of calculation is shown in Supplementary Appendix 4.

## Table S11 Performance comparison with published lithium-ion full batteries work

| Fabrication method | Cathode composition[1] | Max cathode areal loading/capacity (areal loading: rate, areal capacity)[2] | Rate Capacity and retention (discharge rate, specific capacity in mAh g$^{-1}$, capacity retention)[3] | Anode composition[1] | Full cell cycling (number of cycles, rate, capacity retention) | Max areal Energy Density (Wh cm$^{-2}$)/Power Density (W cm$^{-2}$)[4] | Max gravimetric Energy Density (Wh kg$^{-1}$)/Power Density (W kg$^{-1}$)[4] | Ref |
|---|---|---|---|---|---|---|---|---|
| **Co-ESP** | **97.5% Commercial NVPC/CNTF, no binder, no current collector** | **296 mg cm$^{-2}$: 0.1C, 26.5 mAh cm$^{-2}$; 11 mAh cm$^{-2}$, 2C** | **0.1C, 89.4, 76.4%; 2C, 46.5, 39.7%** | **98% Hard carbon/CNTF, no binder, no current collector** | **200 cycles, 0.2C, 84.7%** | **77.7/133.1** | **191.9/328.8** | This work |
| | | **25.4 mg cm$^{-2}$: 0.1C, 2.64 mAh cm$^{-2}$; 50C, 0.47 mAh cm$^{-2}$** | **0.1C, 103.8, 88.7%; 50C, 42.7, 36.5%** | | **200 cycles, 0.2C, 96.2%** | **8.05/248.4** | **231.6/7152.6** | |
| Slurry-casting | 80% LCO/cellulose/CNT, no current collector | 30 mg cm$^{-2}$: 0.1C, 4.6 mAh cm$^{-2}$; 1C, 2.8 mAh cm$^{-2}$ | 0.1C, 152, 89.4%; 1C, 93.3, 54.9% | 80% LTO/cellulose/CNT | N/A | 10.6/10.5 | 141.3/140 | ref |
| Slurry-casting | 94.5% NMC811/biopolymer/CNT/Al foil | 54.4 mg cm$^{-2}$: 0.1C, 9.24 mAh cm$^{-2}$ No rate data | 0.1C, 169.9, 85% No rate data | Graphite/biopolymer/CNT/Cu foil, unknown active content, max 76.9% | 40 cycles, 0.1C, 92.8% | 33.3/3.6 | 367.9/66.2 | ref |
| Slurry-casting | 96.4% NMC811/CNT/Al foil | 156 mg cm$^{-2}$: 1/15C, 29 mAh cm$^{-2}$; 1C, 15 mAh cm$^{-2}$ | 0.05C, 185.9, 93%; 1C, 96.2, 48.1% | 52.1% Si/CNT/Cu foil | 47 cycles, 1/15C, 83% | 98.3/105.8 | 569.3/586.6 | ref |
| Impregnation | 62.2% LFP/Super P/PVDF/CNT-polyester current collector | 168 mg cm$^{-2}$: 1/15C, 26 mAh cm$^{-2}$; No rate data | 1/15C, 155, 91.2% No rate data | 61.1% LTO/Super P/PVDF/CNT-polyester current collector | 33 cycles, 0.1C, 86% | 46.8/3.4 | 85.7/6.2 | ref |
| Extrusion | 93.9 % LFP/carbon/Al foil | 90 mg cm$^{-2}$: 1/24C, 13.5 mAh cm$^{-2}$; 1/12C, 11.1 mAh cm$^{-2}$ | 1/24C, 150, 88.2%; 1/12C, 123, 72.4% | 92.2% LTO/carbon/Cu foil | 100 cycles, 1/12C, 67% | 23.6/14.2 | 121.9/73.3 | ref |
| Slurry-casting | 80% LFP/carbon/EVA, no current collector | 29.4 mg cm$^{-2}$: 0.1C, 4.6 mAh cm$^{-2}$; 1C, 3.5 mAh cm$^{-2}$ | 0.1C, 156.5, 92%; 1C, 120, 70.5% | 80% LTO/carbon/EVA | 60 cycles, 0.5C, 76.9% | 8/8.5 | 108.8/115.6 | ref |
| Slurry-casting | 65% LFP/CB/PVDF, no current collector | 36 mg cm$^{-2}$: 0.05C, 5.5 mAh cm$^{-2}$; 0.1C, 5 mAh cm$^{-2}$ | 0.05C, 152.8, 89.9%; 0.1C, 138.9, 81.7% | 59% LTO/CB/PVDF | 500 cycles, 0.1C, 87% | 9.9/1.1 | 85.3/9.5 | ref |

1. Calculated from the highest areal loading reported in the publication that has rate and cycling data. Estimated 15 µm-thick aluminium current collector weight: 4 mg cm$^{-2}$, 10 µm-thick copper current collector: 9 mg cm$^{-2}$
2. Showing the areal capacity from the lowest and highest cycling rate.
3. The specific capacity of the highest areal loading reported in the publication, capacity retention at different rate; NMC811's theoretical capacity is 200 mAh g$^{-1}$; LCO's theoretical capacity is 170 mAh g$^{-1}$.
4. The energy/power densities are re-calculated considering the weight and volume of all electrode components, including active material, current collector, binder, conductive additive, based on the data reported in the literature. The method of calculation is shown in Supplementary Appendix 4.

# Supplementary Appendix 3: Calculating Energy Density

- The gravimetric energy densities were calculated by: $E_g = \bar{V} C_g c_{active}$. $\bar{V}$ is the average voltage in the discharge process; $C_g$ is the gravimetric specific capacity, considering only the weight of active materials; $c_{active}$ is the active content, calculated by :

- $c_{active} = \dfrac{m_{active}}{m_{active} + m_{binder} + m_{current\ collector} + m_{conductive\ additive}}$. Estimated 15 μm-thick aluminium current collector weight: 4 mg cm$^{-2}$, 10 μm-thick copper current collector: 9 mg cm$^{-2}$

- The gravimetric power densities were calculated by: $P_g = I_g \bar{V} c_{active}$. $I_g$ is the discharge gravimetric current density, calculated by : $I_g = \dfrac{I}{m_{active}}$. $I$ is the discharge current.

- The areal energy densities were calculated by: $E_a = \bar{V} C_a$. $C_a$ is the areal capacity.

- The areal power densities were calculated by: $E_a = \bar{V} I_a$. $I_a$ is the areal current, calculated by $I_a = \dfrac{I}{A_{electrode}}$, $A_{electrode}$ is the geometric area of the electrode.

- The energy densities in all referenced literatures in **Figure 4h, i, Figure 5e, f, Table S4, S5, S10, S11** were recalculated with the above formula, to make consistent comparisons.

- To calculate the gravimetric energy/power density considering the electrolyte and separator in **Figure S34**, we substitute the $c_{active}$ in the equation with $c'_{active} = \dfrac{m_{active}}{m_{active} + m_{binder} + m_{current\ collector} + m_{conductive\ additive} + m_{electrolyte} + m_{separator}}$. $m_{separator}$ is 3.1 mg for a 2.5 cm² Celgard 2400 separator. $m_{electrolyte} = \rho_{electrolyte} v_{composite} P$, P is the porosity of the electrodes/separator composite, $v_{composite}$ is the volume of the composite, $\rho_{electrolyte}$ is the electrolyte's density, 1.26 g cm$^{-3}$.

- To calculate the gravimetric energy density of pouch cells, the $c_{active}^{pouch} = \dfrac{m_{active}}{m_{pouch}}$. The volumetric energy density of the pouch cells were calculated by: $E_g = \dfrac{\bar{V} C}{A_{pouch} L_{pouch}}$. $A_{pouch}$ is the geometric area of the pouch cell, $L_{pouch}$ is the thickness.